\documentclass[aps,prf,preprint,groupedaddress]{revtex4-2}
\usepackage[utf8]{inputenc} 
\usepackage{natbib}
\usepackage{subfig}
\usepackage{graphicx} 
\usepackage[svgnames,dvipsnames]{xcolor}  
\usepackage{fancyhdr}
\usepackage{fancybox}
\usepackage[labelfont=bf]{caption} 
\usepackage[T1]{fontenc} 
\usepackage{comment}
\usepackage{tikz}
\usetikzlibrary{shapes.geometric,calc}
\usetikzlibrary{shapes}
\usepackage{pgfplots} 

\usepackage{amsmath,amsfonts,amssymb}
\usepackage{float} 

\usepackage{colortbl} 
\usepackage[colorlinks=false,linkcolor=blue ,filecolor=red ]{hyperref}

\begin{document}

\title{Inertial settling of an arbitrarily oriented cylinder in a quiescent flow : from short-time to quasi-steady motion}
\author{Jean-Lou Pierson}
\email{jean-lou.pierson@ifpen.fr}
\affiliation{IFP Energies Nouvelles, Rond-point de l'échangeur de Solaize, 69360 Solaize, France}


\begin{abstract}

In this article, we investigate the inertial settling of an arbitrarily oriented cylinder settling under gravity. We focus on two regimes: the very short-time and long-time dynamic. By using the generalized Kirchhoff equations to describe the particle motion,  we demonstrate that during the very short dynamic regime, a cylinder starting from rest behaves with sedimenting velocities and angular velocity proportional to $t$ and $t^3$, respectively. We then explore the long-time behaviour and evaluate the validity of the quasi-steady assumption under which the fluid unsteady term can be neglected. Using a dimensional analysis, we establish that the quasi-steady assumption is only applicable to Reynolds numbers much smaller than one. However, by comparing the results of quasi-steady models to recent experiments and direct numerical simulations, we demonstrate that this assumption is valid for a broader range of Reynolds numbers, particularly for long fibres. We also analyze the effect of particle inertia. We show particle inertia plays no significant role in the magnitude of the sedimenting velocities and angular velocity. However, for sufficiently large inertia we reveal that the quasi-steady model takes the form of a damped oscillator when the particle approaches its equilibrium position, which is broadside on to its direction of motion. We discuss the relevance of this solution in light of direct numerical simulations.

\end{abstract}

\maketitle
\section{Introduction}


The settling of anisotropic particles is a common occurrence in various environmental flows, such as the fall of microplastics in the ocean \citep{poulain2018} or the precipitation of ice crystals in the atmosphere \citep{gustavsson2019,gustavsson2021}. Despite its practical significance, the accurate modeling of anisotropic particle settling in turbulent or quiescent environments is challenging due to the coupling between the particle motion and the surrounding fluid flow. Unlike spherical particles, the orientation of the body has a significant impact on the rate of sedimentation. For example, in Stokes flow and for slender particles, the sedimentation velocity of a particle with its axis aligned with gravity is twice that of a particle with its axis perpendicular \citep{batchelor1970}. Additionally, for inertial flows, the orientation of the body is coupled to the translational equation of motion due to a non-zero hydrodynamic torque \citep{cox1965}. This results in an unsteady problem, as the orientation of the body can change over time in response to torques. In this article, we examine the settling of a cylindrical particle in a quiescent flow as a first step in understanding the effect of an anisotropic shape on particle motion.


The most general equations for studying the gravitational settling of a single body in a quiescent fluid are the generalized Kirchhoff equations originally derived by \citet{howe1995}. Under this framework, added mass and vorticity contributions to the hydrodynamic forces and torque are non-ambiguously separated. However, for most configurations of practical interest, the vorticity contributions cannot be expressed in closed form as they depend on fluid motion history \citep{ern2012}. In the limit of negligible inertia and for a spherical particle translating and rotating, the force and torque can be decomposed into a quasi-steady component and a history term that takes the form of integro-differential equations \citep{kim2013}. To date, there is no equivalent analytical formula for an arbitrary axisymmetric particles, particularly cylinders. The explanation lies in the complexity of the history term for a non-spherical body, whose expression in the frequency domain is often too complicated to allow a closed-form expression in the time domain \citep{loewenberg1993,kabarowski2020}. Moreover, unlike spherical particles, non-spherical particles have distinct high and low frequency expressions for the history term \citet{lawrence1988}.

The situation is even worst for finite Reynolds numbers for which very few results exist for non-spherical particles both for the quasi-steady and history loads. In contrast to Stokesian flow, for which, due to the reversibility of the Stokes equation, an axisymmetric particle with fore-aft symmetry embedded in a uniform flow of velocity $U$ experiences no torque, an inertial torque appears for finite Reynolds number which scales as $U^2$ \citep{cox1965,khayat1989}. This torque naturally induces a coupling between translation and rotation for a sedimenting cylinder. Hence as a cylinder sediment in a fluid with non-negligible inertia, it rotates toward its equilibrium orientation, which is broad-side on to its direction of motion \citet{khayat1989}. There is another nonlinear coupling term in the force balance for a rotating and translating axisymmetric body with fore-aft symmetry which scales as $\Omega U$ where $\Omega$ is the angular velocity \citet{cox1965}. This term is at the origin of the lift force on a spinning sphere translating perpendicularly to its rotation axis \citep{rubinow1961}. This coupling term has not been studied so far in the context of a rotating cylinder settling perpendicular to its rotation axis. The history loads for non-spherical particles in the inertial regime have also received limited attention, with only a few studies providing force expressions for arbitrarily shaped particles in the long-time limit \citep{lovalenti1993}. These expressions require knowledge of the steady velocity field created by the particle in Stokes flow, which is unknown for a moderately long cylinder. Nevertheless scaling arguments indicate a $t^{-2}$ long-time decay of the history force in the finite-inertia regime and a slower $t^{-1/2}$ decay in the Stokes regime \citep{lovalenti1993}.

Based on the previous literature review, it is challenging to make analytical advancements without additional assumptions, primarily due to the absence of closed-form expressions for the history terms and $\Omega U$ load contribution. The quasi-steady assumption, introduced by \citet{cox1965} in his investigation of the settling of a small-eccentricity spheroid, posits that unsteady terms related to fluid motion can be ignored. Through a scaling analysis, \citet{cox1965} demonstrated that this assumption is appropriate as long as the Reynolds number based on the body length is much smaller than unity. This assumption has been widely utilized in various practical configurations, including the settling of fibrous aerosols in quiescent air \citet{newsom1994} and fibres in liquids \citet{roy2019}, yielding satisfactory results when compared to experiments. Also, in those applications, the Reynolds number was not necessarily small. \citet{shin2006} have even shown that the quasi-steady theory remains accurate for Reynolds number based on the body length close to one. Most prior studies \citep{newsom1994,roy2019} have made use of the quasi-steady loads derived by \citet{khayat1989} for slender fibers and the leading-order hydrodynamic torque resisting rotation provided by the slender body theory \citet{batchelor1970}. Recent research has shown that the lift force and inertial torque provided by \citet{khayat1989} and the leading-order expression for the hydrodynamic torque resisting rotation are not accurate for moderately long rods \citep{pierson2021,fintzi2023}, in line with the qualitative but non-quantitative agreement reported by \citet{cabrera2022} between the \citet{khayat1989} theory and their experiments for moderately long rods. Hence, the validity of the quasi-steady assumption must be re-evaluated with accurate formulas for the loads, and the inertial correction to the loads proportional to $\Omega U$ must be considered. Scaling analysis by \citet{cox1965} and \citet{pierson2021} has shown that this term may be small compared to the $U^2$ contribution when the inertia effect is not significant. However, the magnitude of this term for moderate inertia remains a topic of debate \citep{pierson2021}.

There is another limit where analytical progress is possible. In the limit of time shorter than the viscous time scale, added mass effects dominate over viscous contribution \citep{mougin2002}. In this limit, the equations of motion can be solved provided that the added mass coefficients are known. \citet{loewenberg1993} has established the added mass forces through the application of potential flow solutions. However, the added mass torque coefficient for a rotating cylinder has no known solution at present.

The primary objective of this article is to address two key issues in the study of an arbitrarily oriented cylinder settling under gravity. Firstly, analytical solutions are provided for the problem in the short-time limit where the added mass effects are dominant. Secondly, the validity range of the quasi-steady assumption and the neglect of the $\Omega U$ load contribution assumptions are investigated as a function of the relevant dimensionless parameters. These parameters include the cylinder aspect ratio, the Archimedes number which is a Reynolds number based on gravitational velocity and the density ratio. The analysis is conducted using the generalized Kirchhoff equations as proposed by \citet{howe1995} and \citet{mougin2002}, and the quasi-steady load expressions derived in \citet{fintzi2023}, \citet{pierson2021} for moderately long rods and \citet{khayat1989} for very elongated fibers. The proposed model is validated through scaling analysis, experimental measurements from \citet{roy2019} and \citet{cabrera2022} and direct numerical simulations. The article is structured as follows. The governing equations are presented in Section \ref{sec:gov}. In Section \ref{sec:short}, the analytical solutions for the short-time limit are described. The quasi-steady models are derived in Section \ref{sec:scaling}. The comparison between the quasi-steady models and experimental measurements from \citet{roy2019} and \citet{cabrera2022} as well as direct numerical simulations are discussed in Section \ref{sec:exp_dns}. Section \ref{sec:discussion} contains a discussion on the validity of the quasi-steady model as well as our conclusions.

\section{Governing equations}
\label{sec:gov}
\begin{figure}[h]
\centering
\begin{tikzpicture}[scale=1.]
\node[cylinder, 
    draw = black, 
    text = black,
    cylinder uses custom fill, 
    cylinder body fill = black!10, 
    cylinder end fill = black!40,
    aspect = 0.2, 
    shape border rotate = 90, rotate =225, minimum height=3.5cm, minimum width=0.8cm] (c) at (2,2,2) {};


\draw[->] (2,2,2)   -- (3,1,2) node [at end, right]   {$\mathbf{p}$};
\draw[->] (2,2,2)   -- (3,3,2) node [at end, right]   {$\mathbf{q}$};
\draw[->] (2,2,2)   -- (1,2,3) node [at end, left]   {$\mathbf{r}$};


\draw[->, thick] (2,2,2)--(2.5,0,2) node[right]{$\mathbf{U}$};

\draw[-, dashed] (2,2,2)--(2,0,2);

\draw[->] (2.25,1,2) arc (-80:-40:0.8)node[right]{$\theta$};
\draw[->] (2,1.5,2) arc (-90:-45:0.5)node[right]{$\phi$};
\draw[->, thick] (5,2.5,2)--(5,1,2) node[left]{$\mathbf{g}$};


\draw[->] (-1,0,0)--(-1,-1,0)node[below]{$z$};
\draw[->] (-1,0,0)--(0,0,0)node[right]{$y$};

\end{tikzpicture}
\caption{Finite-length cylinder submitted to the gravity acceleration $\mathbf{g}$.}
\label{fig:single_cyl}
\end{figure}
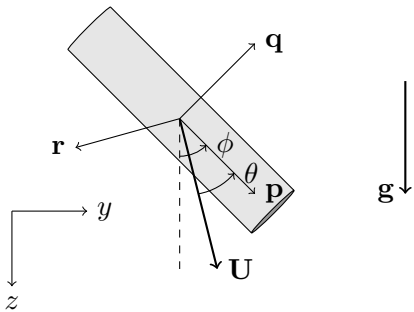
We consider a finite-length cylinder of length $L$ and diameter $D$ settling under gravity with velocity $\mathbf{U}$ and angular velocity $\mathbf{\Omega}$ (Figure \ref{fig:single_cyl}). The main difficulty in studying this problem is the coupling between the body motion and the surrounding flow field which satisfies Navier-Stokes equations \citep{ern2012}.  The equations of motion expressed in a reference frame having its origin fixed with respect to the laboratory but axes rotating with the body read \citep{howe1995,mougin2002}


\begin{equation}
(m \mathbb{I} + \mathbb{A})\frac{\mathrm{d}\mathbf{U}}{\mathrm{d}t} + \boldsymbol{\Omega} \times ((m \mathbb{I} + \mathbb{A})\mathbf{U})  = \mathbf{F}^\omega+(m - \rho V) \mathbf{g},
\label{eq:forcebal}
\end{equation}

\begin{equation}
(\mathbb{J} + \mathbb{D})\frac{\mathrm{d}\boldsymbol{\Omega}}{\mathrm{d}t} + \boldsymbol{\Omega} \times ((\mathbb{J} + \mathbb{D})\boldsymbol{\Omega}) + \mathbf{U}\times (\mathbb{A}\times \mathbf{U})  = \mathbf{T}^\omega,
\label{eq:torquebal}
\end{equation}
where $m$ and $V$ are respectively the cylinder mass and volume, $\mathbb{I}$ is the identity matrix, $\mathbb{J}$ is the inertia tensor, $\mathbb{A}$ and $\mathbb{D}$ are the second-order added inertia tensors. Using indicial notation all those tensors, can be written as $J_{ij} = J_p p_ip_j+J_q(\delta_{ij}-p_ip_j)$, $A_{ij}= A_p p_ip_j+A_q(\delta_{ij}-p_ip_j)$ and $D_{ij} = D_q(\delta_{ij}-p_ip_j)$ where $\mathbf{p}$ is the unit orientation vector and $\mathbf{q}$ is the unit vector perpendicular to $\mathbf{p}$ in the plane $(y,z)$ (Figure \ref{fig:single_cyl}). If the components of the inertia tensor can be readily obtained in a closed form for a finite-length cylinder ($J_p = mD^2/8$, $J_q = m(3D^2/4+L^2)/12$) this is not the case for the components of the added mass tensors $A_p, A_q$ and $D_q$. If $A_p$ and $A_q$ have been already studied in the litterature \citep{loewenberg1993}, to the best of the author knowledge, no expression for $D_q$ has not been published yet. Based on the direct numerical simulation results of \citet{kharrouba2020} and the potential flow results of \citet{loewenberg1993} we derive correlations for $A_p$, $A_q$ and $D_q$ in Appendix \ref{app:added}. In equations \ref{eq:forcebal} and \ref{eq:torquebal}, $\mathbf{F}^{\omega}$ and $\mathbf{T}^{\omega}$ are the force and torque due to the vorticity in the flow. Except in inertia dominated regimes the motion of the cylinder is planar in the $(p,q)$ plane \citep{cabrera2022} and the equation of motion simplifies to  


\begin{align}
    (m + A_p) \frac{d U_p}{dt}
    - (m + A_q)\Omega _r U_q 
    &= F_p^{\omega}
    + (m-\rho V)g\cos \phi, \label{eq:Up}\\
    (m + A_q) \frac{d U_q}{dt}
    + (m + A_p)\Omega _r U_p 
    &= F_q^{\omega}
    - (m-\rho V)g\sin \phi,
\label{eq:Uq}\\
    (J_{q} + D_{q})
    \frac{d\Omega _r}{dt}
    &= -U_pU_q(A_q-A_p) + T_r^{\omega}.
\label{eq:Omegar}
\end{align}

The physical origin of the torque can be exemplified by looking more closely at equation \ref{eq:Omegar}. In steady potential flow, a torque manifests on a cylindrical particle as indicated by the first term on the right-hand side of equation \ref{eq:Omegar} \citep{howe2006}. As the aspect ratio, $\chi = L/D$, exceeds 1, $A_q > A_p$ (see Appendix \ref{app:added}), and the torque is positive, orienting the body broad-side on.  As a result, to leading order the torque expression provided by \citet{khayat1989} for small Reynolds number and large aspect ratio ($\chi \gg 1$) $T_r = -5\pi/24\rho U_pU_qL^3/\ln ^2(\chi)$  is the sum of two positive contribution due to potential flow and vorticity. For $\chi \gg 1$ the potential contribution $-U_pU_q(A_q-A_p)$ scales as $-\rho \pi D^2U_pU_qL$ and is thus negligible to leading order in comparison to the vorticity contribution. The situation is less obvious for moderately large aspect ratio ($\chi \approx 2$) for which both contributions may have the same order of magnitude since the total torque on the body do not scales as $\rho U_pU_qL^3/\ln ^2(\chi)$ \citep{fintzi2023}.

\section{Short-time dynamics}

\label{sec:short}
In the high-frequency limit or equivalently for time shorter than the diffusive scale added mass effect dominates over viscous contribution. This may be proved by deriving the unsteady loads in the Stokes regime \citet{kabarowski2020}, but also by performing a short time-analysis of the Navier-Stokes equation \citet{mougin2002}. The latter is more general as it is not limited to the Stokes flow regime. The proper length scale to be used in the diffusive scale for the problem at hand is unknown \textit{a priori}. However, \citet{kabarowski2020} have shown that for both transverse and longitudinal oscillations, the added mass contributions dominate over the history load if $t \ll D^2 /\nu$. Hence, the proper length scale in the short dynamic is the cylinder diameter, at least in the Stokes flow regime. Assuming $t \ll D^2 /\nu$, equations \ref{eq:Up} - \ref{eq:Omegar} simplify to





\begin{align}
    (m + A_p) \frac{d U_p}{dt}
    - (m + A_q)\Omega _r U_q 
    &= (m-\rho V)g\cos \phi, \label{eq:Upv}\\
    (m + A_q) \frac{d U_q}{dt}
    + (m + A_p)\Omega _r U_p 
    &= - (m-\rho V)g\sin \phi,
\label{eq:Uqv}\\
    (J_{q} + D_{q})
    \frac{d\Omega _r}{dt}
    &= -U_pU_q(A_q-A_p) ,
\label{eq:Omegarv}\\
\frac{d \phi}{dt} &=\Omega _r \label{eq:Phiv},
\end{align}
where we made explicit the equation ruling the dependency of $\phi$ with time. To the best of our knowledge, there is no closed form analytical solution to this non-linearly coupled system of equations, although there is a straightforward analytical treatment in the case of zero gravity \citet{lamb1953}. We have to solve this system numerically but one may get insightful estimates using asymptotic analysis. By balancing the acceleration of gravity with the particle acceleration in equations \ref{eq:Upv} and \ref{eq:Uqv} one get $U \sim gT$ where $U$ is a characteristic velocity scale, and $T$ a characteristic timescale. Injecting this scaling in equation \ref{eq:Omegarv}, and since $(A_q-A_p) / (I_{q} + D_{q}) \sim 1 /L^2$ for $\chi \gg 1$  one get the characteristic angular velocity $\Omega \sim g^2T^3/L^2$. By using this estimates in equations \ref{eq:Upv}, \ref{eq:Uqv}, \ref{eq:Omegarv} and \ref{eq:Phiv} and defining the dimensionless (starred) quantities as $U_p = U U_p^*$, $U_q = U U_q^*$ and $\Omega_r = \Omega ^*\Omega_r$ we obtain






\begin{align}
 \frac{d U_p ^*}{dt^*}
    - \epsilon \frac{\mathcal{A}}{\mathcal{B}}\Omega _r ^* U_q ^* 
    &= \mathcal{A}\cos \phi, \label{eq:Upvd}\\
    \frac{d U_q ^*}{dt^*}
    + \epsilon\frac{\mathcal{B}}{\mathcal{A}}\Omega _r ^* U_p ^* 
    &= -\mathcal{B}\sin \phi, \label{eq:Uqvd}\\
    \frac{d\Omega _r^*}{dt^*}
    &= -\mathcal{C}U_p^*U_q^* ,
\label{eq:Omegarvd}\\
\frac{d \phi}{dt^*} &=\epsilon\Omega _r^* \label{eq:Phivd},
\end{align}
where $\mathcal{A}=(\bar{\rho}-1)/(\bar{\rho}+A_p^*)$, $\mathcal{B}=(\bar{\rho}-1)/(\bar{\rho}+A_q^*)$, $\mathcal{C}=(A_q^*-A_p^*)/(\bar{\rho}J_q^*+D_q^*)$, $A_p^* = A_p / (\rho V), A_q^*= A_q / (\rho V)$, $J_q^* = J_q/ (\rho VL^2)$, $D_q^* = D_q/ (\rho VL^2)$, $\bar{\rho}=\rho _p/\rho$ is the density ratio and $\epsilon = g^2T^4/L^2$ can be understood as the ratio of the characteristic timescale over a gravity time scale. In the following, we consider the small time limit $T \ll (L/g)^{1/2}$   or equivalently $\epsilon \ll 1$. We seek for solutions of equations \ref{eq:Upvd} - \ref{eq:Phivd} in the form of asymptotic expansions in powers of the small parameter $U_p^* = U_p^{*(0)}+\epsilon U_p^{*(1)}+...$, $U_q^* = U_q^{*(0)}+\epsilon U_q^{*(1)}+...$, $\Omega _r^* = \Omega _r^{*(0)}+\epsilon \Omega_r^{*(1)}+...$ and $\phi = \phi^{(0)}+\epsilon \phi^{(1)}+...$. The calculations are straightforward and are detailed in appendix \ref{app:short}. The solutions up to the order 1 with respect to the small parameter $\epsilon$ are 

\begin{align}
U_p ^{*} &=   \mathcal{A}t^*\cos \phi^{(0)} -\frac{\epsilon}{12}\mathcal{A}^2\mathcal{B}\mathcal{C}t^{*5}\cos \phi^{(0)} \sin ^2\phi^{(0)}+\mathcal{O}(\epsilon ^2) \label{eq:upt}, \\
U_q ^{*} &=   -\mathcal{B}t^*\sin \phi^{(0)}-\frac{\epsilon}{12}\mathcal{A}\mathcal{B}^2\mathcal{C}t^{*5}\cos ^2\phi^{(0)} \sin \phi^{(0)}+\mathcal{O}(\epsilon ^2) \label{eq:uqt},\\
\Omega _r^{*} &= \frac{\mathcal{A}\mathcal{B}\mathcal{C}}{3}t^{*3}\cos \phi^{(0)} \sin \phi^{(0)}+\mathcal{O}(\epsilon ^2),\\
\phi^{}&=\phi^{(0)} + \epsilon\frac{\mathcal{A}\mathcal{B}\mathcal{C}}{12}t^{*4}\cos \phi^{(0)} \sin \phi^{(0)} +\mathcal{O}(\epsilon ^2).
\end{align}



\begin{figure}[h!]
    \centering
        \includegraphics[height=0.23\textwidth]{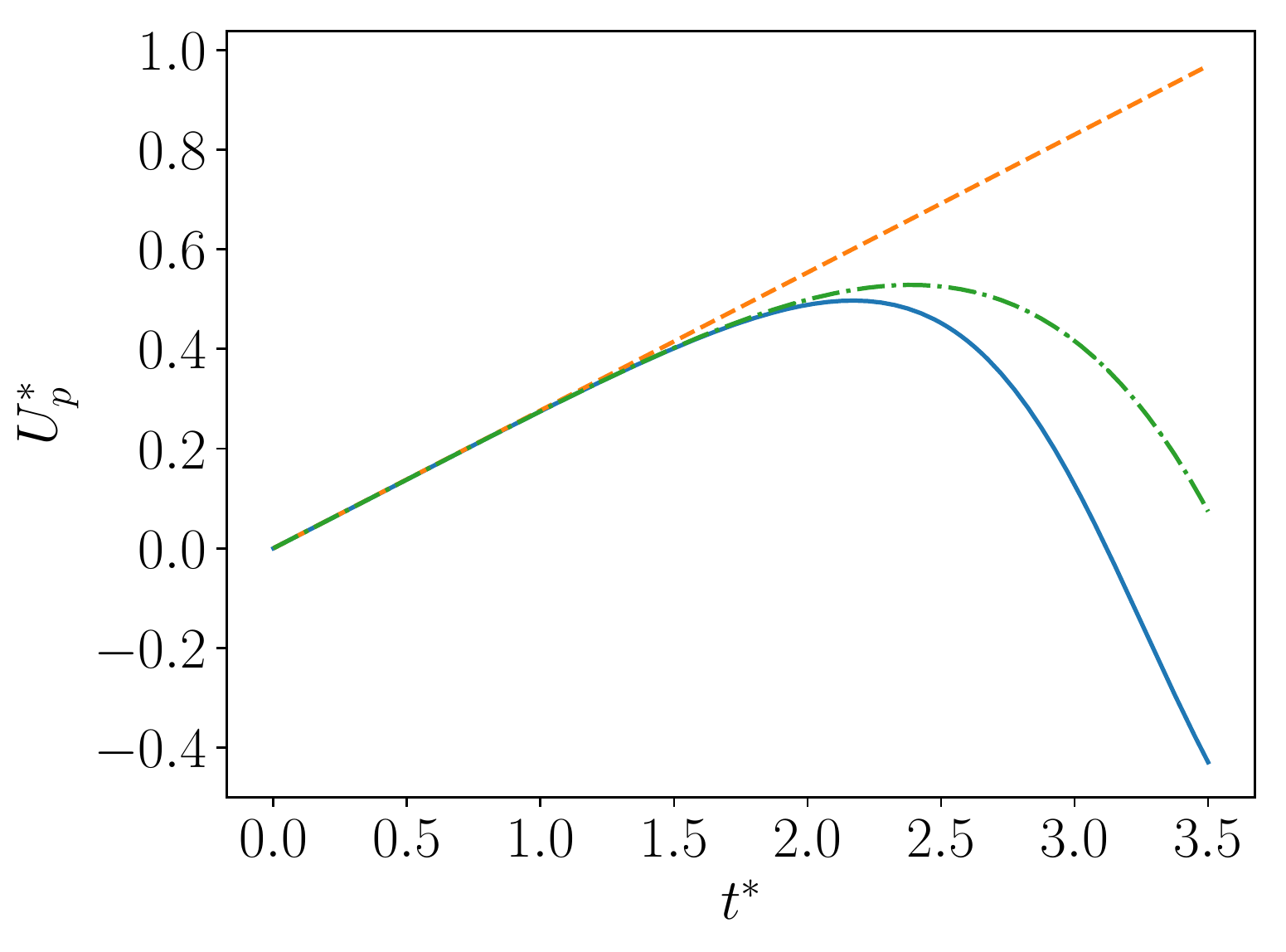}
        \includegraphics[height=0.23\textwidth]{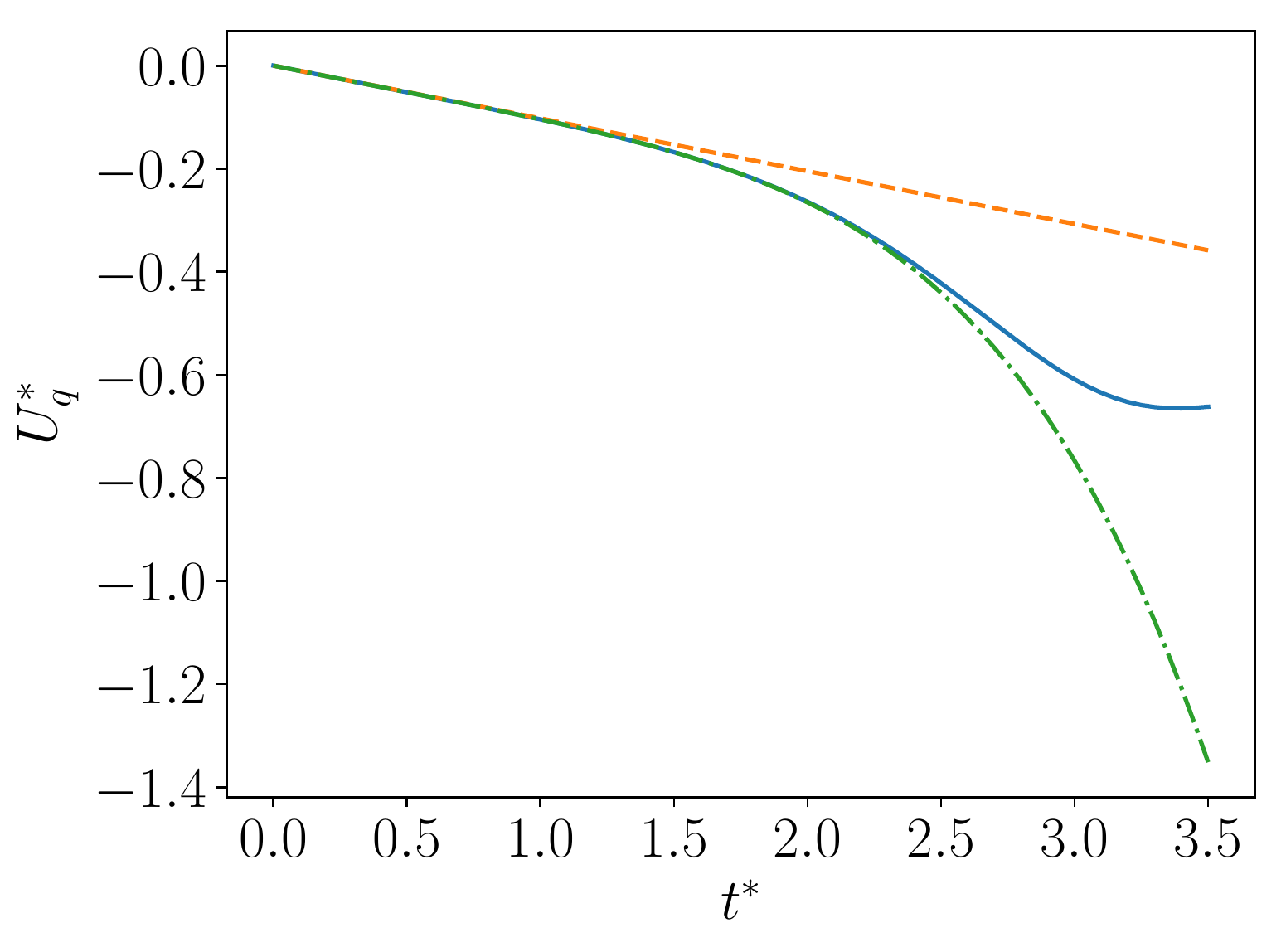}
        \includegraphics[height=0.23\textwidth]{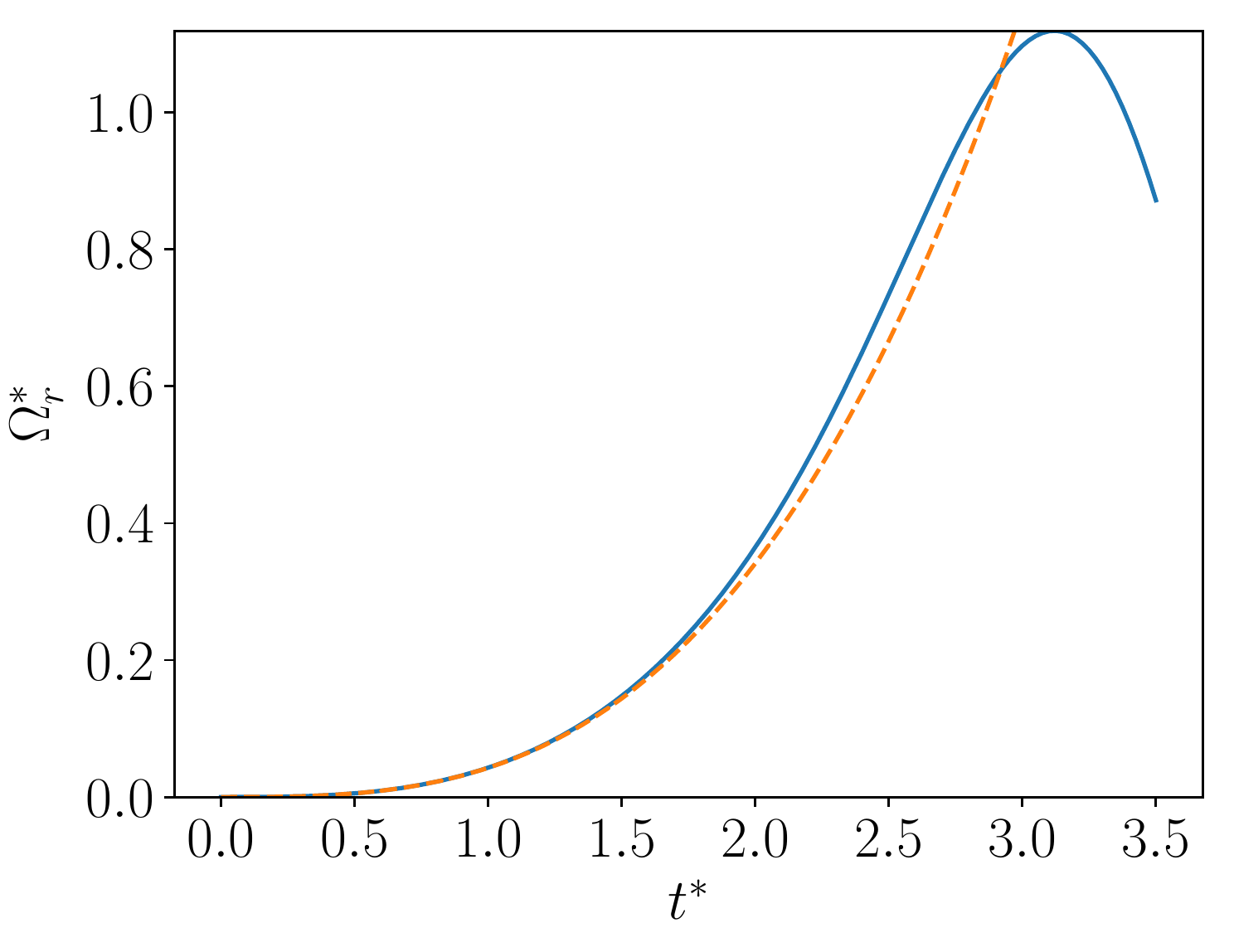}\\
        \vspace{-3mm}
        \hspace{0cm}(a) \hspace{5cm} (b) \hspace{5cm} (c)\\
        \includegraphics[height=0.23\textwidth]{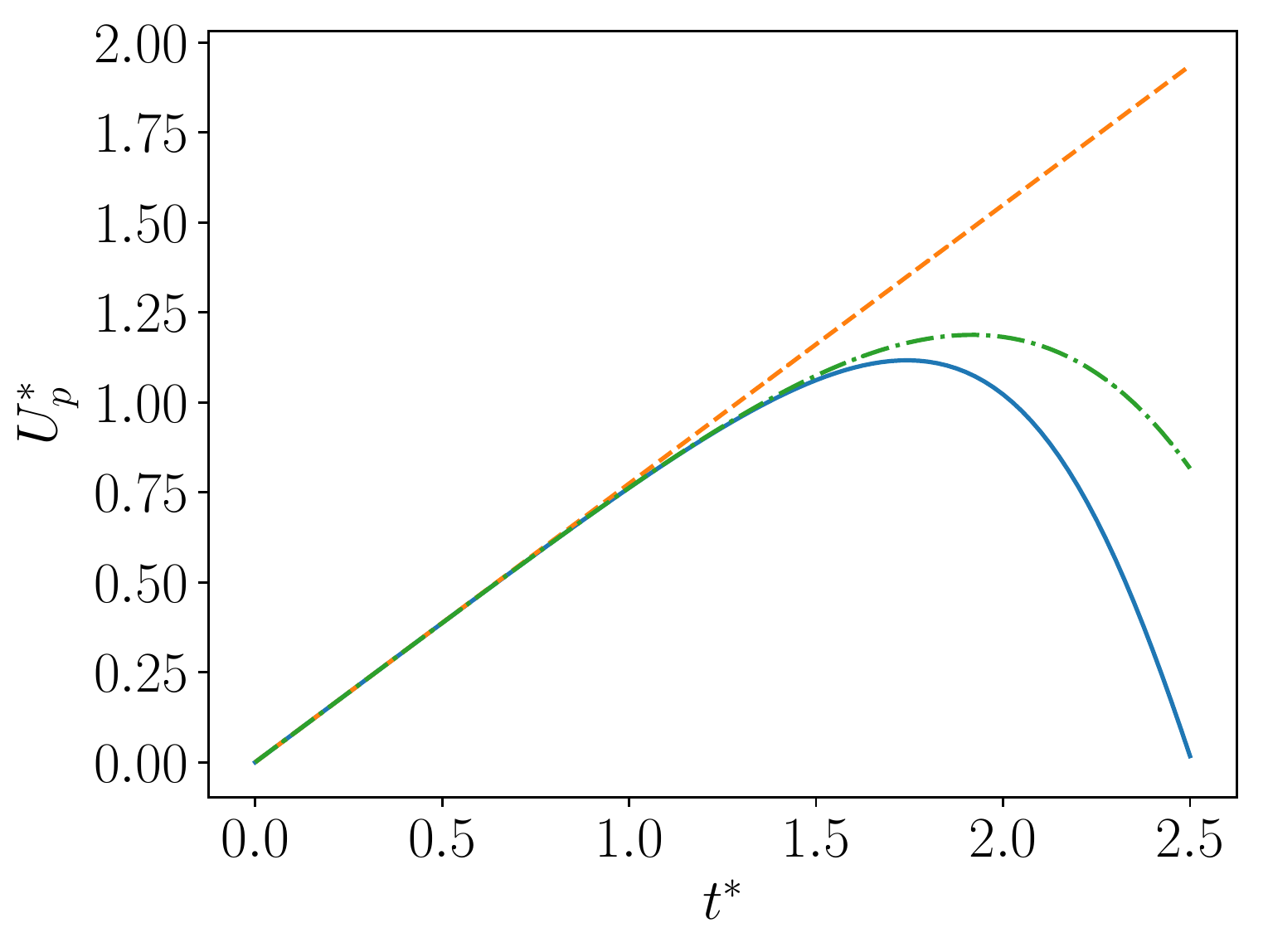}
        \includegraphics[height=0.23\textwidth]{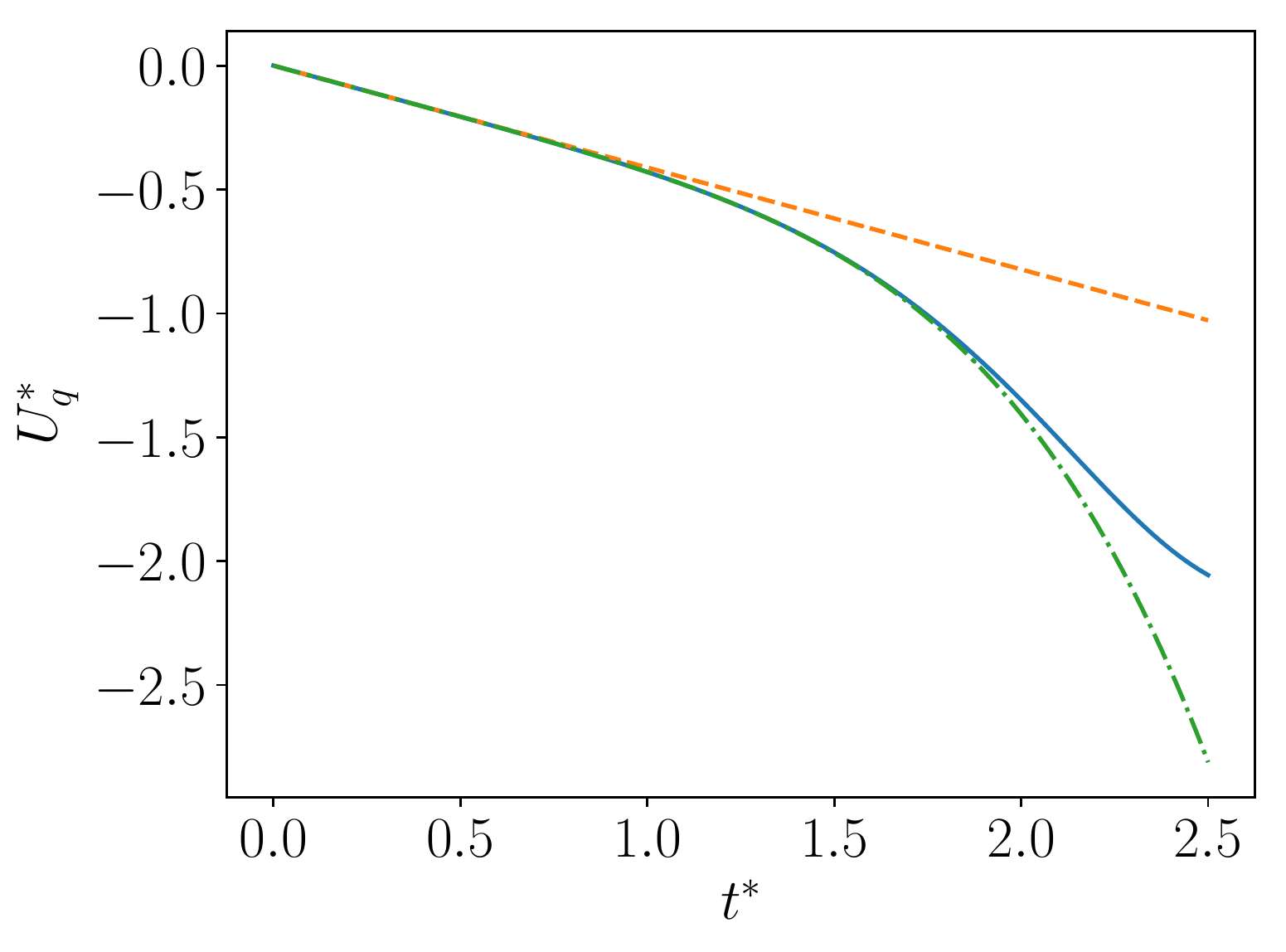}
        \includegraphics[height=0.23\textwidth]{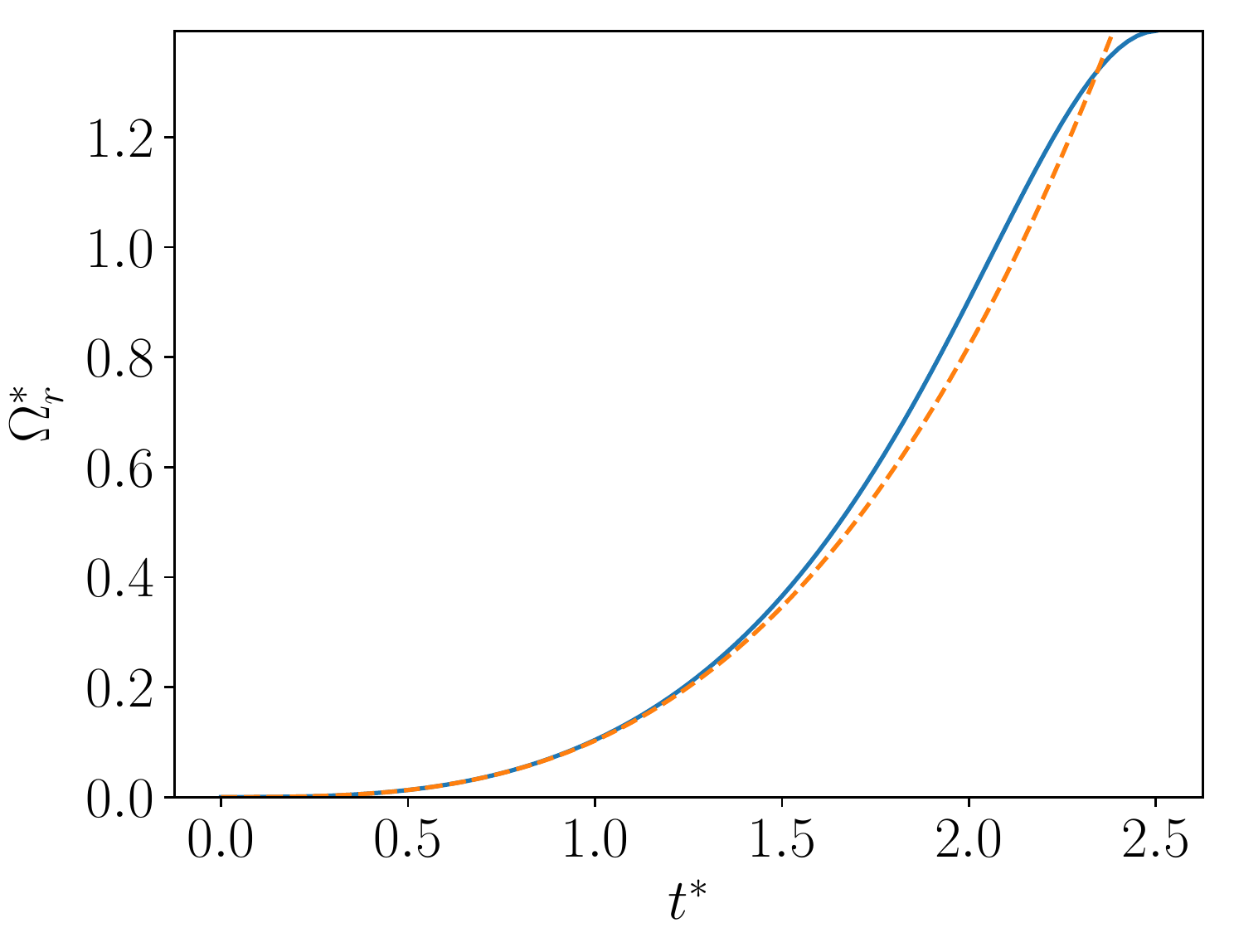}\\
        \vspace{-3mm}
        \hspace{0cm}(d) \hspace{5cm} (e) \hspace{5cm} (f)\\
        \includegraphics[height=0.23\textwidth]{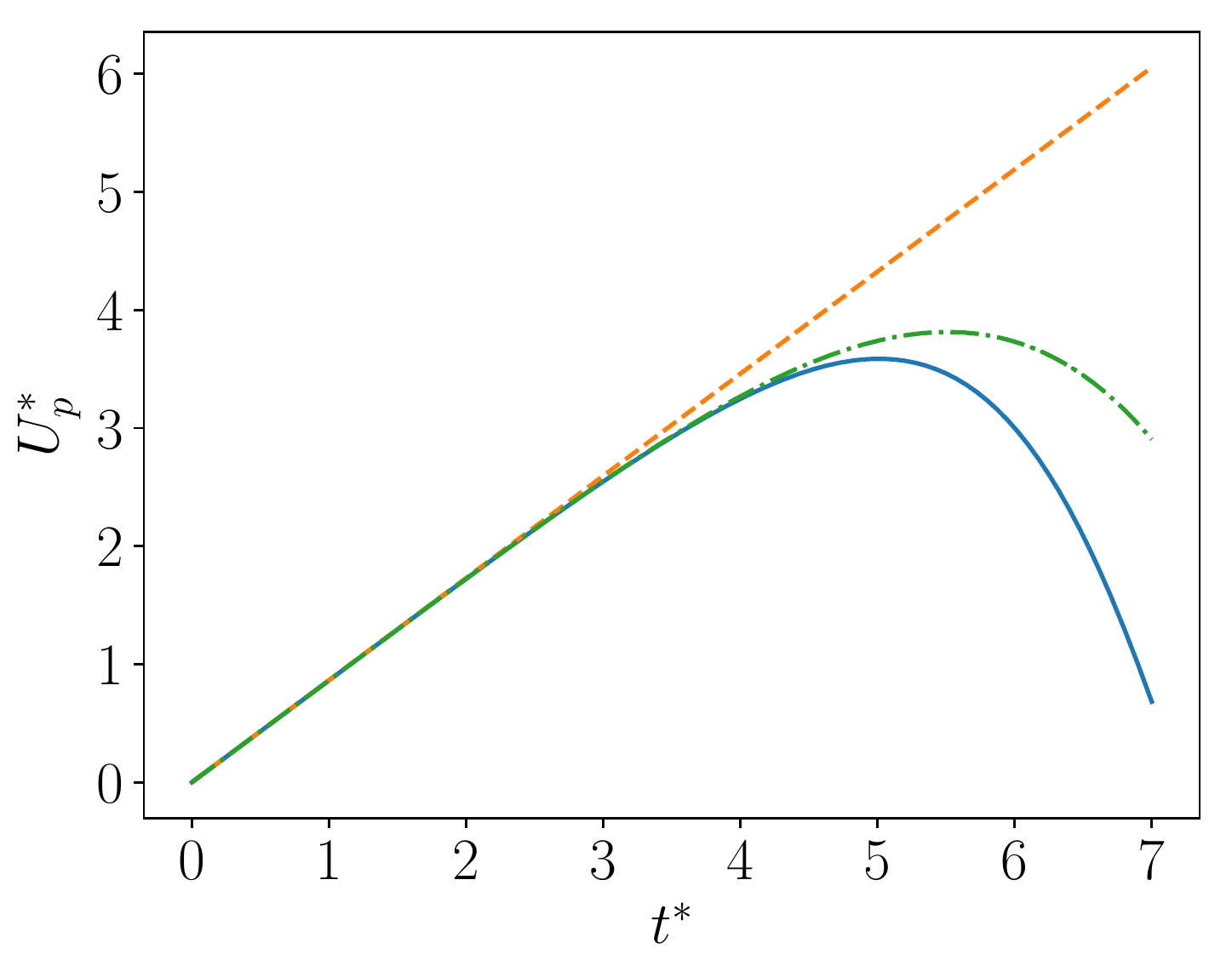}
        \includegraphics[height=0.23\textwidth]{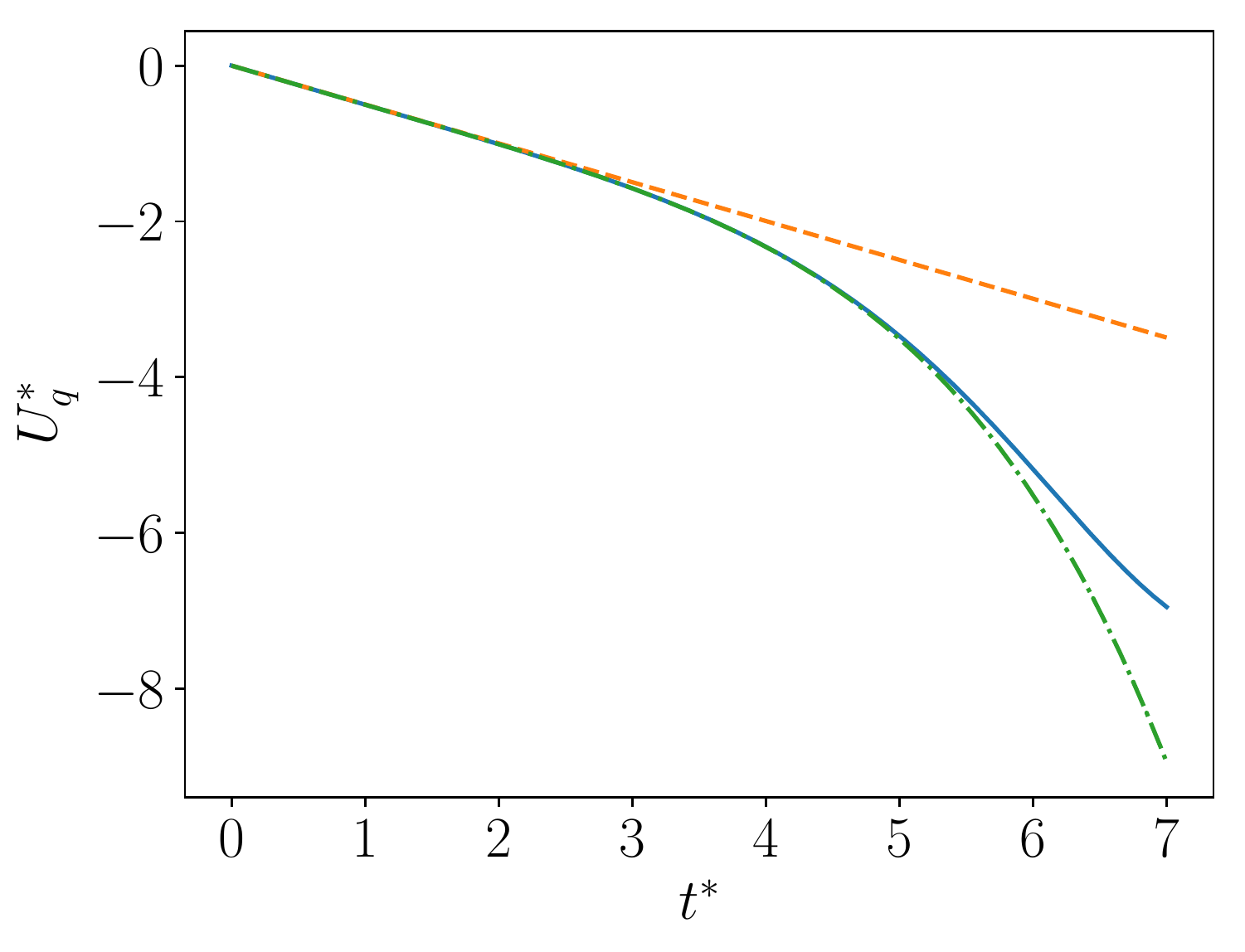}
        \includegraphics[height=0.23\textwidth]{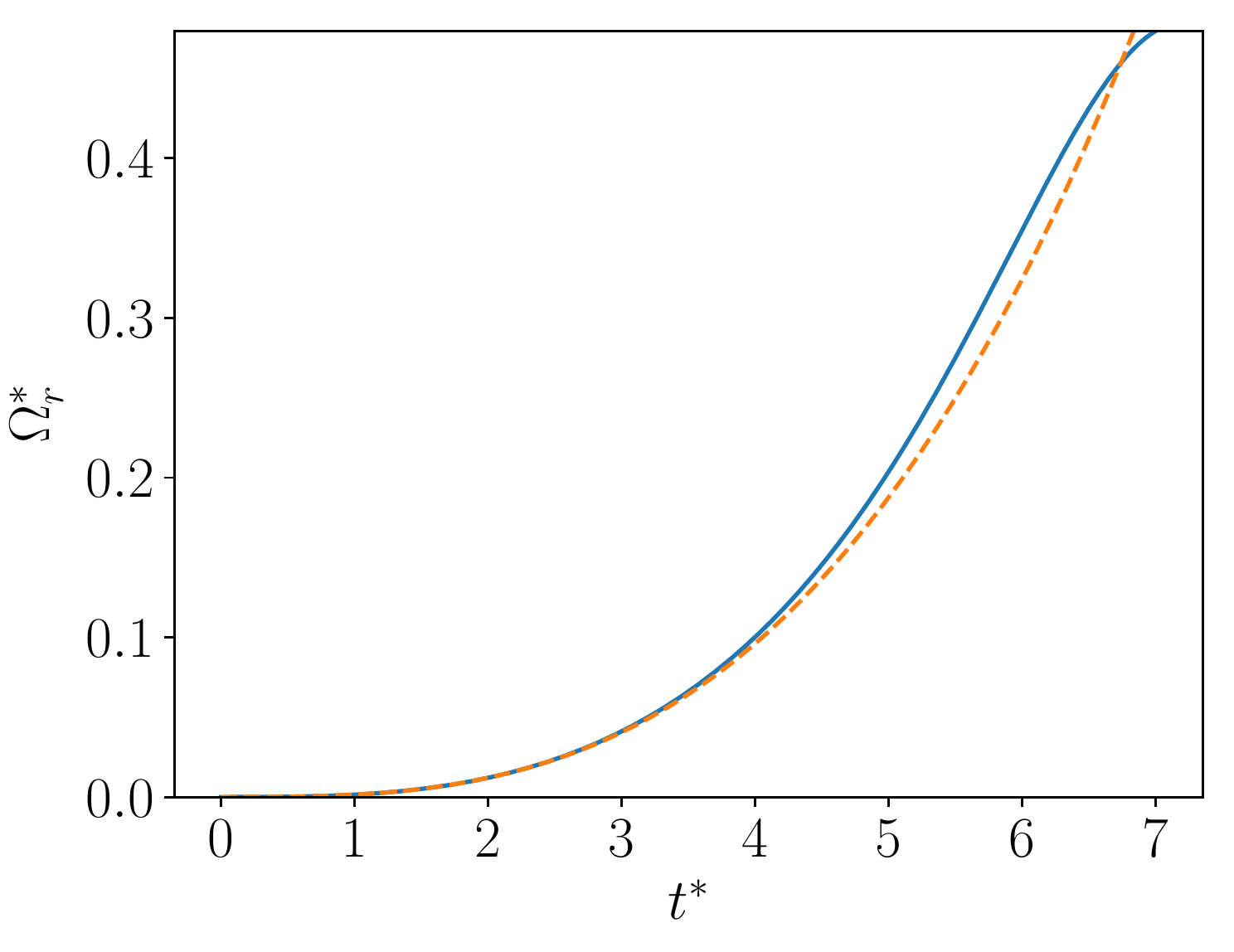}\\
        \vspace{-3mm}
        \hspace{0cm}(g) \hspace{5cm} (h) \hspace{5cm} (i)\\
    \caption{Dimensionless velocities and angular velocity as function of time for a $\chi=10$ cylinder starting from rest with $\phi=30^\circ$ and $\epsilon = 1$. (a), (b), (c) : $\bar{\rho} = 1.5$. (d), (e), (f) : $\bar{\rho} = 10$. (g), (h), (i) : $\bar{\rho} = 1000$. $-$ : numerical solution of equations \ref{eq:Upvd}, \ref{eq:Uqvd}, \ref{eq:Omegarvd} and \ref{eq:Phivd}, $--$ : zero-th order asymptotic expansion, $-\cdot-$ : first order asymptotic expansion.}
    \label{fig:short}
\end{figure}
Figure \ref{fig:short} displays the numerical and analytical solutions for a cylinder of aspect ratio $10$ released with an orientation angle $\phi =30 ^\circ$. We impose $\epsilon = 1$ such that $T = (L/g)^{1/2}$. The numerical solution is obtained by using a Runge-Kutta 4 algorithm. For $\bar{\rho} =1.5$ a good agreement is observed between the numerical and solution and the zeroth order analytical solution up to $t^* \approx 1.5$ and up to $t^*  \approx2.5$ for the first order solution (Figures \ref{fig:short} (a), (b) and (c)). For moderately large density ratios the agreement with the numerical solution is slightly worse. However, for the largest density ratios the first-order analytical solutions remain valid up to $t^*\approx 6$ (Figures \ref{fig:short} (g), (h) and (i)). For all the density ratios one may observe the good agreement between the analytical and numerical solutions out of the range of applicability of the analytical solution. Also, this agreement seems to be dependent on $\bar{\rho}$. The explanation lies in the magnitude of the coupling terms between translation and rotation in equations \ref{eq:Upvd} and \ref{eq:Uqvd}. At first-order this coupling terms are found to be proportional to $\mathcal{A}^2\mathcal{B}\mathcal{C}$ and $\mathcal{A}\mathcal{B}^2\mathcal{C}$ in equations \ref{eq:upt} and \ref{eq:uqt}.






\begin{figure}[h!]
    \centering
        \includegraphics[height=0.26\textwidth]{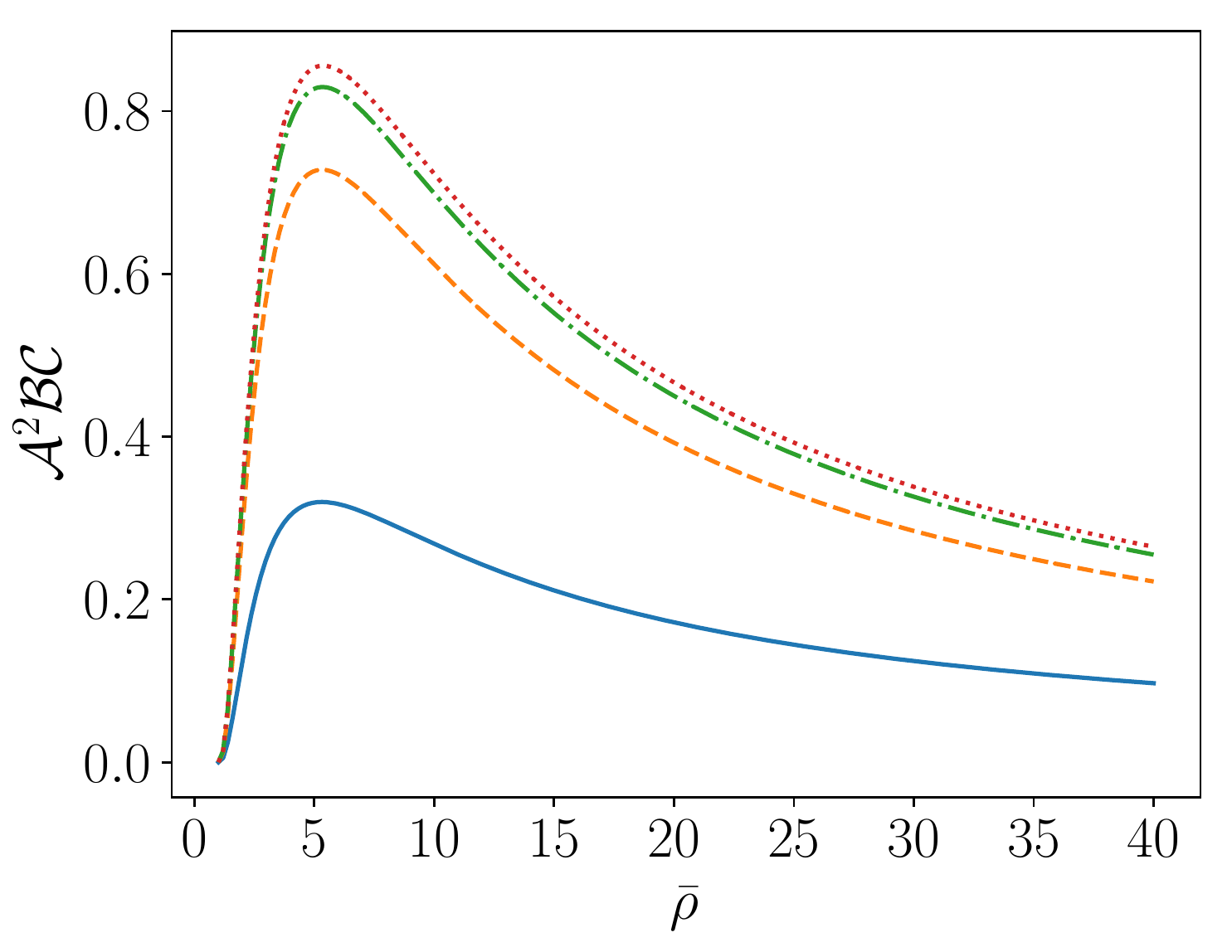}
        \includegraphics[height=0.26\textwidth]{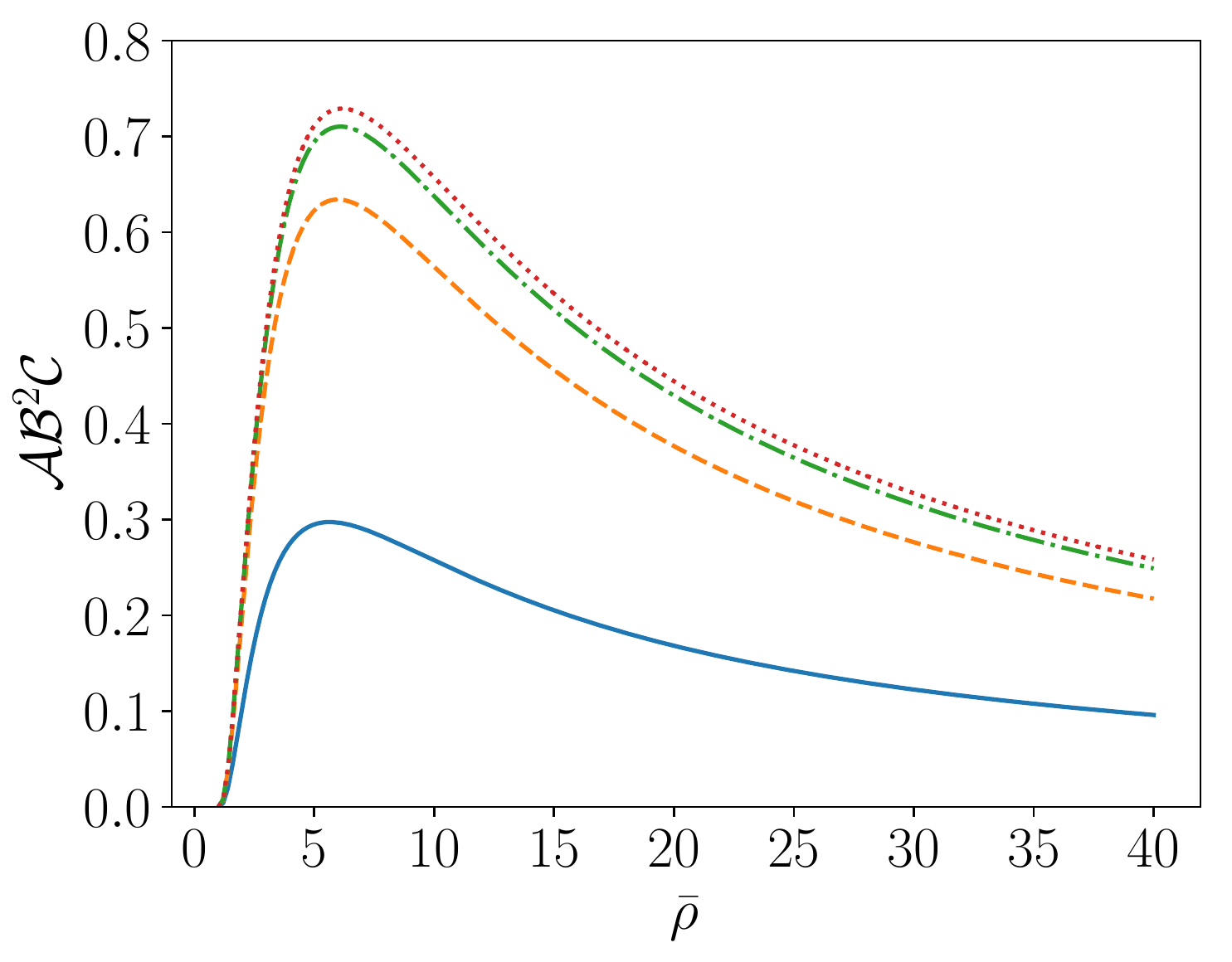}
    \caption{Leading order terms coupling translation and rotation. $-$ : $\chi =2$, $--$ : $\chi =8$, $-\cdot-$ : $\chi =32$, $\cdot \cdot \cdot$ : $\chi \rightarrow \infty$.}
    \label{fig:ABC}
\end{figure}
Figure \ref{fig:ABC} illustrates the variation of $\mathcal{A}^2\mathcal{B}\mathcal{C}$ and $\mathcal{A}\mathcal{B}^2\mathcal{C}$ as a function of $\chi$ and $\bar{\rho}$. Both coupling terms have the same order of magnitude and are smaller than approximatively $0.8$. The behaviour of $\mathcal{A}^2\mathcal{B}\mathcal{C}$ and $\mathcal{A}\mathcal{B}^2\mathcal{C}$ as function of $\chi$ and $\bar{\rho}$ disserve also some comments. Since both quantity evolves comparably, we will focus on $\mathcal{A}^2\mathcal{B}\mathcal{C}$. This quantity increases as a function of $\chi$ for moderate $\chi$. Hence one may expect a better match of the theory for small cylinders. Then $\mathcal{A}^2\mathcal{B}\mathcal{C}$ becomes nearly independent of $\chi$  for $\chi \approx 20$. This is in line with the results of appendix \ref{app:added} in which the added mass coefficients become almost independent of $\chi$ for this aspect ratio. $\mathcal{A}^2\mathcal{B}\mathcal{C}$ behaves non-monotonously with $\bar{\rho}$. It increases for $\bar{\rho} \leq 5$ and then decreases for larger $\bar{\rho}$. The value of $\bar{\rho}$ for which we observed the maxima is weakly dependent on $\chi$. All these trends can be easily obtained by looking more closely at the behaviour of  $\mathcal{A}^2\mathcal{B}\mathcal{C}$ as a function of $\bar{\rho}$. First one may note that for $\bar{\rho}\rightarrow 1$ we have $\mathcal{A}^2\mathcal{B}\mathcal{C} \sim (\bar{\rho} - 1) ^3(A_q^*-A_p^*)/[(1+A_p^*)^2(1+A_q^*)(J_q^*+D_p^*)]$ which explains the strong increase of $\mathcal{A}^2\mathcal{B}\mathcal{C}$ for $\bar{\rho} \leq 5$. In the opposite limit $\bar{\rho} \gg 1$ one obtain $\mathcal{A}^2\mathcal{B}\mathcal{C} \sim 1/ \bar{\rho}\times(A_q^*-A_p^*)/J_q^*$ which explains the decay of $\mathcal{A}^2\mathcal{B}\mathcal{C}$ for large density ratio. 





It is also interesting to discuss the range of applicability of the asymptotic results in experiments. Our observations suggest that the asymptotic expansion is valid up to $t \sim (L/g)^{1/2}$ and even larger value for large density ratios. Thus one may expect this solution to be valid as long as $(L/g)^{1/2} \ll D^2 /\nu$ which can be written $Ar \gg (\bar{\rho}-1)\chi$ where $Ar = (\rho _p -\rho)\rho g D^3/\mu^2$ is the Archimedes number. This implies that for sufficiently inertial regimes, the proposed analytical solution is expected to be observed in experiments. We are not aware of any experiments specifically dedicated to testing the validity of the theory. However, the qualitative observations provided in \citet{toupoint2019} may provide a first comparison with the present results. \citet{toupoint2019} investigated the settling of cylinder in inertia dominated regime $Ar \approx 40000$. Their density ratio was fixed to $\bar{\rho} = 1.16$ and the elongation ratio of the cylinders ranged between $2$ and $20$. In particular, the time related to viscous diffusion in their experiments with a cylinder of diameter $2$mm and $\chi =5$ is $t_\nu \approx 4s$ while $T \approx 0.03s$.  Hence the present theory should apply to their experimental results. They observed that the distance needed for the cylinders to rotate to their equilibrium position was less than $10d$ even when the cylinder was released vertically. From our analytical results, we obtain the time needed for the cylinder to change its orientation from nearly vertical ($\phi^{(0)} \ll 1$) to horizontal ($\phi(t^*) = \pi/2$) $t \sim (L/g)^{1/2} (6\pi/(\phi^{(0)}\mathcal{A}\mathcal{B}\mathcal{C}))^{1/4}$. With $\phi^{(0)} = \pi/10$ one get $t \approx 0.2s$. Injecting this estimate in the velocity equation and integrating one obtains the normalised distance $l/D$ for which the cylinder changes its orientation: $l/D = \mathcal{A}gt^2/2 \approx 11.2$. This estimate is very close to the experimental prediction and gives confidence on the non-viscous flow origin on which this model is grounded.


\section{Quasi-steady models}
\label{sec:scaling}

The particle equations of motion (\ref{eq:Up} - \ref{eq:Omegar}) are coupled to the Navier-Stokes equations \textit{via} the boundary conditions and the hydrodynamic loads. This make the problem very hard to solve and simplifying assumptions are required to make analytical progress as explained in the introduction. Since the characteristic time needed for the body to change its orientation scales like $\Omega ^{-1}$, the unsteady term in the Navier-Stokes equation scales as $\rho U \Omega$ where $\Omega$ and $U$ may be taken as the nominal scales for the angular and settling velocities. Since the inertial term scales as $\rho U^2 / L$ the unsteady term is negligible in comparison to the inertial term as long as $\Omega L /U \ll 1$. This condition is satisfied when the time needed for the vorticity to diffuse from the body for a Reynolds number of unity is much smaller than the rotation time scale \citep{newsom1994}. In this limit, one may consider the fluid unsteady term and as a consequence the history loads to be negligible. To be fully consistent with the neglect of the unsteady term we will also neglect the added mass loads. It remains to consider the inertial correction to the loads proportional to $\Omega U$. One may expect the $\Omega U$ corrections to be smaller than the inertial $U^2$ corrections as long as $\Omega L/U \ll 1$. Hence under the assumption, we postulate that both the history terms and the $\Omega U$ contribution are negligible with respect to the other load contribution as long as $\Omega L/U \ll 1$. The choice of the length of the particles ensures that the particle length is the relevant scale in the limit of a slender fibre \citep{khayat1989,dabade2015}. 
The validity of the assumption $\Omega L/U \ll 1$ will be evaluated through comparisons with experimental results and direct numerical simulations in Section \ref{sec:exp_dns}.



\subsection{Quasi-steady models for an arbitrarily oriented cylinder}

Under the assumption $\Omega L/U \ll 1$, the loads can be approximated as their quasi-steady counterparts, disregarding the $\Omega U$ terms: $F_p^\omega \approx F_p$, $F_q^\omega \approx F_q$ and $-U_pU_q(A_q-A_p) + T_r^{\omega} \approx T_r^i + T_r^\Omega$ where $F_p$ and $F_q$ are the quasi-steady forces in the longitudinal and perpendicular directions to the cylinder. The inertial torque, $T_r^i$, drives the cylinder towards its equilibrium position, while the hydrodynamic torque, $T_r^\Omega$, resists rotation. We will distinguish two different configurations depending on the cylinder aspect ratio: moderately long rods $ 2\leq \chi \leq 30$ and long fibres $\chi > 30$. For $\chi \leq 30$, expressions derived through slender-body theory and direct numerical simulations by \citet{kharrouba2021,pierson2021,fintzi2023} will be used, as they provide more accurate predictions compared to those by \citet{khayat1989}. Conversely, the expression by \citet{khayat1989} will be used for $\chi > 30$. To the author's knowledge, finite-inertia effects have not been considered for a slender body rotating in a fluid at rest at infinity. Thus, the most accurate expression for this case under the Stokes flow assumption by \citet{pierson2021} will be used. The expressions for $F_p$, $F_q$, $T_r^i$, and $T_r^\Omega$ are outlined in Appendix \ref{app:qsloads}. In both configurations, the linearized approximation is employed to describe the force as a function of the particle velocity. This approximation, which is exact in the Stokes flow regime, has been demonstrated to be accurate up to $Re \approx 1$ for $10 \leq\chi \leq 30$ and up to $Re \approx 10$ for $\chi < 10$ by \citet{fintzi2023} where $Re = \rho D U/\mu$ is the Reynolds number based on the body diameter. The experimental works of \citet{lopez2017} with $11.5 \leq \chi \leq 34.5$ and \citet{roy2019} with $20 \leq \chi \leq 100$ have also validated its relevance for larger aspect ratios.







Equations \ref{eq:Up} - \ref{eq:Omegar} are normalized by defining dimensionless (starred) quantities as $U_p = U U_p^*$, $U_q = U U_q^*$, $\Omega_r = \Omega \Omega ^*$ and $t = \Omega ^{-1}t^*$ where the characteristic angular and velocity scales are \textit{a priori} unknown. In the small inertia limit making use of the linearized approximation, the forces can be expressed as $F_p = -\mu U U_p^* L F_p^*(Re_L^*,\chi) $ and $F_q = -\mu U U_q^*L F_q^*(Re_L^*,\chi) $ where the expressions for $F_p$ and $F_q$ can be obtained from appendix \ref{app:qsloads}. $Re_L^* = Re_L(U_p^{*2}+U_q^{*2})^{1/2}$ is the Reynolds number based on the instantaneous settling velocity while $Re_L = \rho U L/(2\mu)$ is the characteristic Reynolds number based on the body half length. The inertial torque reads $ -\rho U^2U_p^*U_q^* L^3T_i^*(Re_L^*,\chi, \theta)$ while the torque resisting rotation reads $-\mu \Omega \Omega _r^* L^3 T_\Omega^*(Re_\Omega ^*,\chi)$ where $Re_\Omega^* = Re_\Omega|\Omega_r^*|$ and $Re_\Omega = \rho \Omega D^2/\mu$. Injecting all those scaling in equations \ref{eq:Up}, \ref{eq:Uq} and \ref{eq:Omegar} one obtain

\begin{align}
    \bar{\rho} Re_\Omega  \left(\frac{d U_p^*}{dt^*}
    - \Omega_r^* U_q^* \right)
    &= -\frac{4}{\pi}F_p^*(Re_L^*,\chi) U_p^*
    + \frac{(\rho _p-\rho)gD^2}{\mu U}\cos \phi, \label{eq:Up*}\\
    \bar{\rho} Re_\Omega \left(\frac{d U_q^*}{dt^*}
    + \Omega_r^* U_p^* \right)
    &= -\frac{4}{\pi}F_q^*(Re_L^*,\chi) U_q^*
    - \frac{(\rho _p-\rho)gD^2}{\mu U}\sin \phi, 
\label{eq:Uq*}\\
  \bar{\rho}\left(\frac{Re_\Omega}{Re}\right)^2 J_q^*
    \frac{d\Omega _r ^*}{dt ^*}
    &= -\frac{4}{\pi}\left(T_i^*(Re_L^*,\chi, \theta)U_p^*U_q^* + \frac{\mu \Omega }{\rho U^2}T_\Omega^*(Re_\Omega ^*,\chi)\Omega _r^*\right).
\label{eq:Omegar*}
\end{align}

From the above set of equations and for moderate density ratios $\bar{\rho} \sim 1$ it appears that all the unsteady terms vanish if $Re_\Omega \ll 1$ and $Re_\Omega \ll Re$. In contrast for large density ratios, one may take into account the particle inertia \citep{newsom1994}. We have to recall that to derive equations \ref{eq:Up*} - \ref{eq:Omegar*}  we have to assume $\Omega L /U \ll 1$, which is equivalent to $Re_\Omega \ll \chi Re$. This a rather restrictive condition for the applicability of the model but we shall see hereafter that it is valid for a non-negligible range of dimensionless parameters. Since the last terms in equation \ref{eq:Up*} - \ref{eq:Omegar*} must be of order one the velocity and angular velocity scales can be readily obtained as $U\sim D^2(\rho _p-\rho)g/\mu$ and $\Omega = \rho D^4(\rho _p-\rho)^2g^2/\mu^3$. Inserting these scalings yields

\begin{align}
    \bar{\rho} Ar^2  \left(\frac{d U_p^*}{dt^*}
    - \Omega_r^* U_q^* \right)
    &= -\frac{4}{\pi}F_p^*(Re_L^*,\chi) U_p^*
    + \cos \phi, \label{eq:UpA**}\\
    \bar{\rho} Ar^2 \left(\frac{d U_q^*}{dt^*}
    + \Omega_r^* U_p^* \right)
    &= -\frac{4}{\pi}F_q^*(Re_L^*,\chi) U_q^*
    - \sin \phi, 
\label{eq:UqA**}\\
  \bar{\rho}Ar^2 J_q^*
    \frac{d\Omega _r ^*}{dt ^*}
    &= -\frac{4}{\pi}\left(T_i^*(Re_L^*,\chi, \theta)U_p^*U_q^* + T_\Omega^*(Re_\Omega^*,\chi)\Omega _r^*\right).
\label{eq:OmegarA**}
\end{align}

The dimensionless nonlinear system of coupled differential equations, represented by Equations \ref{eq:UpA**} - \ref{eq:OmegarA**}, describe the velocity and angular velocity of a cylinder under an external force. Analytical solutions to this system can only be obtained in the limit of low Archimedes number, as reported by \citet{cox1965}. Otherwise, numerical solutions must be sought. Note that $\bar{\rho} Ar^2$ can be seen as a Stokes number ($St$) which is a dimensionless measure of particle inertia. The system exhibits two distinct regimes depending on $St$. For moderate particle inertia ($St \sim 1$ or equivalently $\bar{\rho}\sim 1/Ar^2$), as in the case of small solid particles settling in air, the left-hand sides (LHS) of equations \ref{eq:UpA**} - \ref{eq:OmegarA**} are non-negligible. The resulting system then consists of three ordinary differential equations which are solved thanks to a Runge-Kutta 4 algorithm.  Conversely, for small particle inertia ($St \ll 1$) as for small particles with densities close to that of the fluid the 
(LHS) can be neglected. The simplified system is a non-linear function composed of three equations, which can be solved using classical root finding methods. It reads 
\begin{align}
0
    &= -\frac{4}{\pi}F_p^*(Re_L^*,\chi) U_p^*
    + \cos \phi, \label{eq:UpA**qs}\\
0
    &= -\frac{4}{\pi}F_q^*(Re_L^*,\chi) U_q^*
    - \sin \phi, 
\label{eq:UqA**qs}\\
 0
    &=  T_i^*(Re_L^*,\chi, \theta)U_p^*U_q^* + T_\Omega^*(Re_\Omega^*,\chi)\Omega _r^*.
\label{eq:OmegarA**qs}
\end{align}
The validity of the two sets of equations is restricted to the condition $\Omega L /U \ll 1$. Since $\Omega L /U\sim Ar \chi$ we obtain the very restrictive condition $Ar \ll 1/\chi$. The next section aim to demonstrate that the present set of equations remain valid up to moderate Archimedes numbers, despite its seemingly limited range of validity. It is also important to note that these scalings are \textit{a priori} limited to small fluid inertial effects as the forces and resistive torque are viscous in nature. However, \citet{fintzi2023} and \citet{pierson2021} have demonstrated that this viscous based laws remain valid for moderate inertial effects ($Re_L \leq 1$). The behavior for larger inertial effects is complex and non-linearly dependent on the Reynolds number, as described in Appendix \ref{app:qsloads}.

\subsection{Damped oscillations around $\phi = \pi/2$}


Upon reaching its equilibrium position $\phi = \pi/2$, which is perpendicular to its direction of motion, an additional regime of interest can be expected for the particle. We linearize the equation of motion around this position. By writing $\phi(t) = \pi/2 + \epsilon \phi^*(t)$, where $\epsilon \ll 1$, the leading-order expressions for $\cos \phi$ and $\sin \phi$ are found to be $-\epsilon \phi^*$ and $1$, respectively. In order for the velocity terms in the momentum equation to be non-trivial, they must scale as $\epsilon U$ and $U$ for the parallel and perpendicular velocity, respectively, such that $U_p = \epsilon U U_p^*$ and $U_q = U U_q^*$. By balancing the viscous torque with the inertial torque, we get $\Omega_r = \epsilon \Omega \Omega_r^*$ with $\Omega _r^* = d\phi^*/dt^*$, resulting in the following system of equations


\begin{align}
    St\left(\frac{d U_p^*}{dt^*}
    - \frac{d\phi^*}{dt^*} U_q^* \right)
    &= -\frac{4}{\pi}F_p^*(Re_L^*,\chi) U_p^*
    - \phi^*,\\
    St\left(\frac{d U_q^*}{dt^*}
    + \epsilon^2\Omega_r^* U_p^* \right)
    &=-\frac{4}{\pi}F_q^*(Re_L^*,\chi) U_q^*
    - 1, \label{eq:Uqo}\\
  StJ_q^*
    \frac{d^2\phi ^*}{dt ^{*2}}
    &= -\frac{4}{\pi}\left(T_i^*(Re_L^*,\chi, \theta)U_p^*U_q^* +  T_\Omega^*(Re_\Omega ^*,\chi)\frac{d\phi^*}{dt^*}\right). \label{eq:Omegaro}
\end{align}



Despite assuming a small angle around the equilibrium position, the resulting system comprises three coupled nonlinear equations, rendering analytical solutions for arbitrary Reynolds numbers impractical. However, various simplifications can be made. First, the second term on the left-hand side of equation \ref{eq:Uqo} becomes negligible in the limit of $\epsilon \ll 1$. Second, using the velocity estimate mentioned above, $Re_L^* \sim Re_L|U_q^*|$ at leading order, resulting in equation \ref{eq:Uqo} being decoupled from the other two equations. Third, in the long-time limit $t^* \gg St$, the left-hand side of equation \ref{eq:Uqo} becomes negligible, leading to $U_q^* \sim - \pi/4 /F_q^*(Re_L^*,\chi)$. Fourth, we consider the small inertia limit $Ar \ll 1$. Under this limit, the system reads 



\begin{align}
    St\left(\frac{d U_p^*}{dt^*}
    + \frac{\pi}{4} \frac{1}{F_q^*} \frac{d\phi^*}{dt^*} \right)
    &= -\frac{4}{\pi}F_p^* U_p^*
    - \phi^* \label{eq:Uqos},\\
  StJ_q^*
    \frac{d^2\phi ^*}{dt ^{*2}}
    &=  \frac{T_i^*}{F_q^*}U_p^* -\frac{4}{\pi}T_\Omega^* \frac{d\phi^*}{dt^*}. \label{eq:Omegaros}
\end{align}

The resulting set of equations is made of two coupled linear ordinary differential equations (ODE) since in the limit $Ar \ll 1$, $T_i^*$, $F_q^*$, $T_\Omega^*$ are independent of $Re_L^*$, $Re_\Omega^*$ and $\theta$. By substituting the expression for $U_p$ from Equation \ref{eq:Omegaros} into Equation \ref{eq:Uqos}, the resulting equation becomes a third-order linear ODE with constant coefficients. Although an analytical solution may be obtained by summing three exponential functions, the coefficients inside these functions are complex and obey a third-order polynomial. Instead, we assume that the left-hand side of Equation \ref{eq:Uqos} is negligible compared to the right-hand side. This assumption is not satisfactory from a pure asymptotic perspective but allows for clearer disentanglement of the physical mechanism underlying the particle dynamic. 
In this limit, $U_p^* \approx -\pi/4/F_p^*(Re_L^,\chi)\phi^*$ and


\begin{equation}
 StJ_q^*\frac{d^2\phi ^*}{dt ^{*2}}+ \frac{4}{\pi}T_\Omega^* \frac{d\phi^*}{dt^*}+\frac{\pi}{4}\frac{T_i^*}{F_q^*F_p^*}\phi^*= 0.
\label{eq:dampedeq}
\end{equation}



\begin{figure}[h]
\centering
\includegraphics[width=6cm]{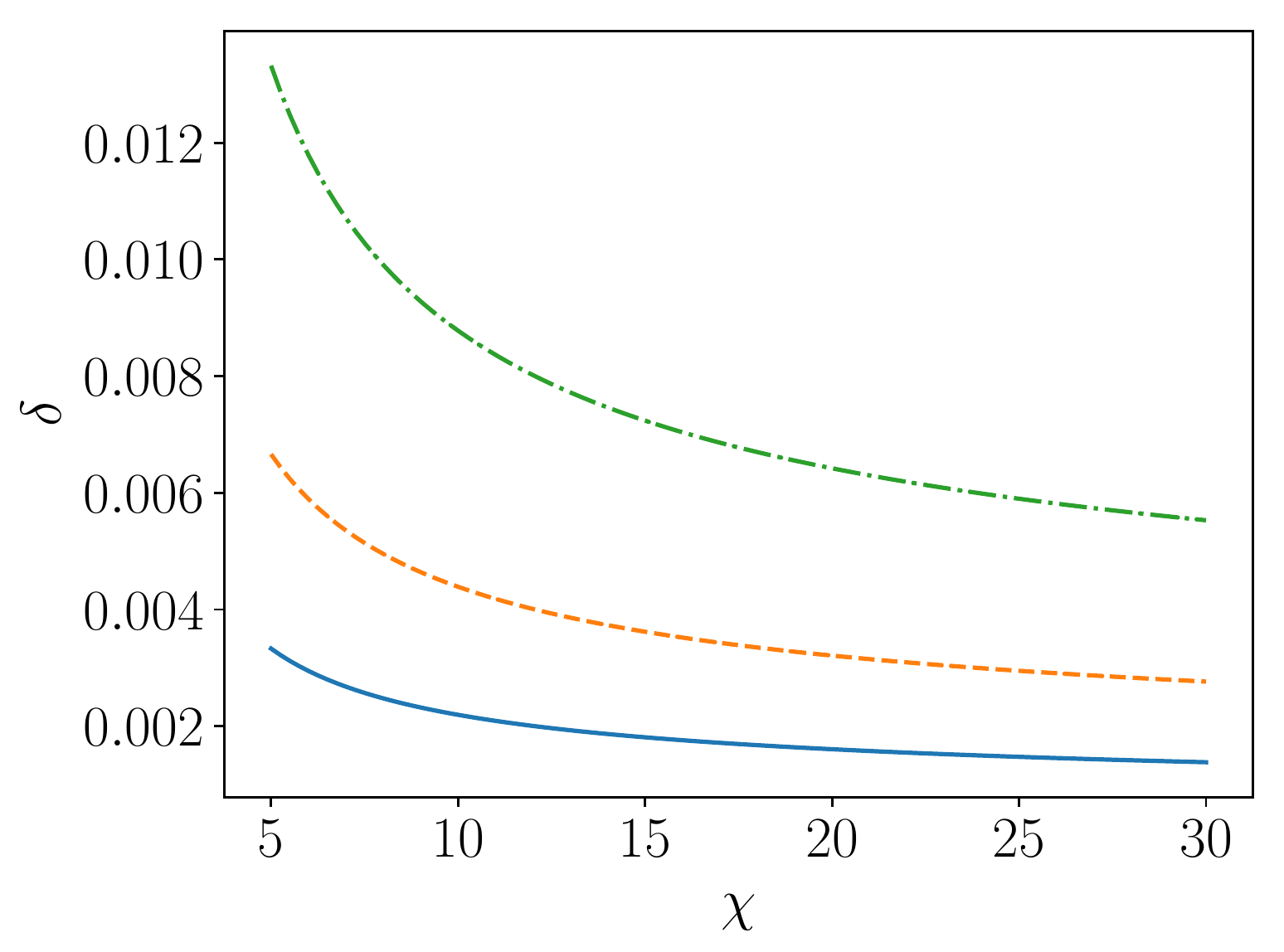}\includegraphics[width=6cm]{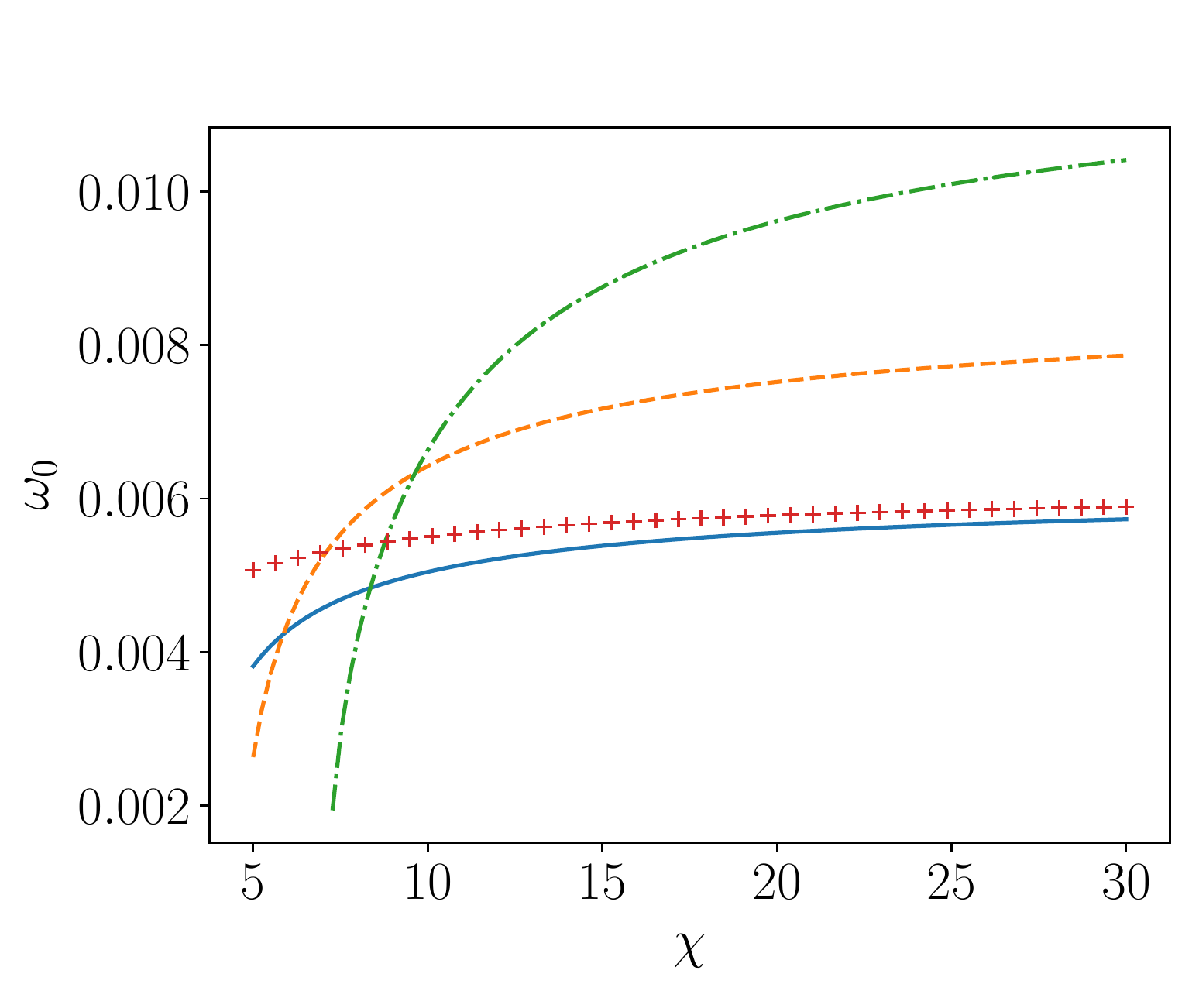}\\
\hspace{0.5cm}$(a)$  \hspace{5.5cm} $(b)$ \\
\caption{(a) : Evolution of $\delta$ (equation \ref{eq:delta}) as function of $\chi$. $(b):$ Evolution of $\omega _0$ (equation \ref{eq:delta}) as function of $\chi$. $-$:$St = 2000$, $--$:$St=1000$, $-\cdot$:$St=500$, $+$ : equation \ref{eq:omega0} for $St = 2000$.}
\label{fig:damped_omega_delta}
\end{figure}

The above equation is a classical damped harmonic oscillator. This equation shares many similarities with the one used by \citet{gustavsson2019,gustavsson2021} who studied the effect of particle inertia on the orientation of spheroids in turbulent flows. Although the derivation presented in \citet{gustavsson2021} is different from the present one they nicely complement each other. Two natural solutions exist depending on the sign of discriminant related to the second order polynomial of this linear ODE. If the discriminant is positive the solution is overdamped and leads to a decay of the motion without oscillation while for a negative discriminant the solution is underdamped and oscillatory motion is observed. Hence to obtain underdamped solution the particle must obey the following criteria
\begin{equation}
St > \frac{16}{\pi^3}\frac{T_\Omega^{*2}F_q^*F_p^*}{J_q^*T_i^*},
\label{eq:damped}
\end{equation}
which means that the particle inertia must be sufficiently large to lead to an oscillating regime. Under the condition $Ar \ll 1$, one may expect underdamped oscillation only in the case where the density ratio is very high. Specifically, oscillatory motion can only be expected for high-density particles, such as steel, settling in air. 
\begin{figure}[h]
\centering
\includegraphics[width=6cm]{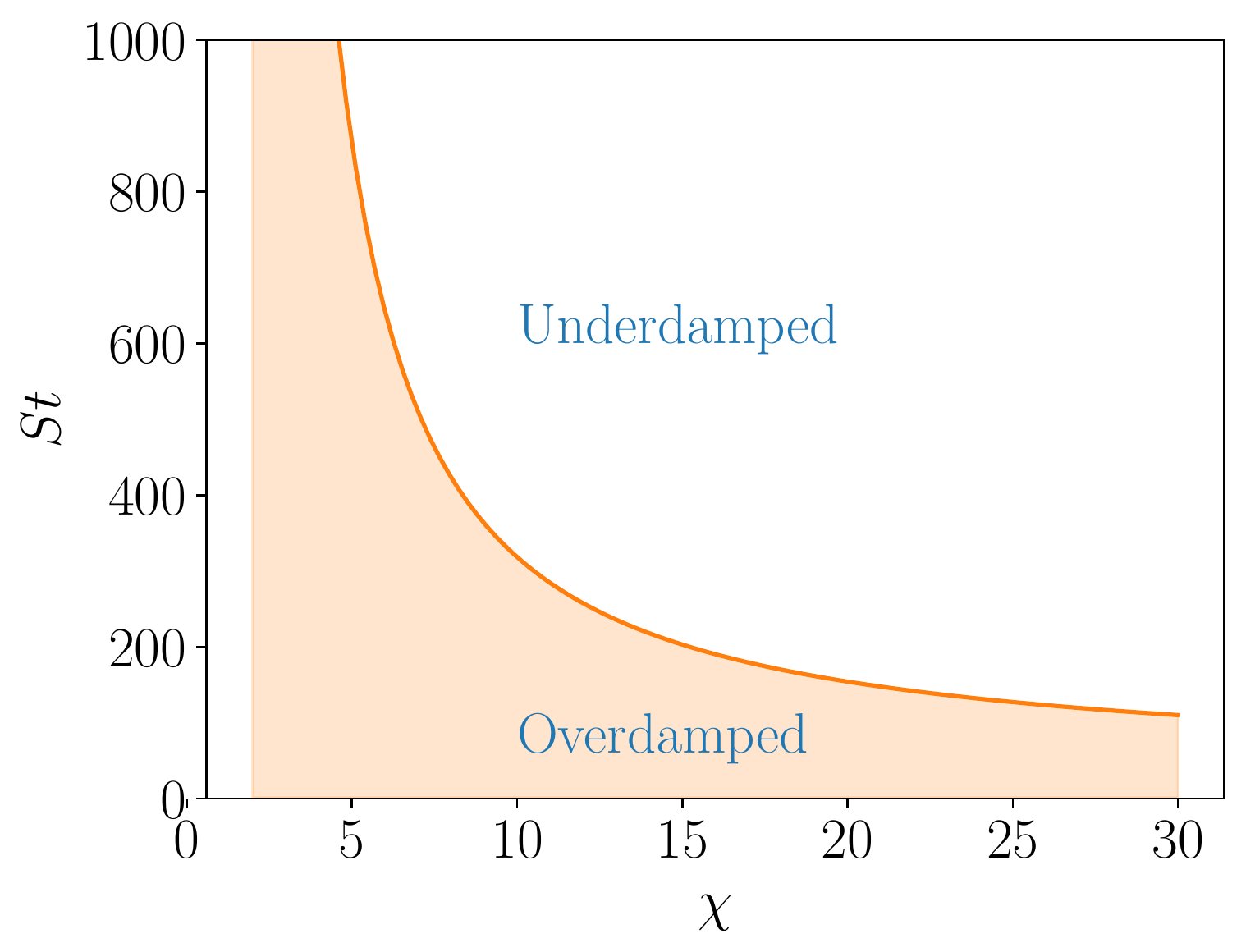}
\caption{Regime map showing the overdamped versus underdamped configuration in the $(St,\chi)$ plane. Solid line : analytical criterion \ref{eq:damped}.}
\label{fig:damped}
\end{figure}
Figure \ref{fig:damped} presents a map of the two regimes. The critical Stokes number separating the two regimes is a decreasing function of $\chi$. This may be proved using scaling arguments for $\chi \gg 1$. Indeed under this assumption $F_p^*\sim F_q^* \sim 1/\ln(\chi)$, $T_\Omega^*\sim 1/\ln(\chi)$ and $T_i^*\sim 1/\ln^2(\chi)$. Hence for $\chi \gg 1$ the critical Stokes number decreases as $1/\ln^2(\chi)$. Hence for a given Stokes number, the underdamped regime will be promoted by using longer particle. If the criterion \ref{eq:damped} is satisfied the solution of equation \ref{eq:dampedeq} is given by
\begin{equation}
\phi^*(t) = A e^{-\delta t^*}\cos(\zeta+\omega _0t^*) \quad \text{where} \quad \delta = \frac{2}{\pi}\frac{1}{St}\frac{T_\Omega^*}{J_q^*} \quad \text{and}  \quad \omega _0 = \left(\frac{\pi}{4}\frac{1}{St}\frac{T_i^*}{F_q^*F_p^*J_q^*}-\delta^2\right)^{1/2}.
\label{eq:delta}
\end{equation}
$A$ and $\zeta$ are constants given by the initial boundary conditions while $\omega _0$ is the dimensionless natural frequency of the oscillator and $\delta$ is the dimensionless damping coefficient. In the limit $St\gg 1$ which is relevant limit to expect underdamped oscillations one get

\begin{equation}
\omega _0 = \left(\frac{\pi}{4}\frac{1}{St}\frac{T_i^*}{F_q^*F_p^*J_q^*}\right)^{1/2}.
\label{eq:omega0}
\end{equation} 
As the value of the Stokes number increases, both $\omega_0$ and $\delta$ decrease, albeit with a less pronounced decrease observed for $\omega_0$. Figure \ref{fig:damped_omega_delta} displays the behavior of both quantities, which approach a near-constant value for $\chi \gg 1$. Specifically, as $\chi$ increases, $\delta \sim 1/(St\ln\chi)$ while $\omega_0 \sim (1/St)^{1/2}$. The results for $\chi$ larger than 30 were not plotted; however, it is expected that the behavior for larger $\chi$ will remain consistent as all expressions for the loads converge to the slender body theory results for $\chi \gg 1$.


One should recall the many assumptions required to derive the model. In particular, we have assumed that the left-hand side (LHS) of Equation \ref{eq:Uqos} is negligible. Upon substitution of our estimate for $U_p$, the LHS of Equation \ref{eq:Uqos} simplifies to $\pi/4 St d\phi^*/dt(1/F_q^*-1/F_p^*)$. For $\chi \leq 10$, it has been reported that $F_q^* \approx F_p^*$ \citep{fintzi2023}, leading to a near-cancellation of the LHS. However for longer fiber this assumption is discutable. Additionally, the model assumes that both the history effect and the coupling effect between translation and rotation can be neglected. The former is questionable as the current model exhibits a constant period of oscillation while $U_p$ is a decreasing function of time. Hence, in the long-time limit, the unsteady term in the Navier-Stokes equation may be much larger than the advective term. 
 Therefore, the proposed model is not designed to provide an exact value for the damping coefficient and natural frequency for arbitrary dimensionless parameters especially larger fluid inertia, but rather serves as a useful framework for qualitatively understanding the origin of the damped oscillations observed in the subsequent section. \citet{newsom1994} investigated the settling of elongated graphite particles in air ($\bar{\rho} \approx 1442, Ar \approx 0.054$, $\chi \geq 100$) and did not observe oscillations around their stable orientation. However, in their study $St \approx 4.2$, which is probably too low to expect such a phenomenon (see Figure \ref{fig:damped}). The recent study of \citet{bhowmick2023} (not published yet) who investigated the settling of ellipsoidal particle with larger inertia in air indicate that this solution is relevant. 
\section{Comparison of the quasi-steady models with experiments and direct numerical simulations}
\label{sec:exp_dns}
\subsection{Experiments for $Re \ll 1$} 

In this section, we conduct a comparison between the experimental results presented in \citet{cabrera2022} and \citet{roy2019} with those provided by the models presented in the previous section. Despite having previously compared our quasi-steady results to those of \citet{cabrera2022} in \citet{fintzi2023}, we present the comparison again as we aim to further discuss the effect of the particle inertia.
\begin{figure}[h]
\centering
\hspace{1cm}$\chi=8$ \hspace{5cm} $\chi=8$\\
\includegraphics[width=6cm]{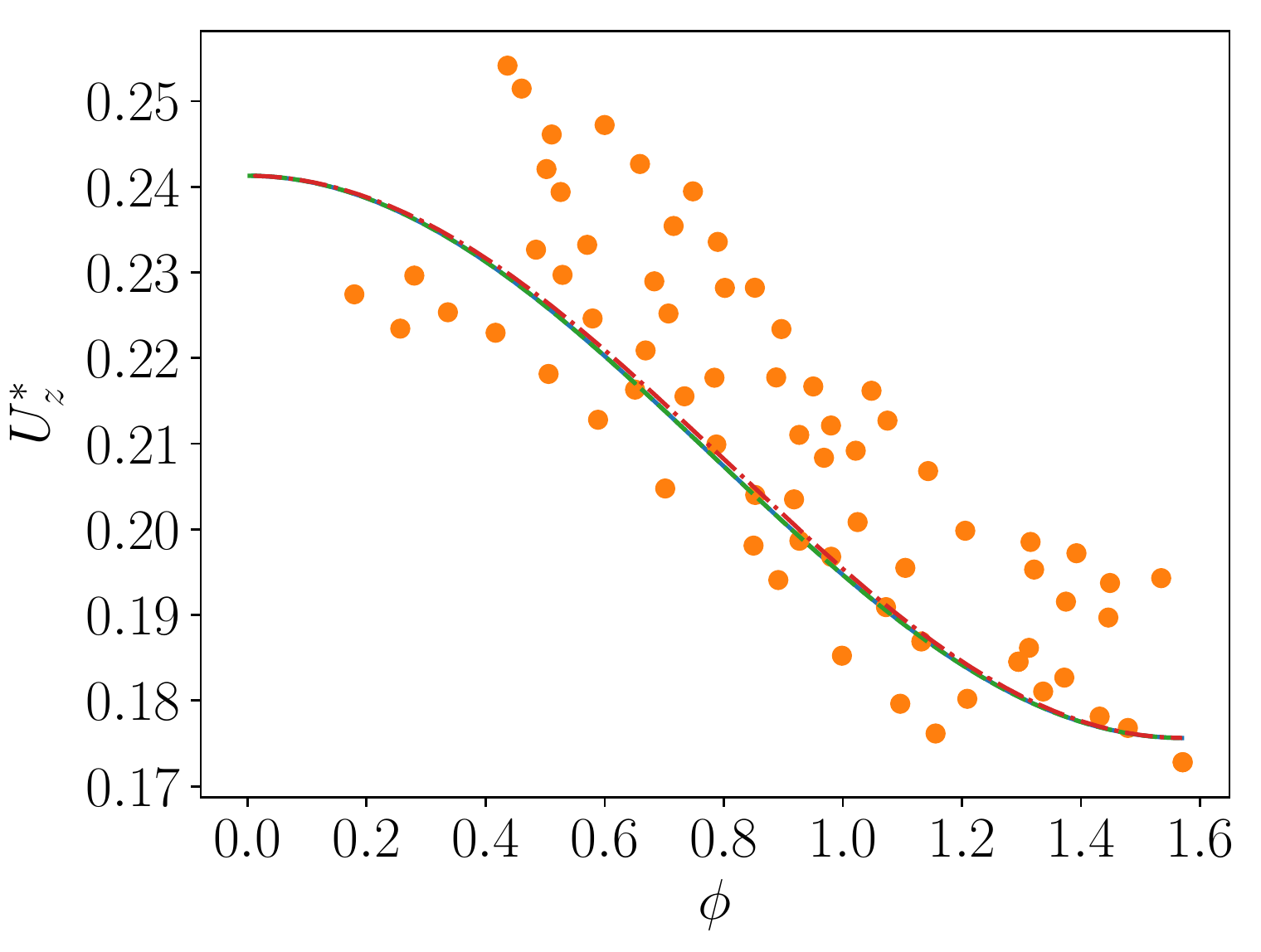}
\includegraphics[width=6cm]{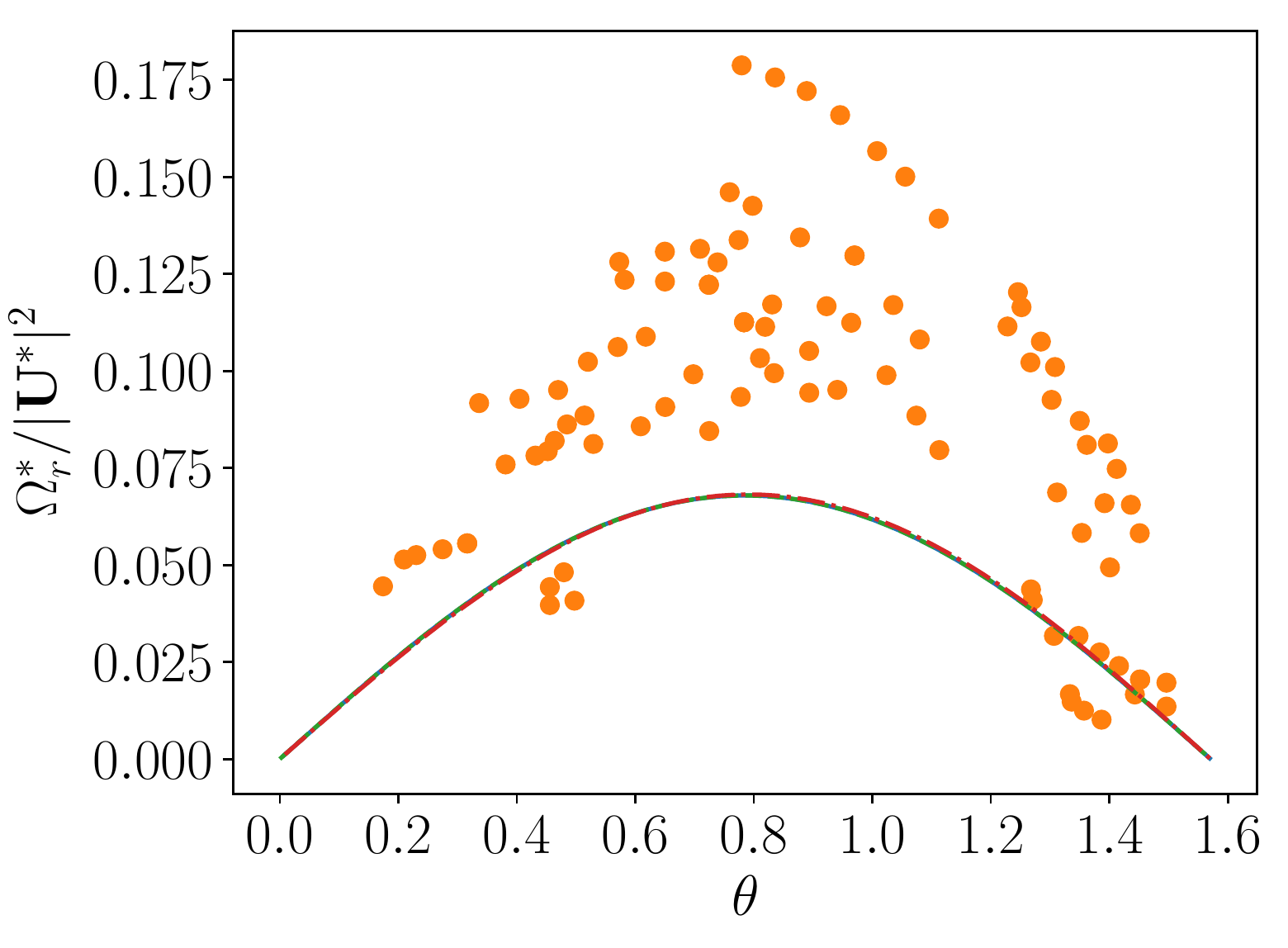}\\
\hspace{0.5cm}$(a)$ \hspace{5.5cm} $(b)$\\
\hspace{1cm}$\chi=16$ \hspace{5cm} $\chi=16$\\
\includegraphics[width=6cm]{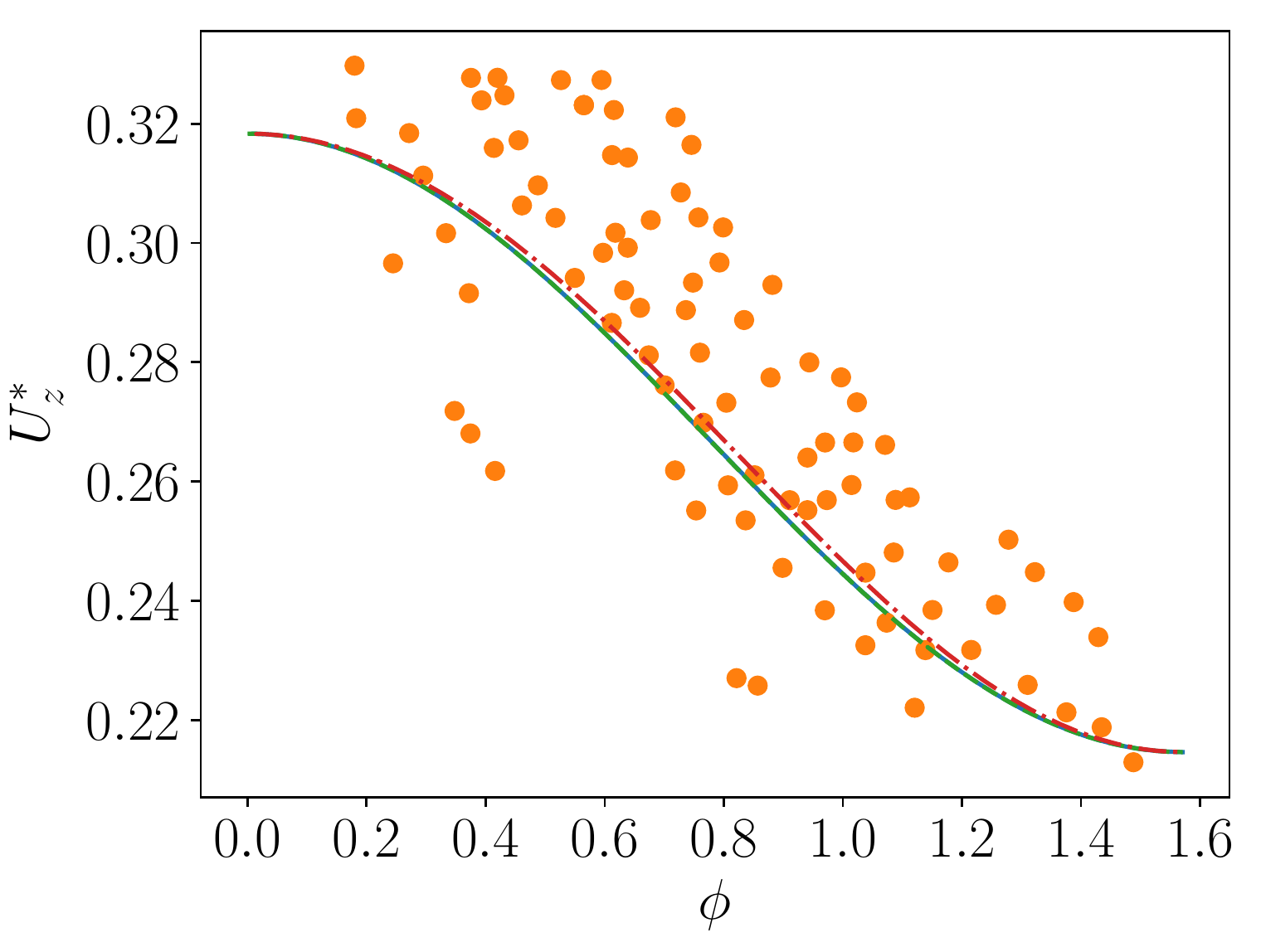}
\includegraphics[width=6cm]{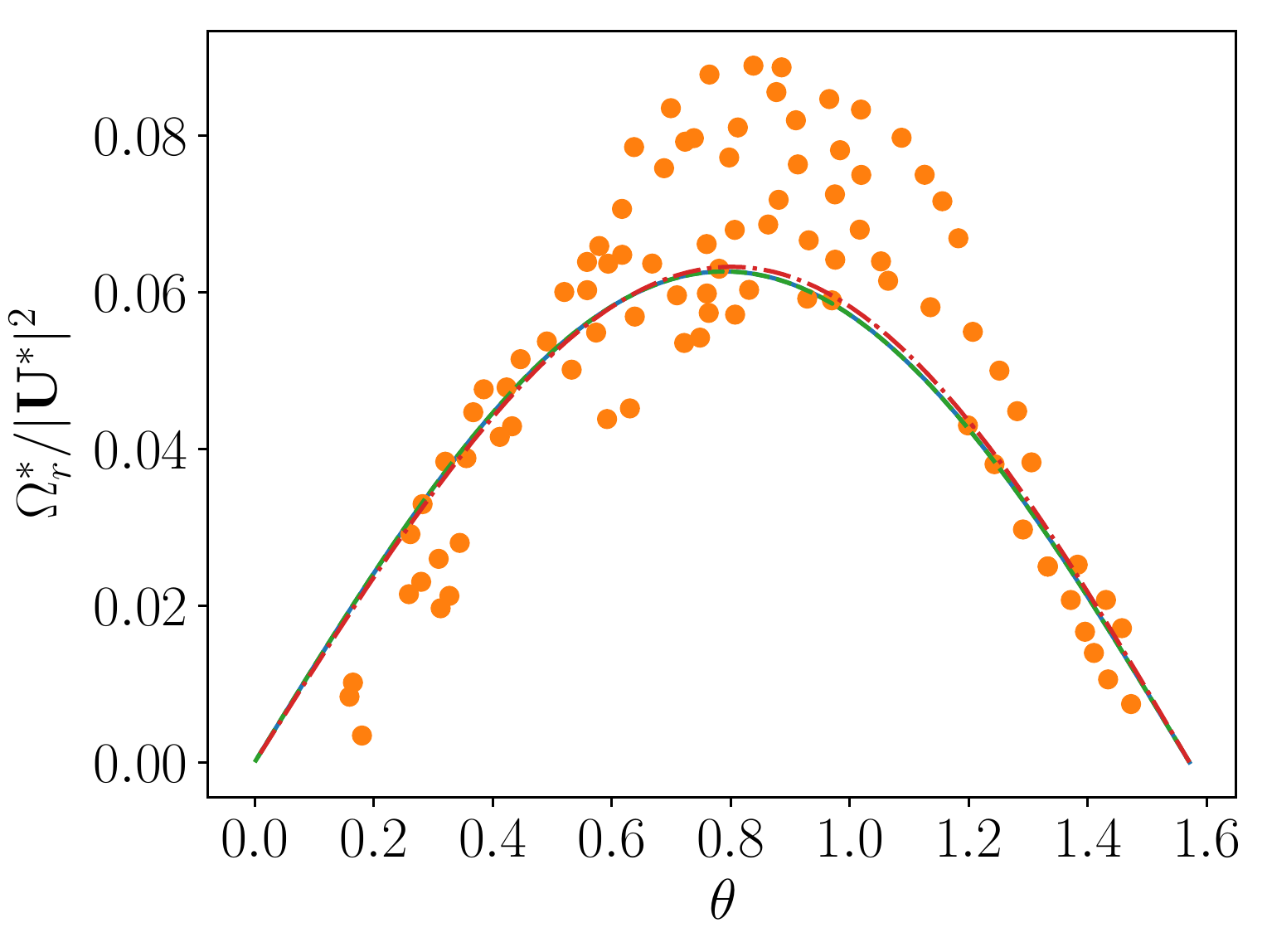}\\
\hspace{0.5cm}$(c)$ \hspace{5.5cm} $(d)$\\
\caption{Dimensionless sedimenting velocities. (a), (c) : Dimensionless sedimenting velocity as function of $\phi$. (b), (d) : Dimensionless angular velocity divided by the square of the particle velocity as function of $\theta$ (Figure \ref{fig:single_cyl}). $\bullet$ : \citet{cabrera2022} experiments for $Ar \approx 0.147$, $\bar{\rho} \approx 12.2$. $-$ : prediction from the unsteady equations \ref{eq:UpA**} - \ref{eq:OmegarA**} for $Ar \approx 0.147$, $-\cdot-$ : prediction from the unsteady equations \ref{eq:UpA**} -\ref{eq:OmegarA**} for $Ar \approx 0.147$, $\bar{\rho} = 1000$, $--$ : prediction from the steady equations \ref{eq:UpA**qs} - \ref{eq:OmegarA**qs} for $Ar \approx 0.147$, $\bar{\rho} \approx 12.2$. In the unsteady computations, we have not displayed the transient behaviour of the particles starting from rest. Top panel corresponds to $\chi = 8$ and bottom panel to $\chi=16$.}
\label{fig:cabrera}
\end{figure}


The sedimentation velocity and angular velocity measured in the study of \citet{cabrera2022} for particles with aspect ratios of $\chi=8$ and $\chi=16$ are presented in Figures \ref{fig:cabrera} (a) and (c). For $\chi =8$ the Reynolds number varies in the range $0.026 \leq Re \leq 0.035$ and is slightly higher for $\chi=16$ ($0.032 \leq Re \leq 0.047$). The model predictions are consistent with the experimental data, with good agreement for the angular velocity for $\chi=16$ but less so for $\chi=8$ (Figures \ref{fig:cabrera} (b) and (d)). The experimental data for $\chi=8$ exhibits considerable scatter, which the authors possibly attribute to the presence of particle defects, particularly mass inhomogeneities. The gravitational torque caused by mass inhomogeneities along the body axis, $T_g$, scales as $T_g \sim mgL$ \citep{roy2019}. The ratio of inertial torque to gravitational torque, scales as $T_i/T_g\sim(\bar{\rho}-1)Ar\chi$. Therefore, for low Archimedes number and moderate aspect ratio, as in the present case, mass inhomogeneities may impact the angular velocity. Additionally, the results obtained from both sets of equations, (\ref{eq:UpA**} - \ref{eq:OmegarA**}) and (\ref{eq:UpA**qs} - \ref{eq:OmegarA**qs}), are indistinguishable, thus particle inertia can be safely neglected in the experiments of \citet{cabrera2022}. To further examine the impact of particle inertia, we conducted computations of the same physical configuration reported by \citet{cabrera2022} (with $Ar \approx 0.147$ and $\chi=8$ or $\chi=16$), but with a significantly larger density ratio $\bar{\rho} = 1000$, which is representative of particles settling in air (represented by the dashed-dotted line in Figures \ref{fig:cabrera}). Our results indicated that particle inertia has minimal effect on the outcome. This can be explained by the fact that the maximum value of $Re_\Omega^*$ was found to be approximately $\text{max}(Re_\Omega^*) \approx 9\cdot 10^{-5}$ implying that the inertial term in the momentum equations which scales as $\bar{\rho} Re_\Omega$ remains much smaller than unity even with $\bar{\rho} = 1000$.





\begin{figure}[h]
\centering
\includegraphics[width=5cm]{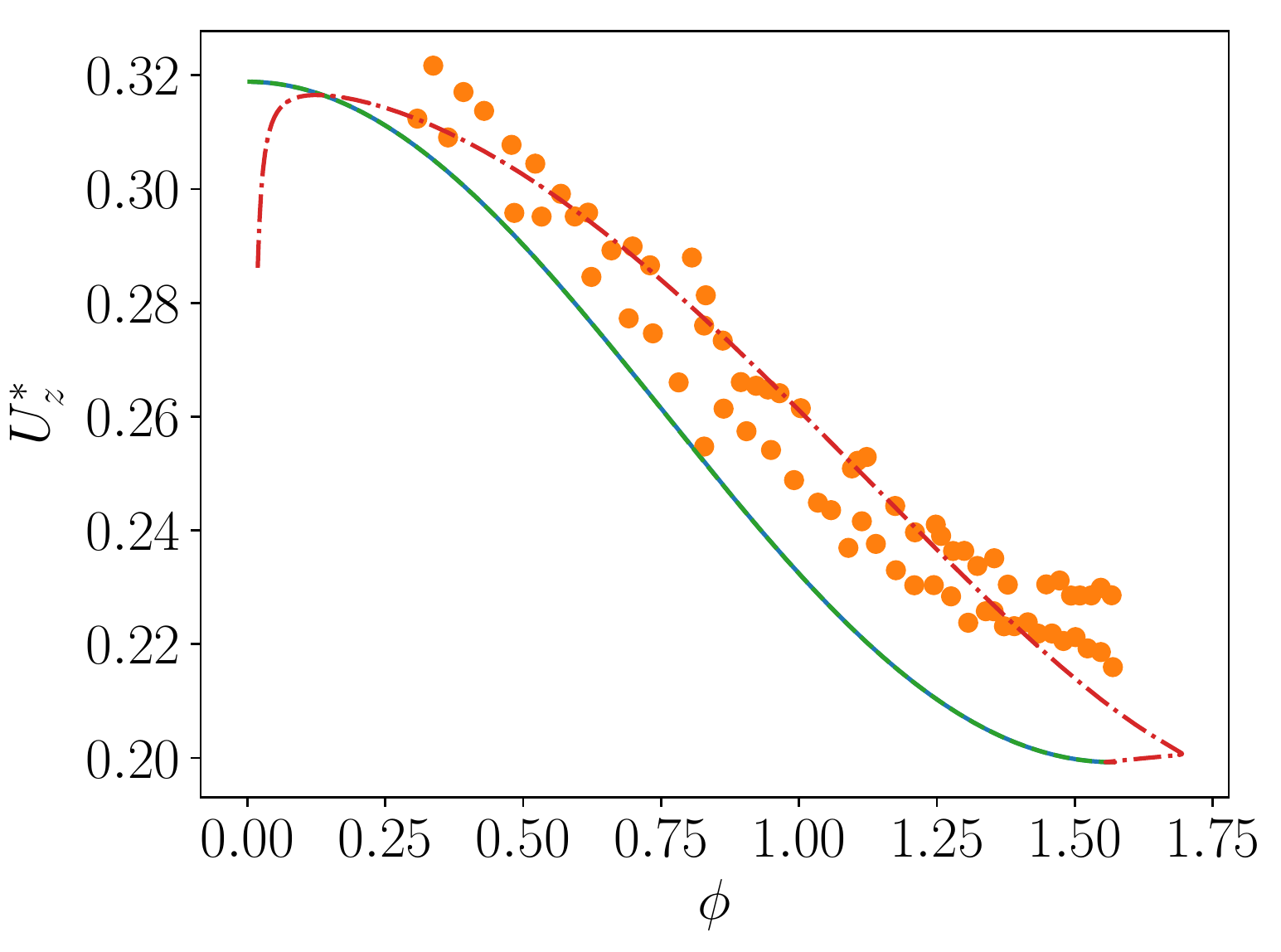}
\includegraphics[width=5cm]{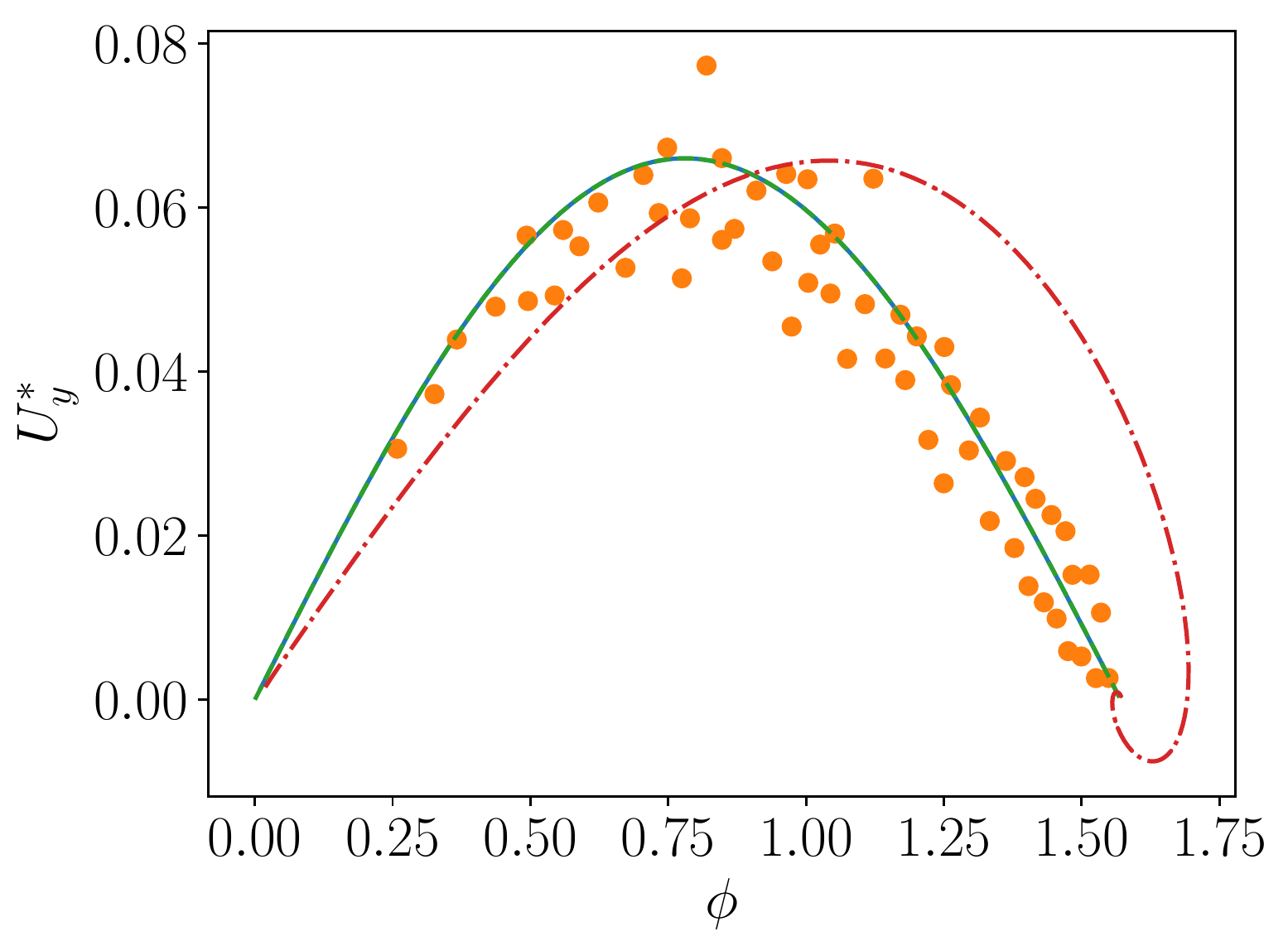}
\includegraphics[width=5cm]{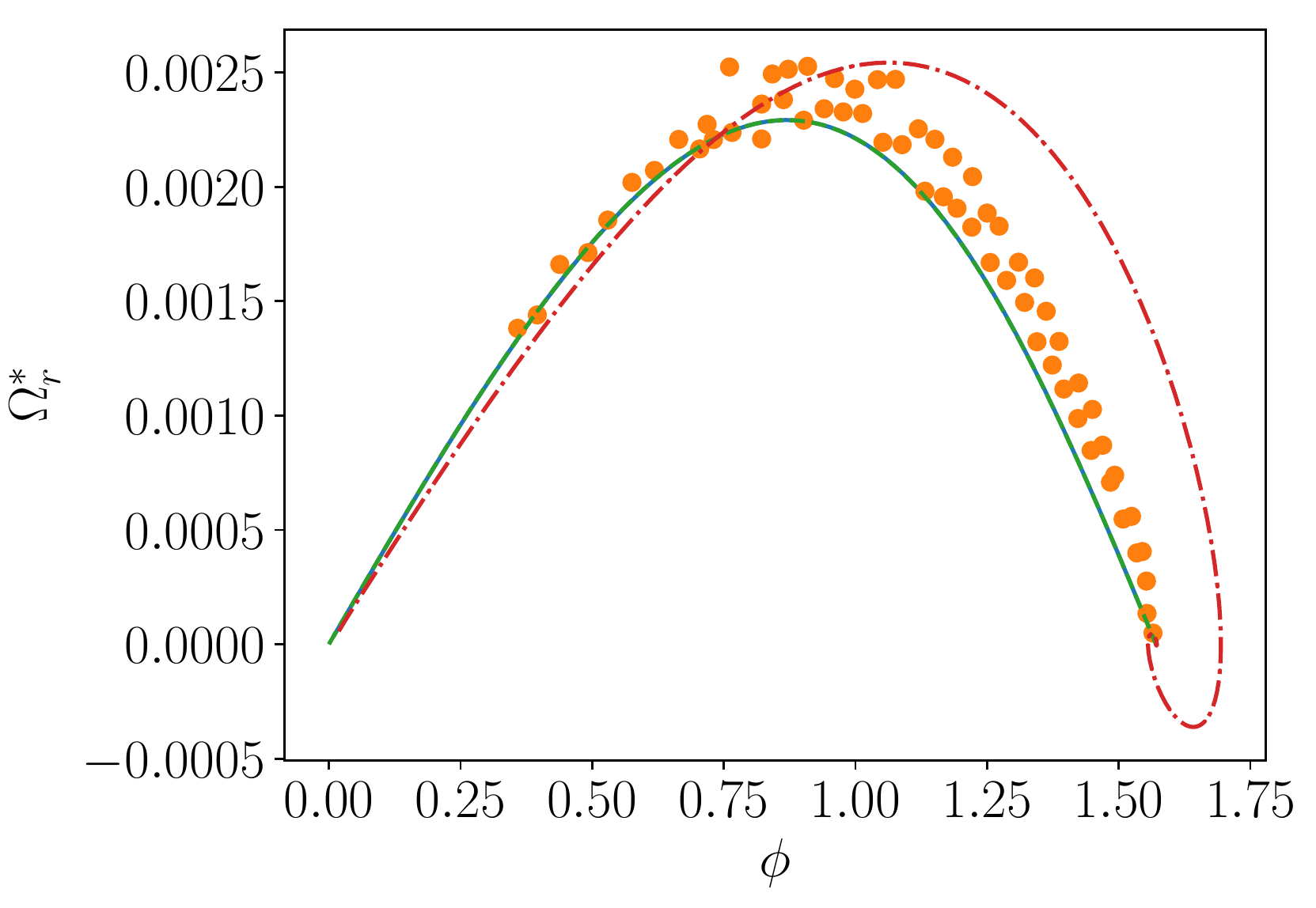}\\
\hspace{0.5cm}$(a)$ \hspace{4.5cm} $(b)$ \hspace{4.5cm} $(c)$\\
\includegraphics[width=5cm]{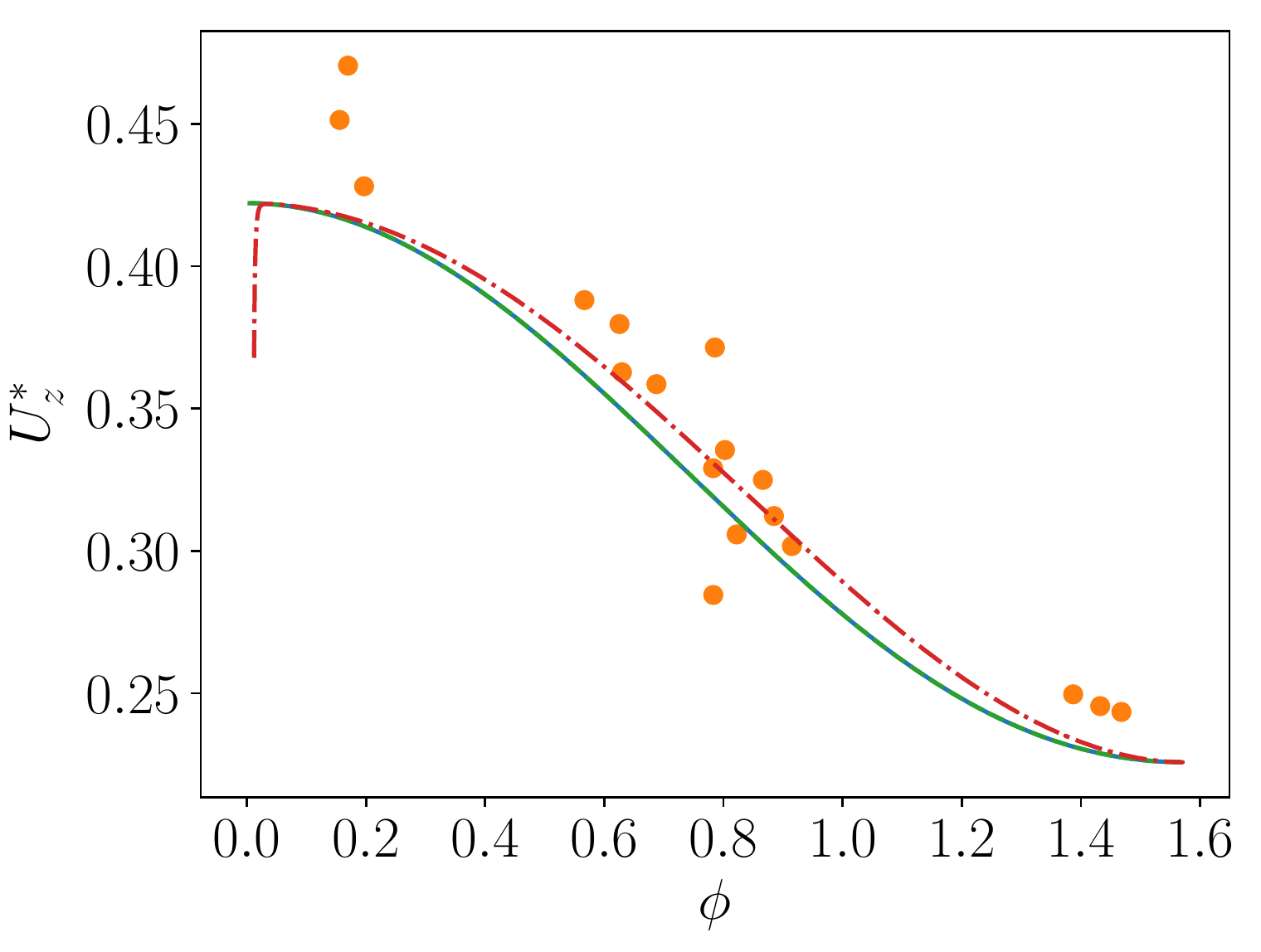}
\includegraphics[width=5cm]{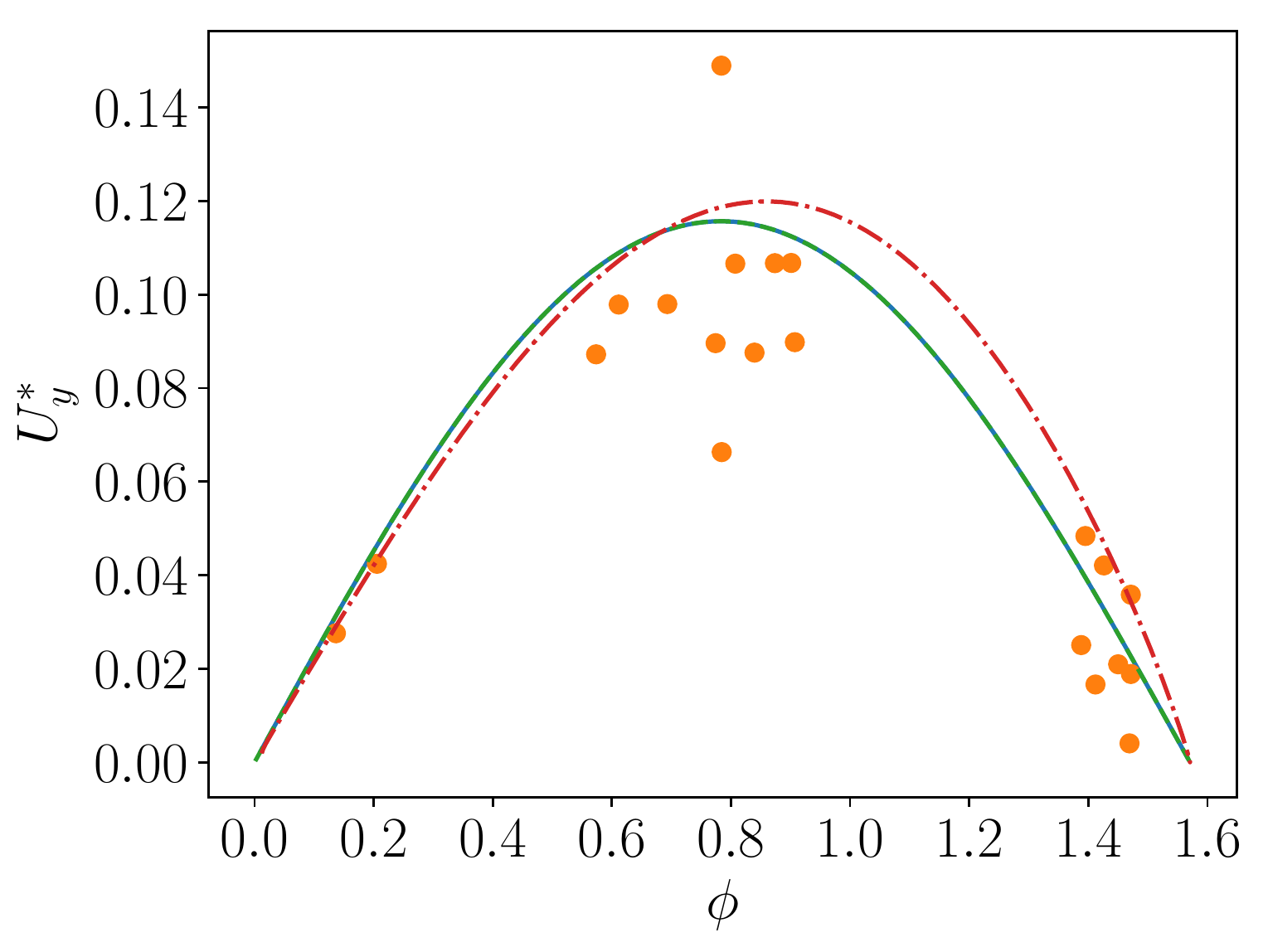}
\includegraphics[width=5cm]{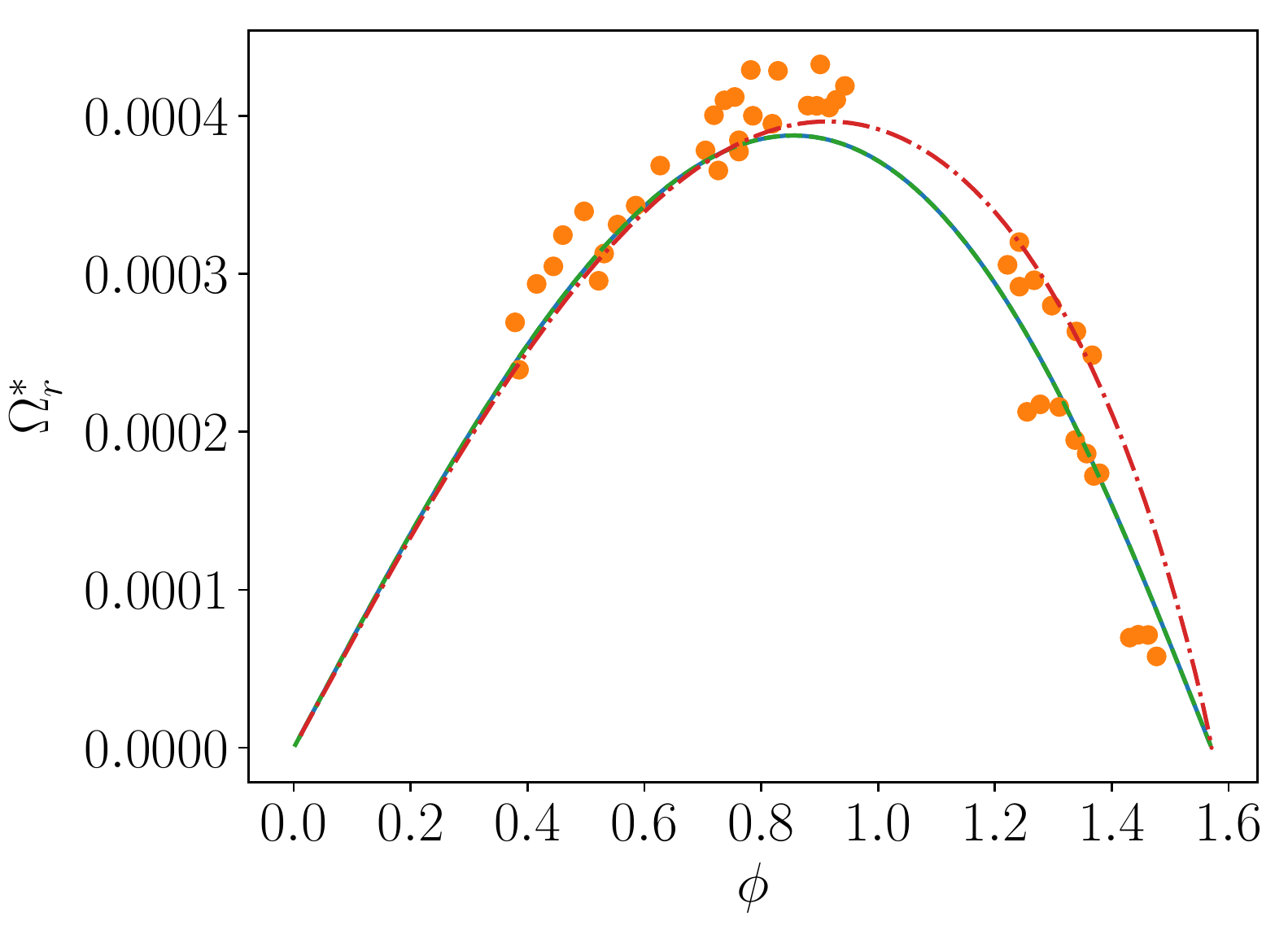}\\
\hspace{0.5cm}$(d)$ \hspace{4.5cm} $(e)$ \hspace{4.5cm} $(f)$
\caption{Dimensionless sedimenting velocities. $(a),$ $(d)$ : vertical sedimenting velocity, $(b)$, $(e)$ : drift velocity and $(c)$, $(f)$ : angular velocity. $\bullet$ : \citet{roy2019} experiments for $Ar \approx 0.76$ and $\bar{\rho} \approx 1.16$. $-$ : prediction from equations \ref{eq:UpA**} - \ref{eq:OmegarA**}, $--$ : prediction from equations \ref{eq:UpA**qs} - \ref{eq:OmegarA**qs}, $-\cdot$ : prediction from the unsteady equations \ref{eq:UpA**} - \ref{eq:OmegarA**} for $Ar \approx 0.76$, $\bar{\rho} = 1000$. Top panel corresponds to $\chi \approx 20.5$ and bottom panel to $\chi \approx 101$.}
\label{fig:roy}
\end{figure}

The sedimenting velocities and angular velocities as a function of $\phi$, from the experiments of \citet{roy2019}, are shown in Figures \ref{fig:roy}. Although the angular velocity is not directly given in their work, it can be deduced from their plot of the inertial torque, as they utilize a quasi-steady balance for angular momentum. The results obtained in this study perfectly match the experimental data. Furthermore, the difference between the sets of equations (\ref{eq:UpA**} - \ref{eq:OmegarA**}) and (\ref{eq:UpA**qs} - \ref{eq:OmegarA**qs}) is once again negligible, indicating that particle inertia is negligible in the experiments of \citet{roy2019}. However, when considering more inertial particles ($\bar{\rho}=1000$), deviations from the experimental data are observed. Specifically, the dimensionless velocities remain largely unchanged in amplitude, but a noticable shift to larger $\phi$ is observed. The system of ODEs represented by Equations \ref{eq:UpA**} - \ref{eq:OmegarA**} can be visualized as a phase portrait, as shown in Figures \ref{fig:roy} (c) and (f). Notably, the stable fixed point located at $\phi = \pi/2$ is identified as a focus or spiral.


\begin{figure}[h]
\centering
\includegraphics[width=6cm]{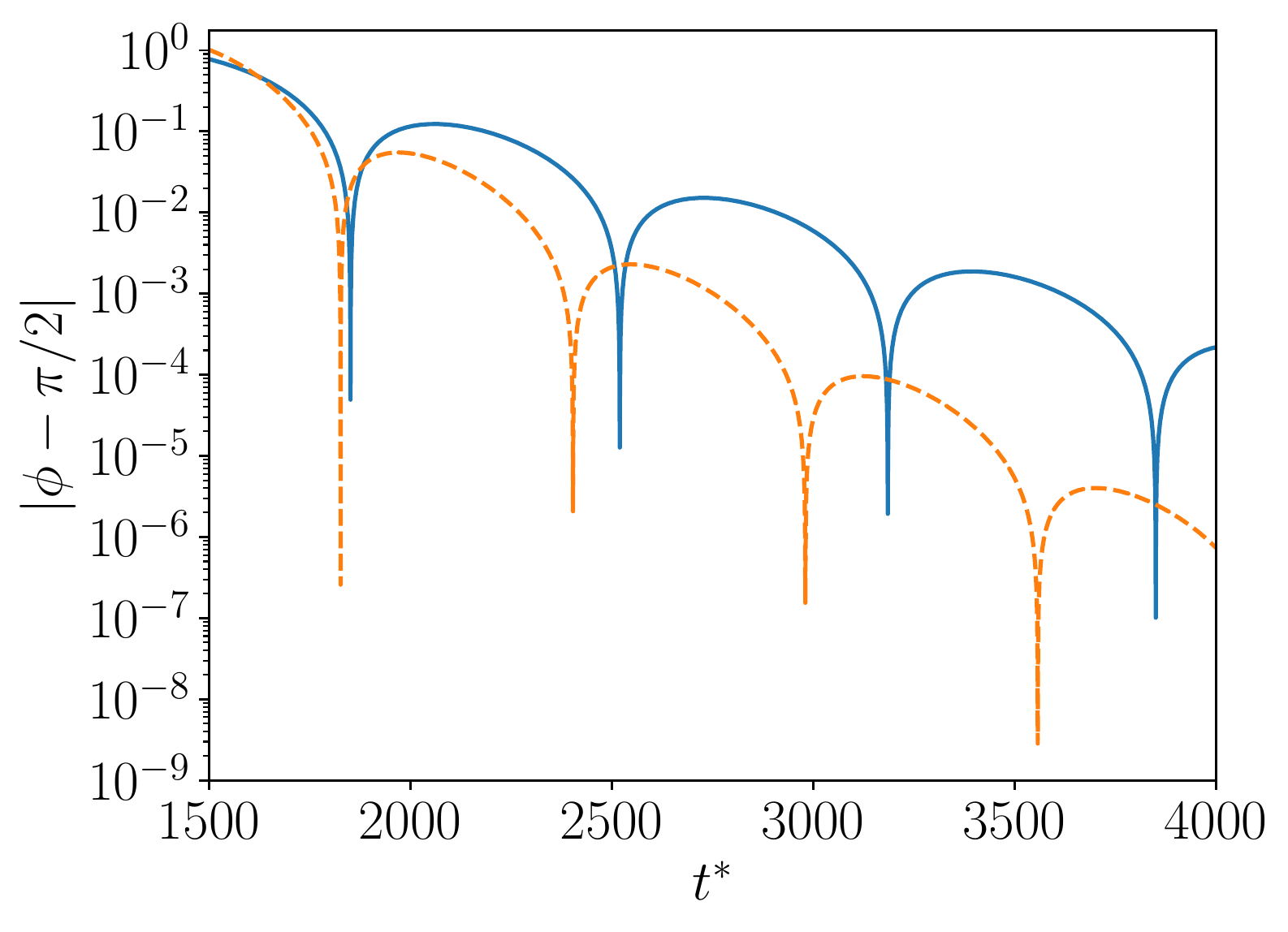}
\caption{Evolution of $|\phi -\pi/2|$ as function of time with $Ar \approx 0.76$, $\bar{\rho} = 1000$ and $\chi \approx 20.5$. $-$ : prediction from equations \ref{eq:UpA**} - \ref{eq:OmegarA**}. $--$ : prediction from equation \ref{eq:delta} with $Re\approx 0.156$ and $Re_\Omega \ll 1$.}
\label{fig:roy_phi}
\end{figure}

The oscillatory behaviour of the angle as a function of time is evidenced in figure \ref{fig:roy_phi} for $\chi \approx 20.5$. Exponential decay of $\phi$ over time is observable for both the simplified model, expressed in Equation \ref{eq:delta}, and the complete numerical solution. However, one may observe discrepancies between the two models both for the decrease slope and oscillation period. This is not surprising due to the many assumptions required to derive the model. The analytical solution only offers a qualitative depiction of the underdamped behaviour in this inertial configuration.

\subsection{Direct numerical simulations for $Re \sim 1$} 


In the preceding subsection, we have demonstrated the unexpected agreement of the quasi-steady equations with various literature findings for $Re \ll 1$. However, no experiments of such nature have been conducted for higher Reynolds numbers, and to the best of our knowledge no available simulations. Therefore, we conducted direct numerical simulations to validate the applicability of the model at higher Reynolds numbers ($Re \sim 1$).

\subsubsection{Numerical methodology}

Computations are carried out with the PeliGRIFF code \citet{wachs2015}. This code was used in the past to investigate the settling of spherical and angular particles \citep{rahmani2014,seyed2019}. In brief, the code solves the three-dimensional Navier-Stokes equations using a finite-volume discretization on a staggered grid. The time-stepping strategy for the fluid phase is done thanks to a second-order time-accurate Crank-Nicolson and Adams-Bashforth schemes. In order to enforce the rigid-body motion inside the solid region a Lagrange Multiplier/Fictitious Domain (DLM/FD) is used. We make use of a uniform distribution of the lagrangian points along the surface of the cylindrical body as detailed in \citep{pierson2019}. More details on the numerical methods can be found in \citet{wachs2015}.

\begin{figure}[h]
\centering
\begin{tikzpicture}[scale=2.5]

\def\opacity{0.8}

\draw (-0.5,-0.5,-0.5) -- ++(1,0,0) -- ++(0,2,0) -- ++(-1, 0, 0) -- cycle;
\draw (-0.5,-0.5,-0.5) -- ++(0,2,0) -- ++(0,0,1) -- ++(0, -2, 0) -- cycle;
\draw[] (0.5,1.5,0.5) -- ++(-1,0,0) -- ++(0,-2,0) -- ++(1, 0, 0) -- cycle;
\draw[] (0.5,1.5,0.5) -- ++(0,-2,0) -- ++(0,0,-1) -- ++(0, 2, 0) -- cycle;

\draw[orange!60, opacity=0.2 ,fill=orange!60, fill opacity=0.2](-0.5,0.25,-0.5) -- ++(1,0,0) -- ++(0,0.5,0) -- ++(-1, 0, 0) -- cycle;
\draw[orange!60, opacity=0.2 ,fill=orange!60, fill opacity=0.2](-0.5,0.25,-0.5) -- ++(0,0.5,0) -- ++(0,0,1) -- ++(0, -0.5, 0) -- cycle;
\draw[orange!60, opacity=0.2 ,fill=orange!60, fill opacity=0.2] (0.5,0.75,0.5) -- ++(-1,0,0) -- ++(0,-0.5,0) -- ++(1, 0, 0) -- cycle;
\draw[orange!60, opacity=0.2 ,fill=orange!60, fill opacity=0.2] (0.5,0.75,0.5) -- ++(0,-0.5,0) -- ++(0,0,-1) -- ++(0, 0.5, 0) -- cycle;

\node[cylinder, 
    draw = black, 
    text = black,
    cylinder uses custom fill, 
    cylinder body fill = black!10, 
    cylinder end fill = black!40,
    aspect = 0.2, 
    shape border rotate = 90, rotate =225, minimum height=0.8cm, minimum width=0.2cm] (c) at (0,0.45,0) {};

\draw[->, thick] (2.,1.5,2)--(2,1,2) node[left]{$\mathbf{g}$};

\draw[<->, thick] (-0.5,-0.45,-0.5)-- node[above]{$\mathcal{L}$} ++(1,0,0) ;
\draw[<->, thick] (-0.55,-0.5,-0.5) -- node[left]{$2\mathcal{L}$} ++(0,2,0);
\draw[<->, thick] (-0.55,-0.5,-0.5) -- node[left]{$\mathcal{L}$} ++(0,0,1);


\draw[->] (-1,0,0)--(-1,0,-0.5)node[right]{$x$};
\draw[->] (-1,0,0)--(-0.5,0,0)node[right]{$y$};
\draw[->] (-1,0,0)--(-1,-0.5,0)node[below]{$z$};

\end{tikzpicture}
\caption{Scheme of the computational domain (not to scale).}
\label{fig:domain}
\end{figure}
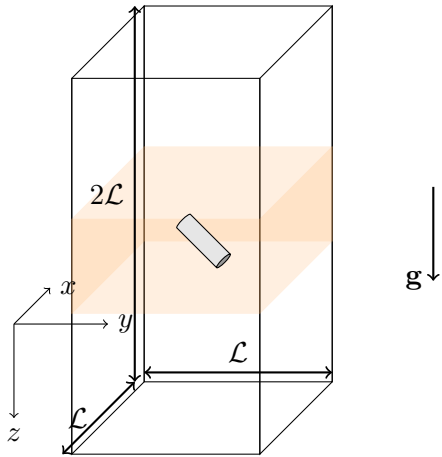
We explore the effect of $\chi$, $Ar$ and $\bar{\rho}$ within the range $\chi = \{2,4\}$, $Ar = \{24,96\}$ and $\bar{\rho} = \{1.5,10\}$. The cylinder is initiated with an angle $\phi =5 ^\circ$ and it is released at rest. In addition, a case with a larger density ratio was computed, starting at $\phi=60^\circ$, to investigate the damped oscillation regime. The computational domain for this problem is depicted in Figure \ref{fig:domain}. Its characteristic size length depends on the aspect ratio: for $\chi=2$, $\mathcal{L} = 12D$ while for $\chi=4$ $\mathcal{L} = 15D$. Note that the value prescribed for $\chi=2$ is larger than the one used in \citep{seyed2019}. We have checked that the present domain was sufficiently large by increasing its size of $25\%$ in all directions finding less than $2\%$ error on $\Omega L /U$ for the more challenging configuration ($Ar=24$, $\chi=4$). Obviously, larger domains might be used to avoid this small effect of the boundaries at the expense of a substantially larger numerical cost. However, we have to recall that the present paper is not aimed to provide domain size perfectly independent simulations of the problem but rather to provide bound to the quasi-steady assumptions. To this aim, we compare the numerical results to the model of section \ref{sec:scaling} which heavily relies on fit and also contains some error with respect to the simulation they are extracted from. Regarding the boundary conditions, the domain is biperiodic in the lateral directions while zero velocity and outflow conditions are imposed on the upstream and downstream boundaries, respectively. A uniform cell distribution is imposed in a rectangular region depicted in Figure \ref{fig:domain} as an orange volume. This region is located  $2D$ below the middle of the domain and extends up to $6D$ in the direction of gravity. In this flow region, $25$ cells are distributed per body diameter which is sufficient to accurately compute the dynamic for the range of Reynolds number investigated \citep{pierson2019}. The number of cells per mesh is 26 millions for $\chi=2$ and 46 millions for $\chi=4$. The time step is imposed such that the CFL always fall below $0.25$. To prevent the cylinder to exit the numerical domain the computational domain is moved in the vertical direction so as to keep the particle at least at a distance of $10D$ of the upstream boundary. More details on the domain translation technique can be found in \citet{rahmani2014,seyed2019} The magnitude of this domain translation is chosen to be the grid size ($D/25$). The simulations are run up until the cylindrical particle reaches its equilibrium position.

\subsubsection{Numerical results}



To investigate the impact of moderate inertia on the rod motion we first consider the situation $Ar=24$. In this scenario, the Reynolds number will fluctuate between $1.7$ and $1.9$ when $\chi$ equals $2$, and between $2$ and $2.7$ when $\chi$ is $4$.

\begin{figure}[h]
\centering
\includegraphics[width=5cm]{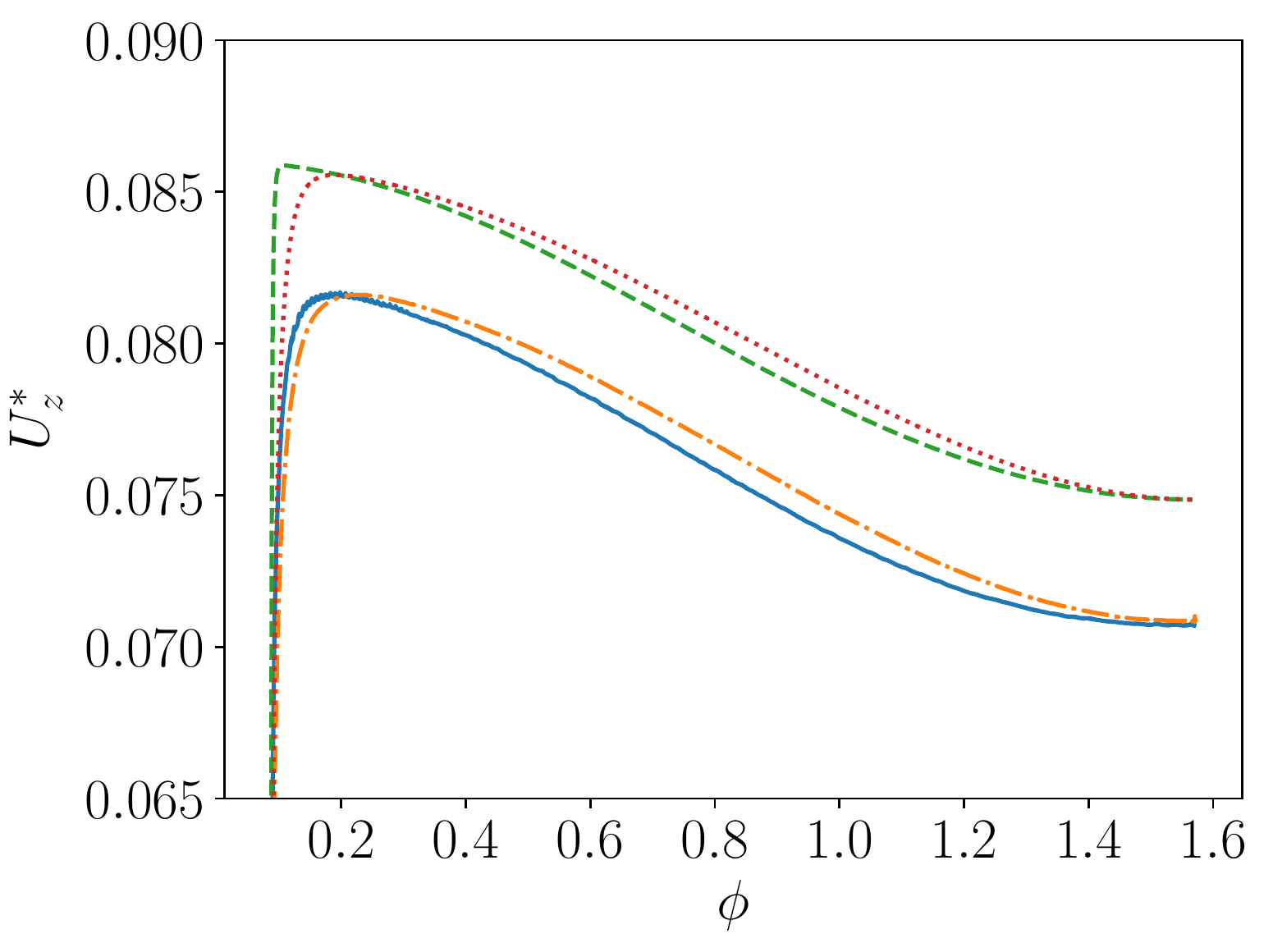}
\includegraphics[width=5cm]{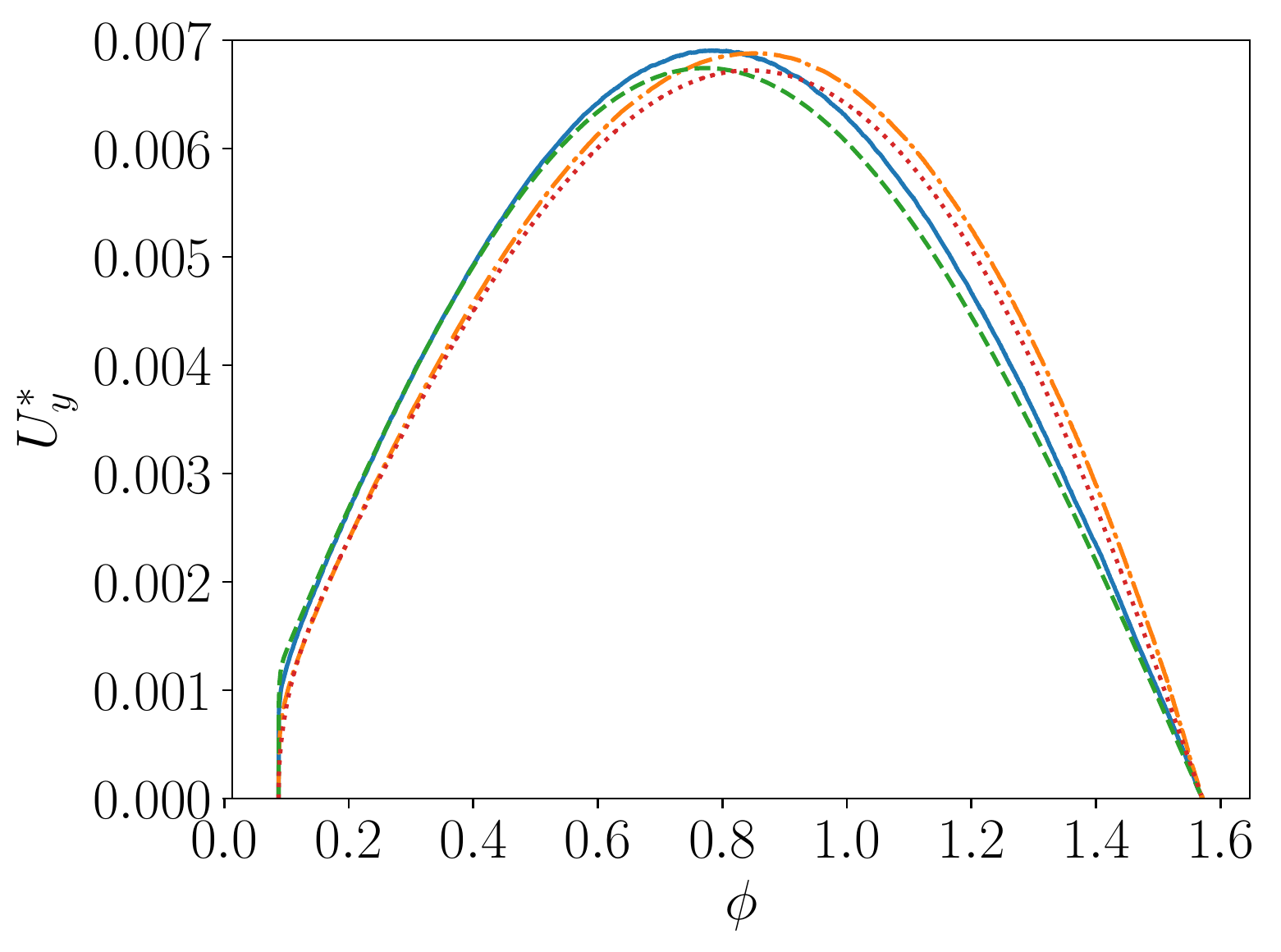}
\includegraphics[width=5cm]{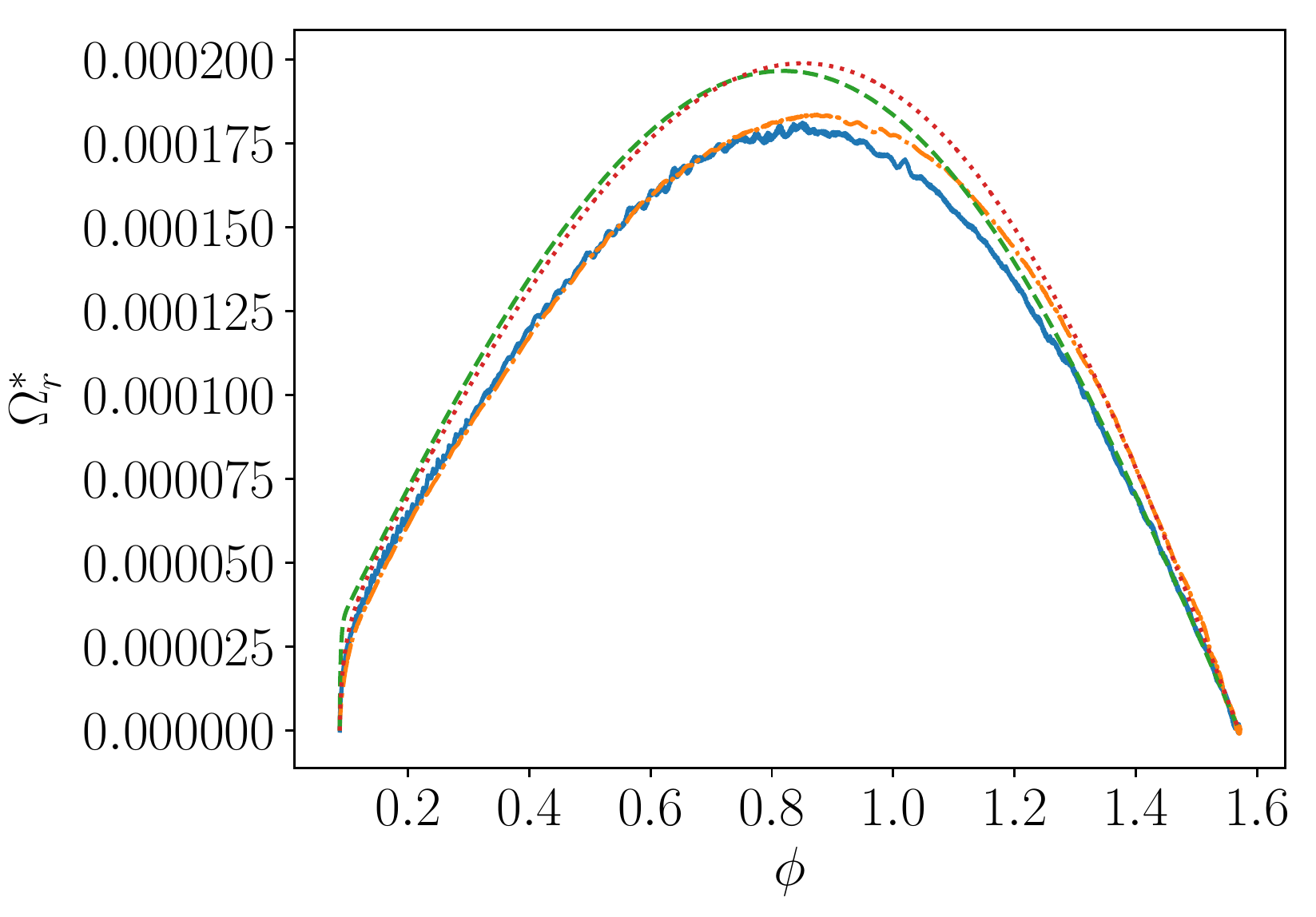}\\
\hspace{0.5cm}$(a)$ \hspace{4.5cm} $(b)$ \hspace{4.5cm} $(c)$\\
\includegraphics[width=5cm]{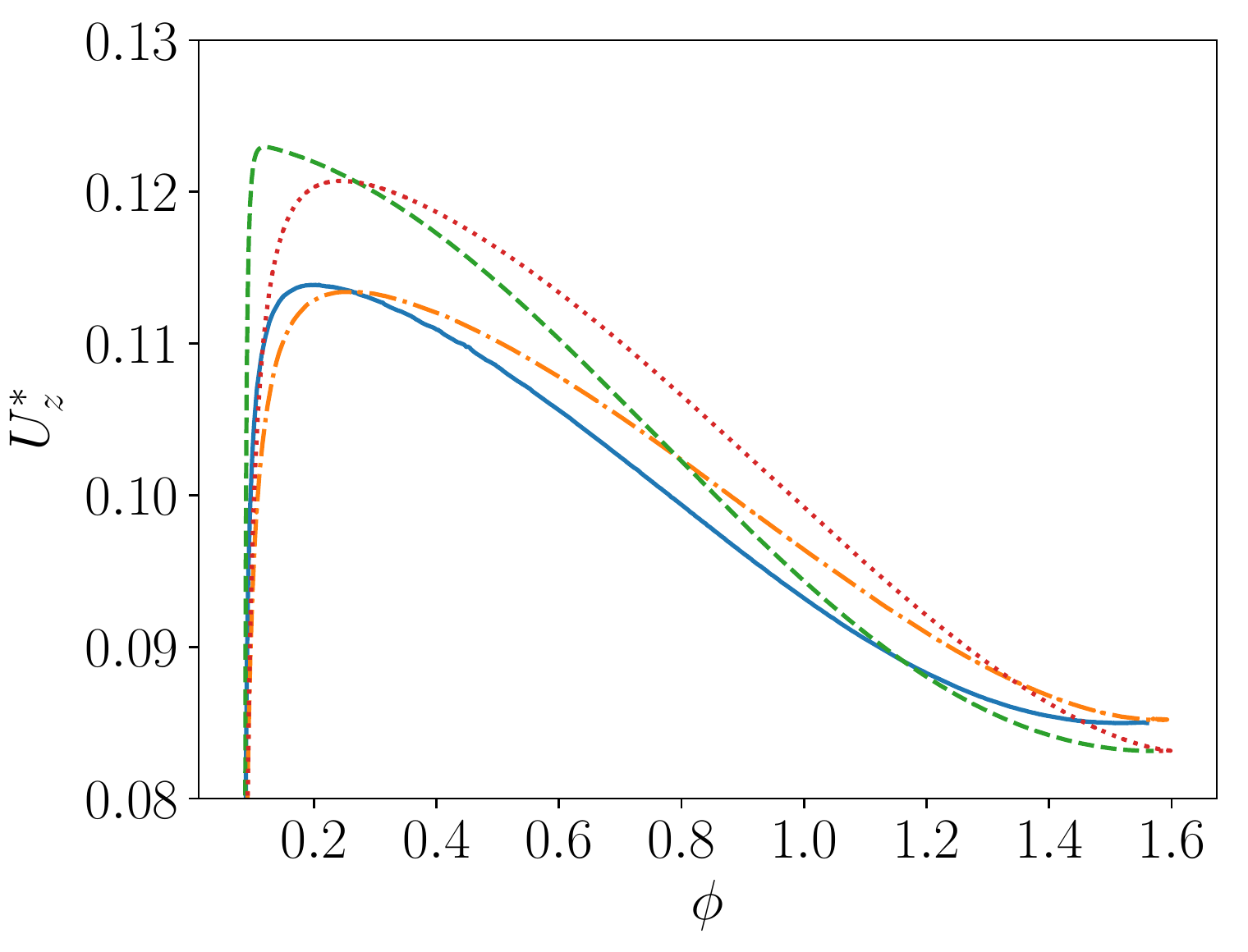}
\includegraphics[width=5cm]{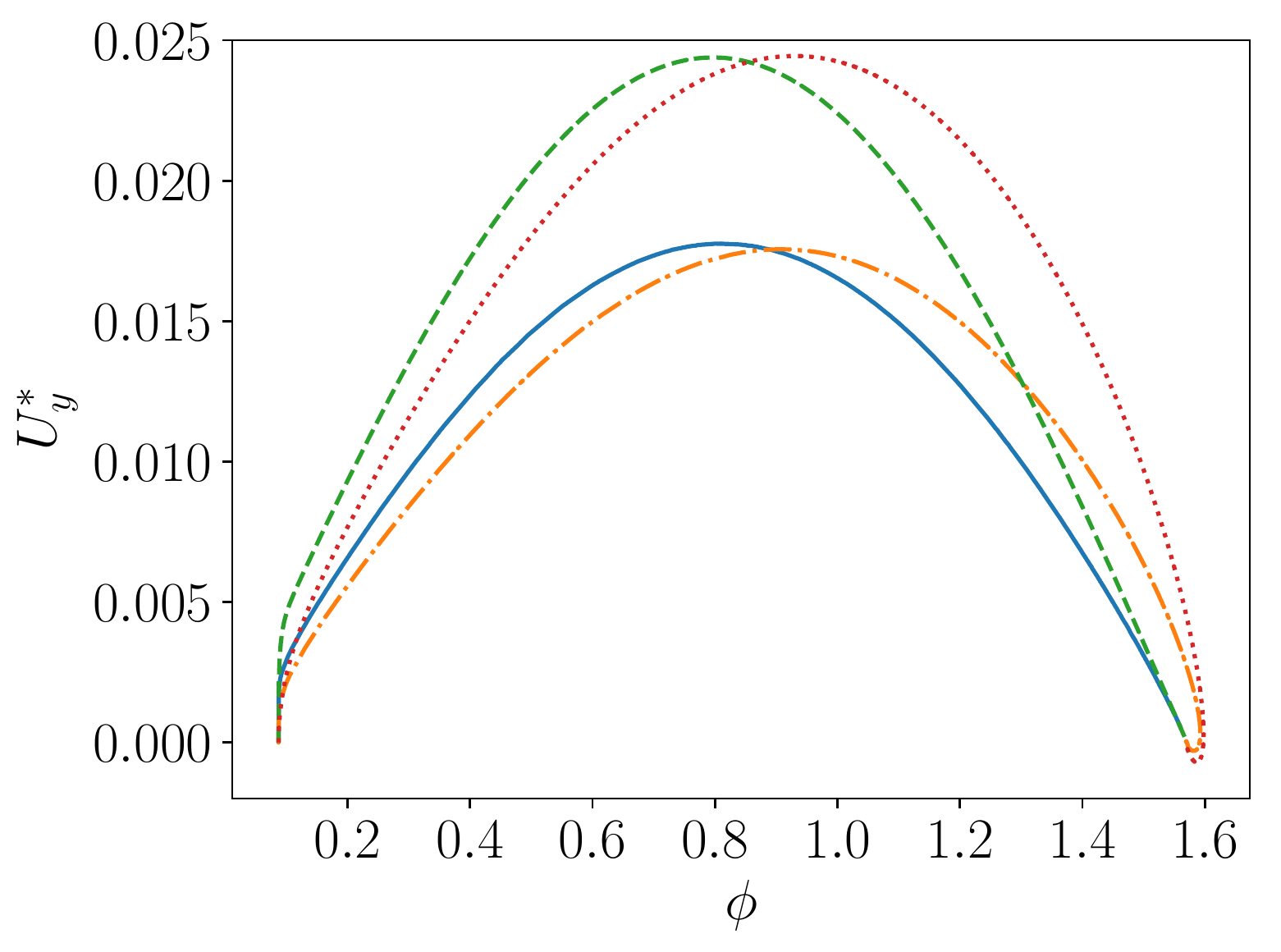}
\includegraphics[width=5cm]{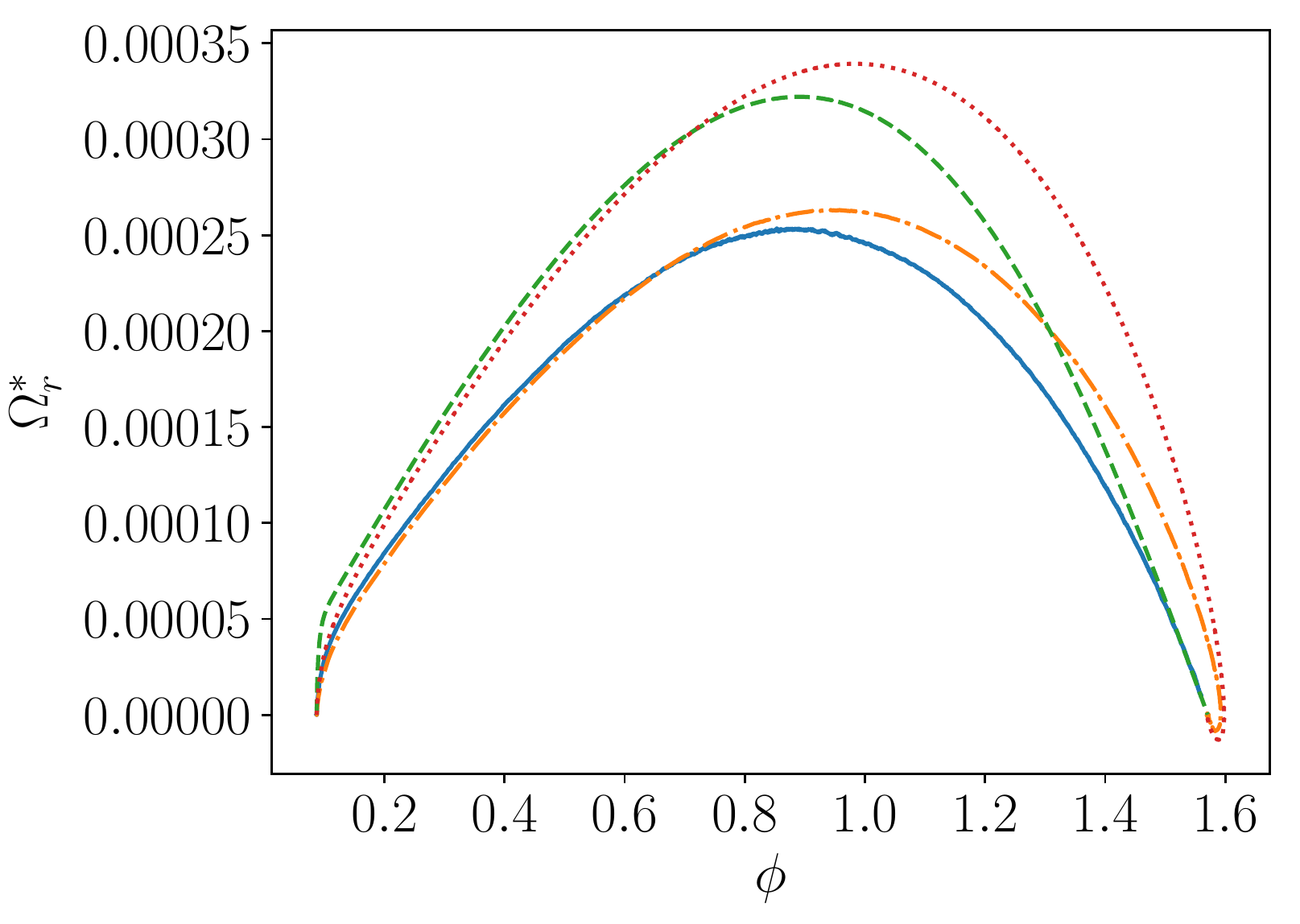}\\
\hspace{0.5cm}$(d)$ \hspace{4.5cm} $(e)$ \hspace{4.5cm} $(f)$
\caption{Dimensionless sedimenting velocities for $Ar = 24$. $(a),$ $(d)$ : vertical sedimenting velocity, $(b)$, $(e)$ : drift velocity and $(c)$, $(f)$ : angular velocity. $-$ : direct numerical simulation results with $\bar{\rho} = 1.5$, $-\cdot-$: direct numerical simulation results with $\bar{\rho} = 10$, $--$ : prediction from equations \ref{eq:UpA**} - \ref{eq:OmegarA**} with $\bar{\rho} = 1.5$, $\cdot\cdot$ : prediction from equations \ref{eq:UpA**} - \ref{eq:OmegarA**} with $\bar{\rho} = 10$. Top panel corresponds to $\chi = 2$ and bottom panel to $\chi = 4$.}
\label{fig:DNS_Ar24}
\end{figure}


For the shortest cylinder with aspect ratio $\chi=2$, there is a noticeable agreement between the results obtained from equations \ref{eq:UpA**} - \ref{eq:OmegarA**} (including particle inertia) and those from direct numerical simulations, as illustrated in Figure \ref{fig:DNS_Ar24} (a) - (c). Varying the density ratio from 1.5 to 10 has a negligible effect on the sedimentation velocities in this particular case. In Figure \ref{fig:DNS_Ar24} (d) - (f), we present the results for $\chi=4$. Although there is a qualitative agreement between the model and the numerical results, the model overestimates the drift velocity and the angular velocity. Moreover, for $\chi=4$, the effect of particle inertia on the sedimentation velocities is more pronounced. In particular, this case exhibits underdamped motion. To investigate this oscillating behavior further, we have carried out one simulation of the settling of a rod with the same dimensionless parameters ($Ar=24$, $\chi=4$), but with a much larger density ratio ($\bar{\rho}=50$). In order to maintain reasonable computational time, the simulation is initiated with $\phi=60^\circ$.
\begin{figure}[h]
\centering
\includegraphics[height=5cm]{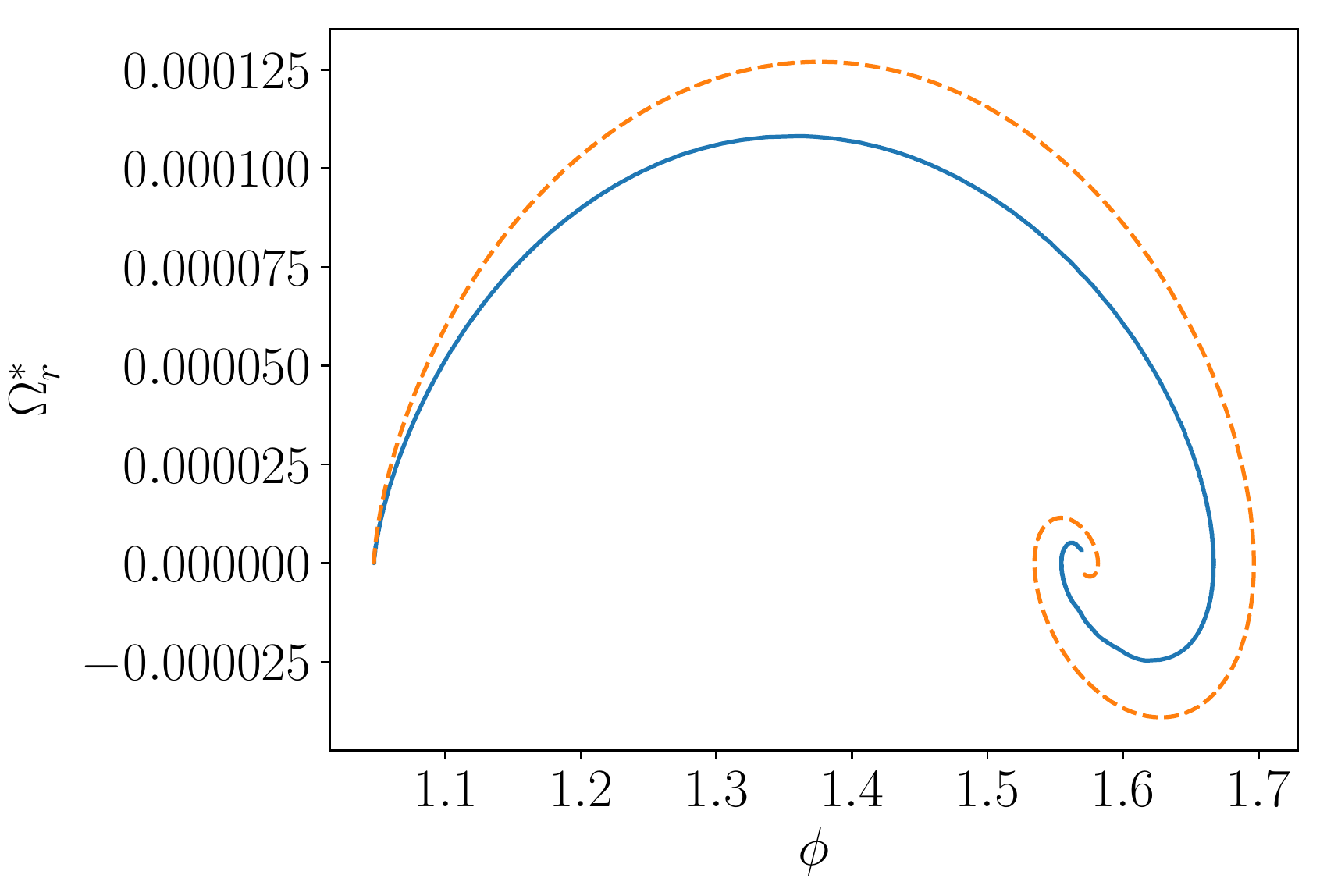}
\includegraphics[height=5cm]{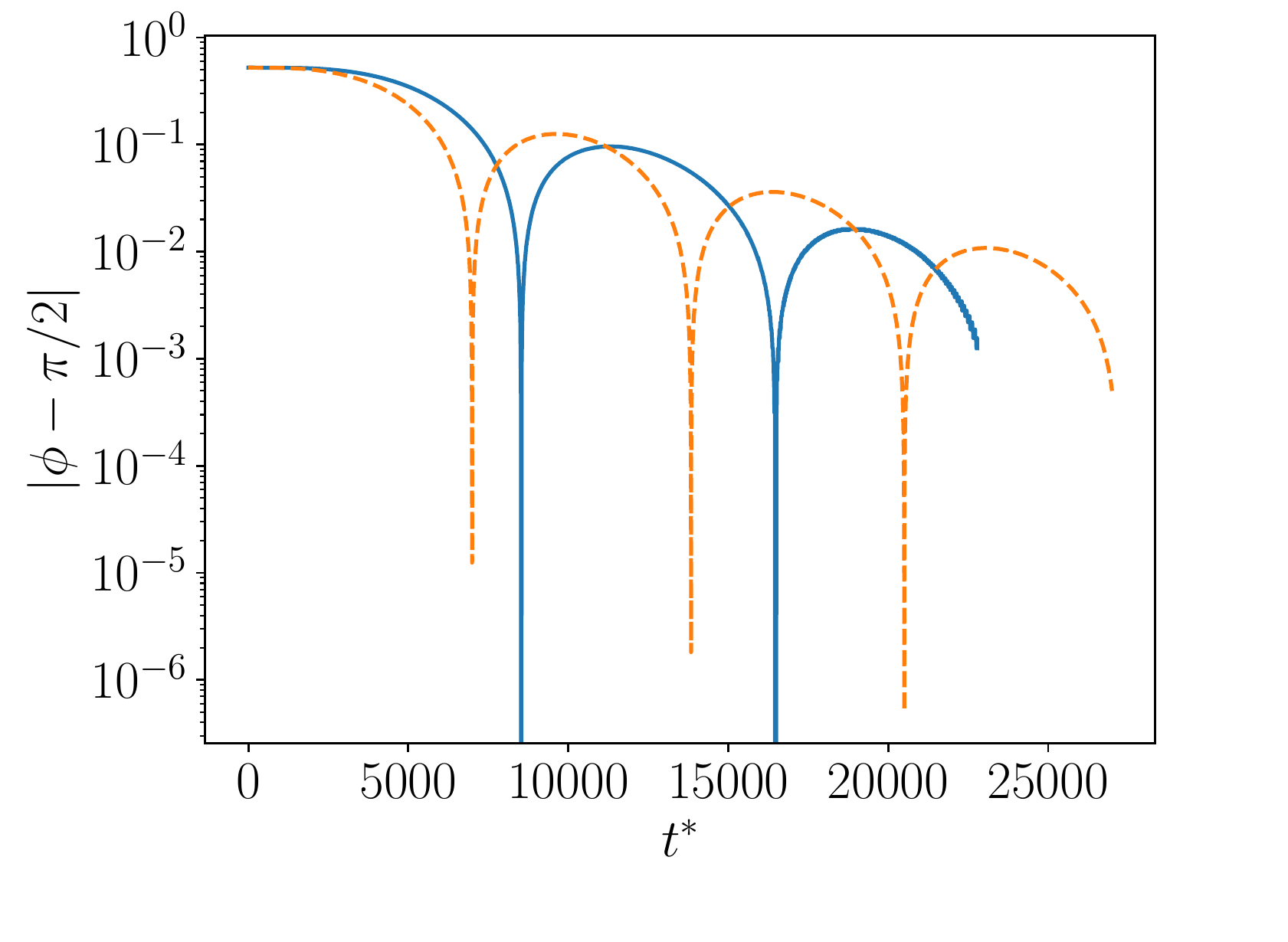}\\
\hspace{0.5cm}$(a)$ \hspace{5.5cm} $(b)$
\caption{Comparison between the simulation results and the quasi-steady model with $Ar=24$, $\chi=4$ and $\bar{\rho}=50$. (a) : Phase space diagram of the oscillator. (b) : Time evolution of the inclination angle. $-$ : direct numerical simulation results, $--$ : prediction from equations \ref{eq:UpA**} - \ref{eq:OmegarA**}. }
\label{fig:DNS_Ar24_rho50}
\end{figure}
There is a good agreement between the simulation and the model even if the model overpredicts the angular velocity (Figure \ref{fig:DNS_Ar24_rho50}). We recall that we neglect the history loads in equations \ref{eq:UpA**} - \ref{eq:OmegarA**} which may affect the initial transient since the rod starts from rest. One may also observe a good agreement between the oscillating period, the decrease in amplitude given by the numerical results and the model made of equations \ref{eq:UpA**} - \ref{eq:OmegarA**} (Figure \ref{fig:DNS_Ar24_rho50} (b)). This is somehow surprising since there is \textit{a priori} no reason for neglecting the history loads in the underdamped regime. They appear to have a negligible effect on this regime.

\begin{figure}[h]
\centering
\includegraphics[width=5cm]{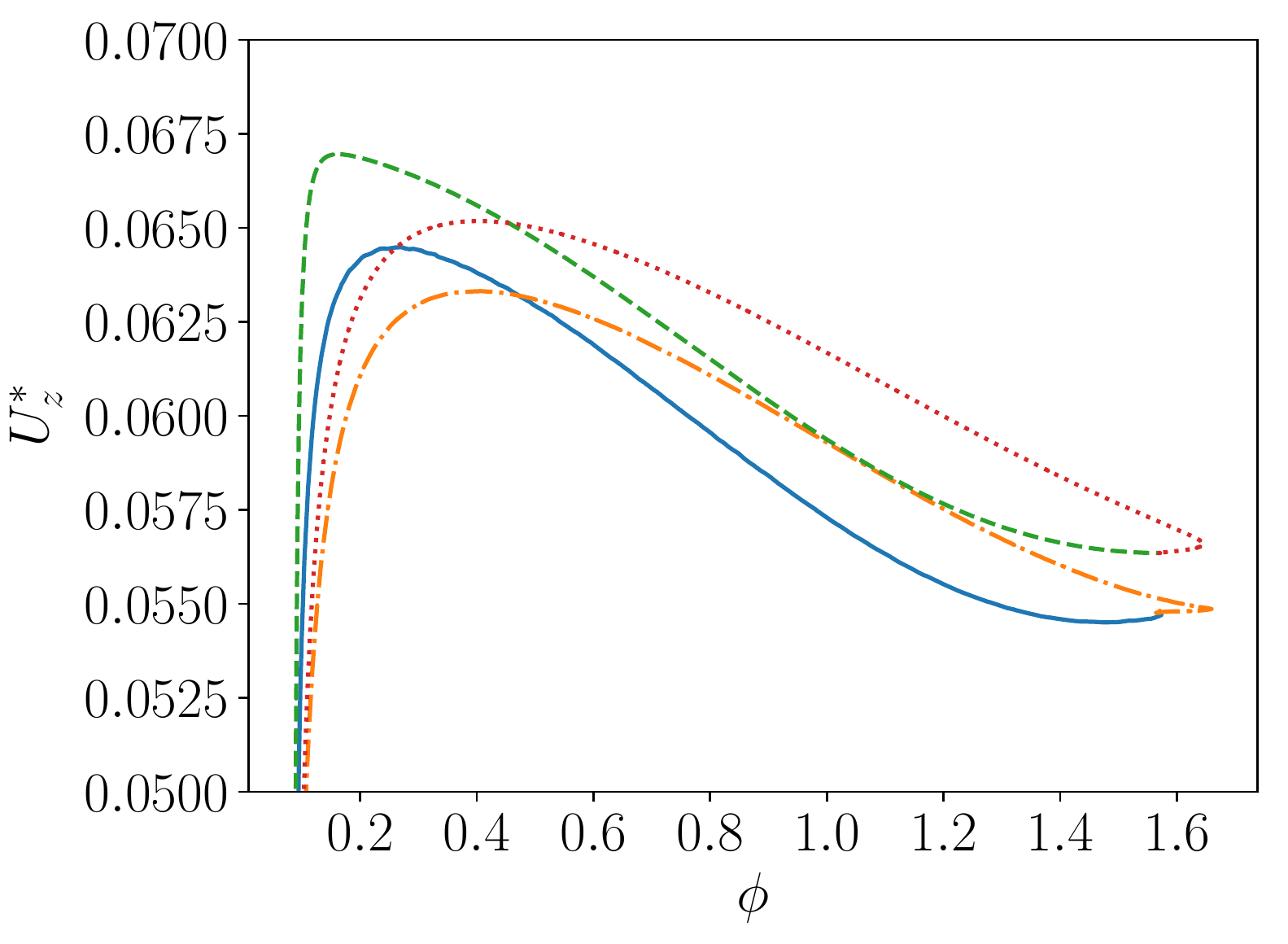}
\includegraphics[width=5cm]{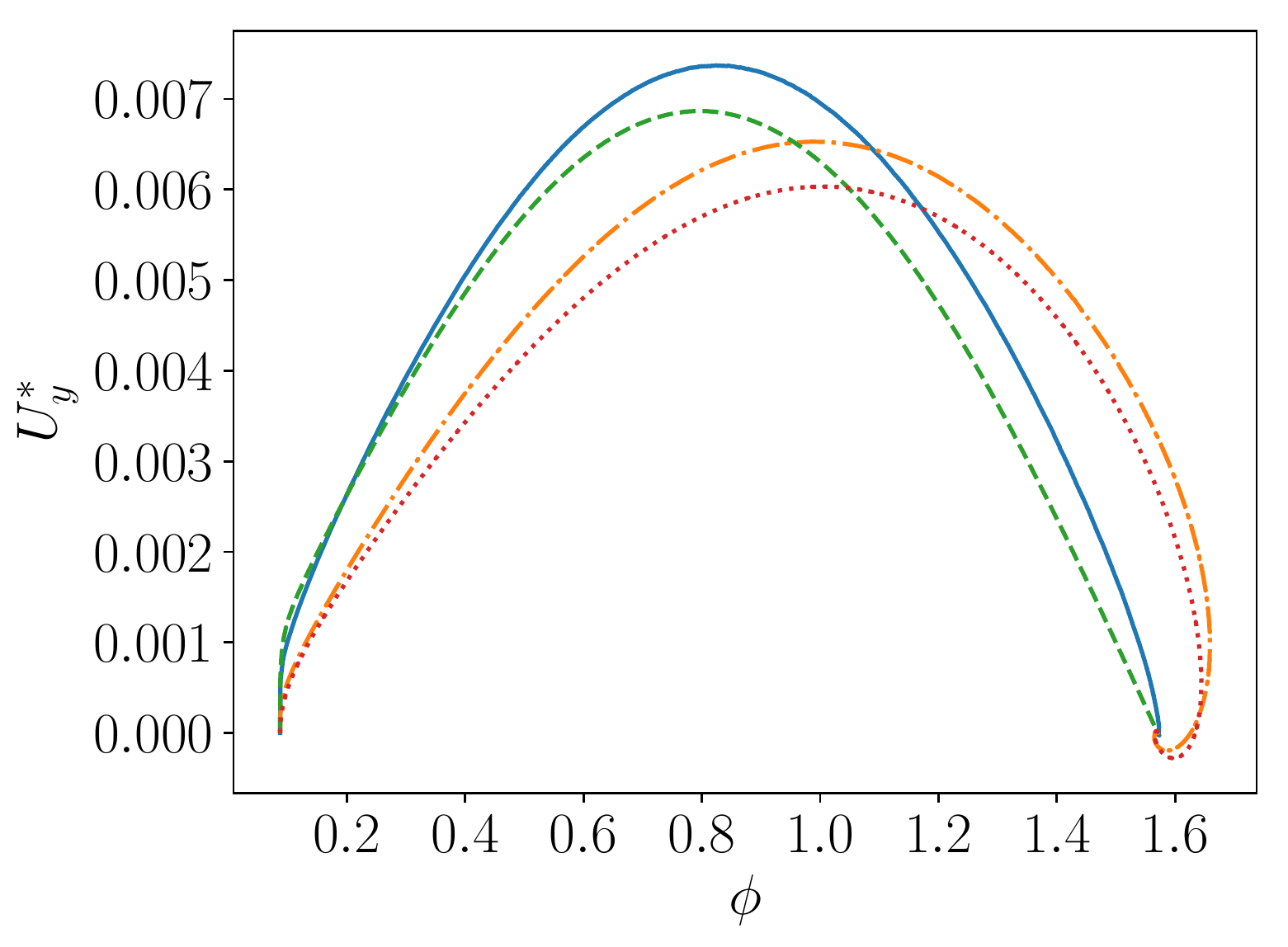}
\includegraphics[width=5cm]{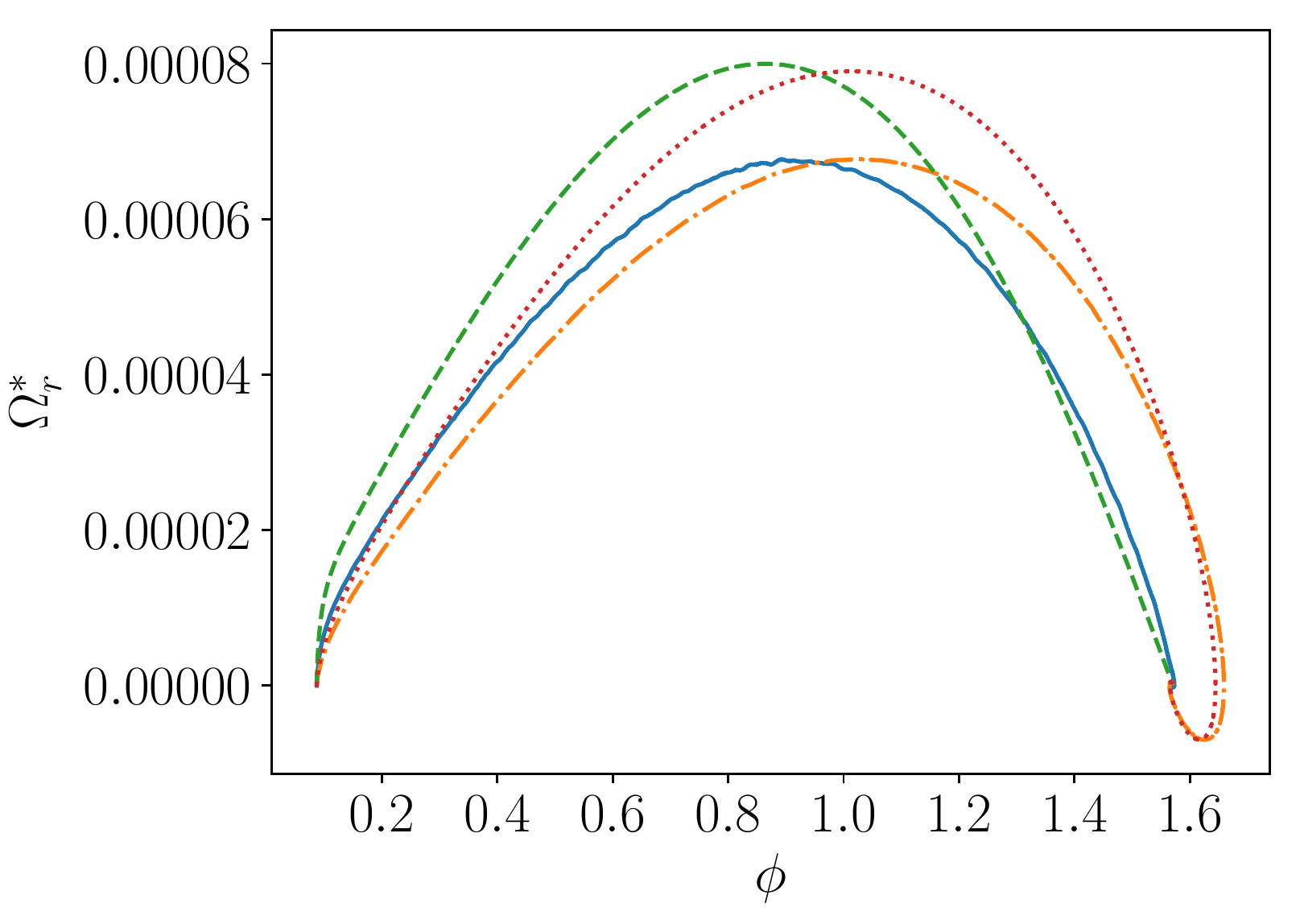}\\
\hspace{0.5cm}$(a)$ \hspace{4.5cm} $(b)$ \hspace{4.5cm} $(c)$\\
\includegraphics[width=5cm]{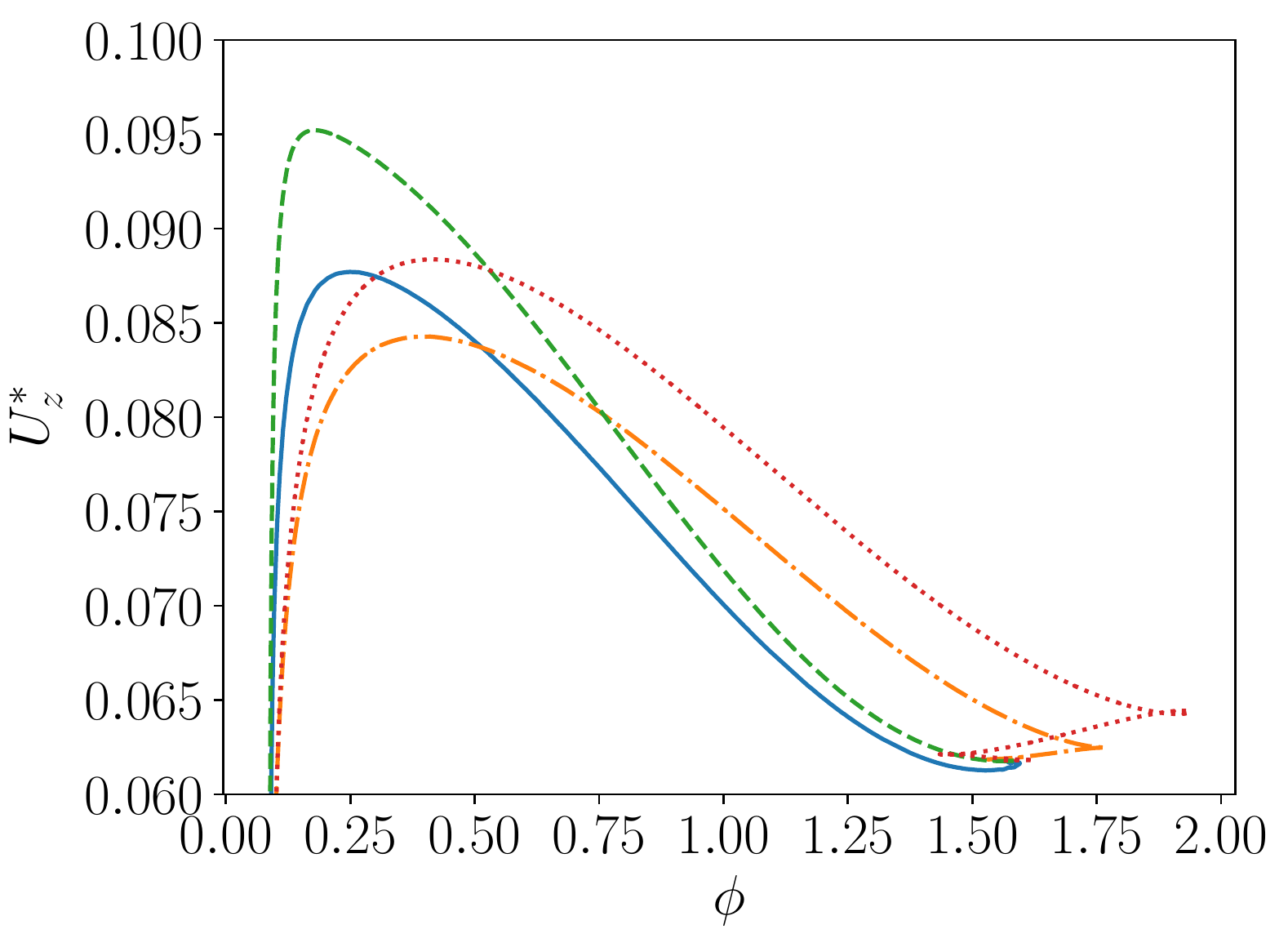}
\includegraphics[width=5cm]{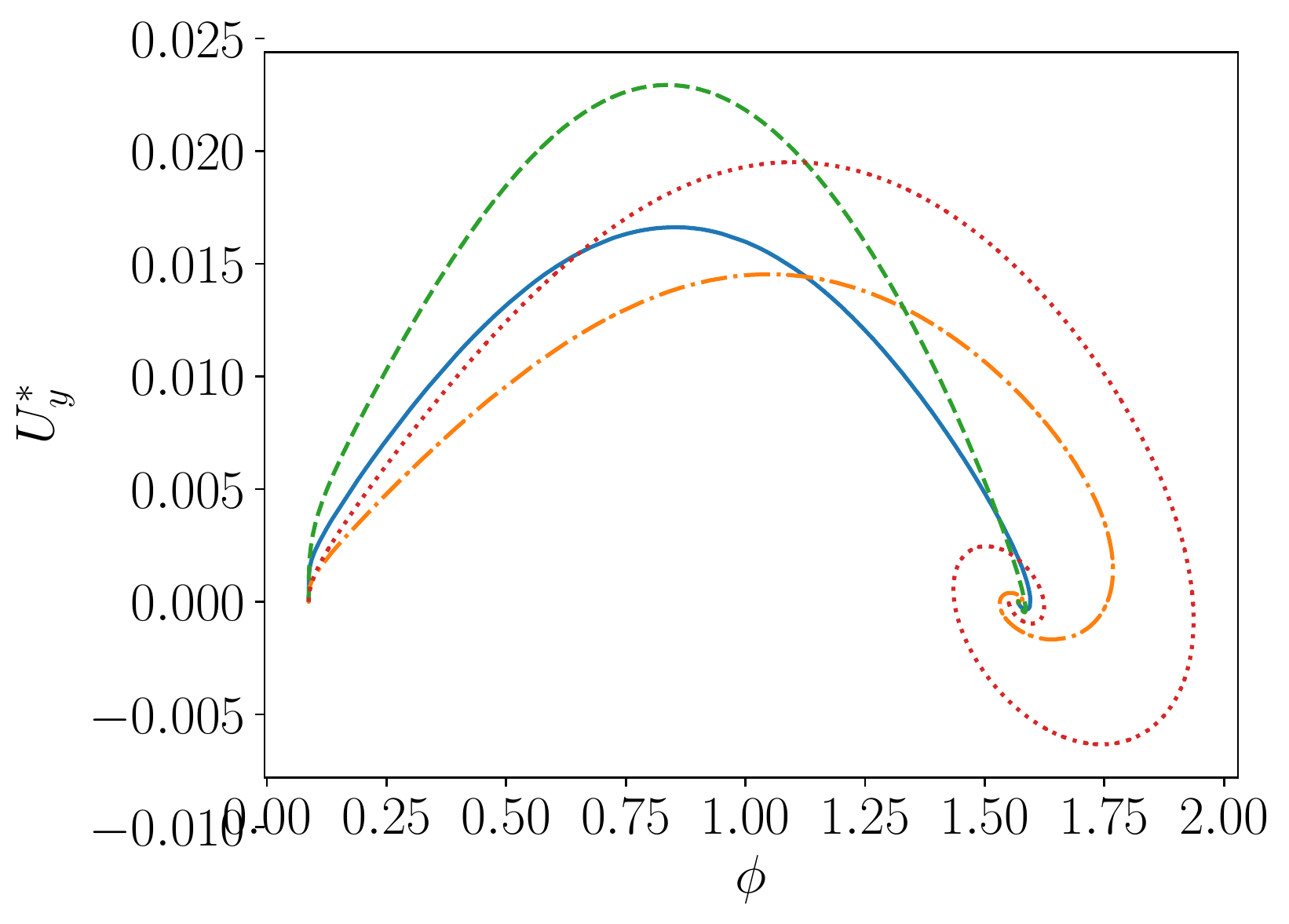}
\includegraphics[width=5cm]{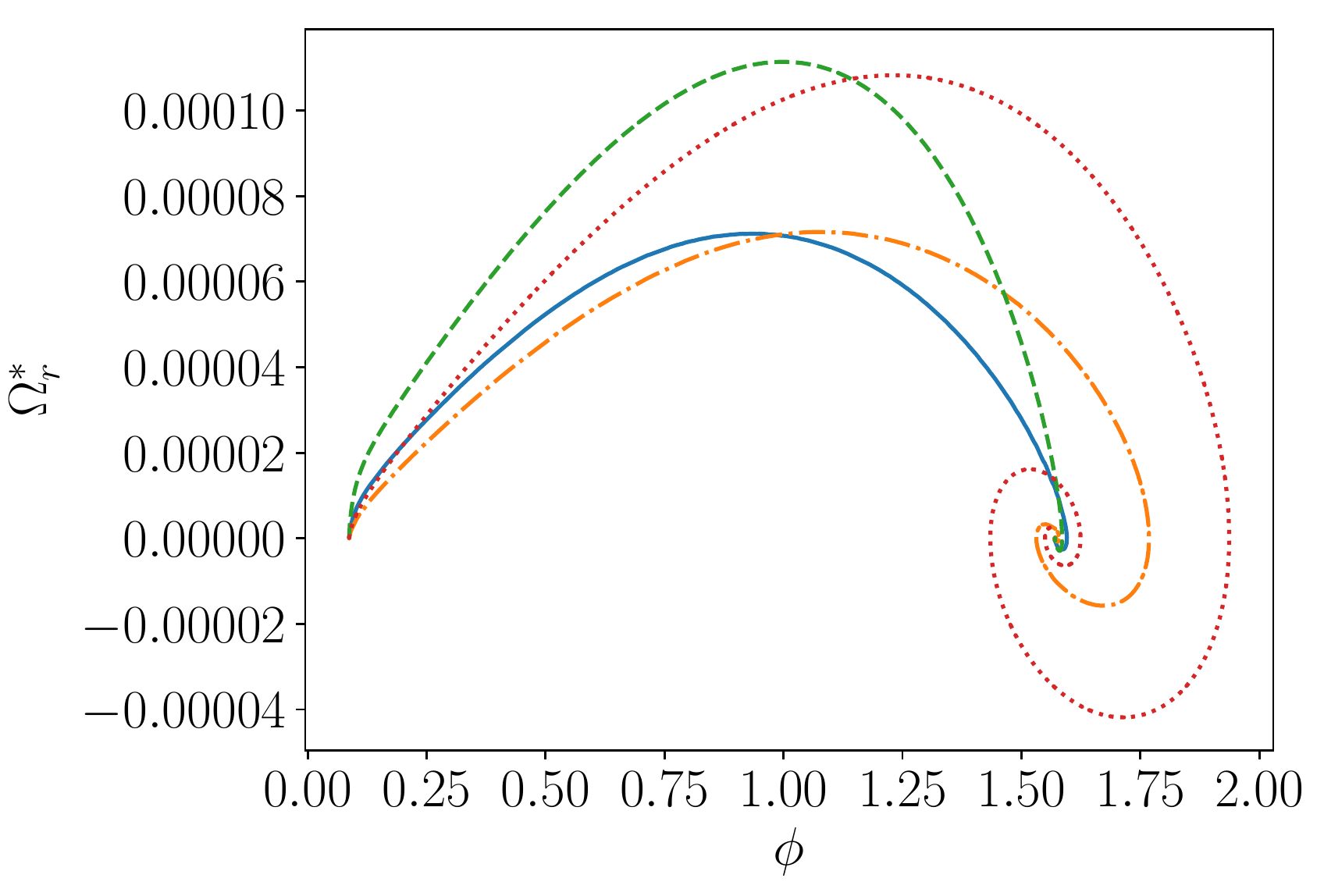}\\
\hspace{0.5cm}$(d)$ \hspace{4.5cm} $(e)$ \hspace{4.5cm} $(f)$
\caption{Dimensionless sedimenting velocities for $Ar = 96$. $(a),$ $(d)$ : vertical sedimenting velocity, $(b)$, $(e)$ : drift velocity and $(c)$, $(f)$ : angular velocity. $-$ : direct numerical simulation results with $\bar{\rho} = 1.5$, $-\cdot-$: direct numerical simulation results with $\bar{\rho} = 10$, $--$ : prediction from equations \ref{eq:UpA**} -\ref{eq:OmegarA**} with $\bar{\rho} = 1.5$, $\cdot\cdot$ : prediction from equations \ref{eq:UpA**}, \ref{eq:UqA**} and \ref{eq:OmegarA**} with $\bar{\rho} = 10$. Top panel corresponds to $\chi = 2$ and bottom panel to $\chi = 4$.}
\label{fig:DNS_Ar96}
\end{figure}

\begin{figure}[h]
\centering
\includegraphics[height=5cm]{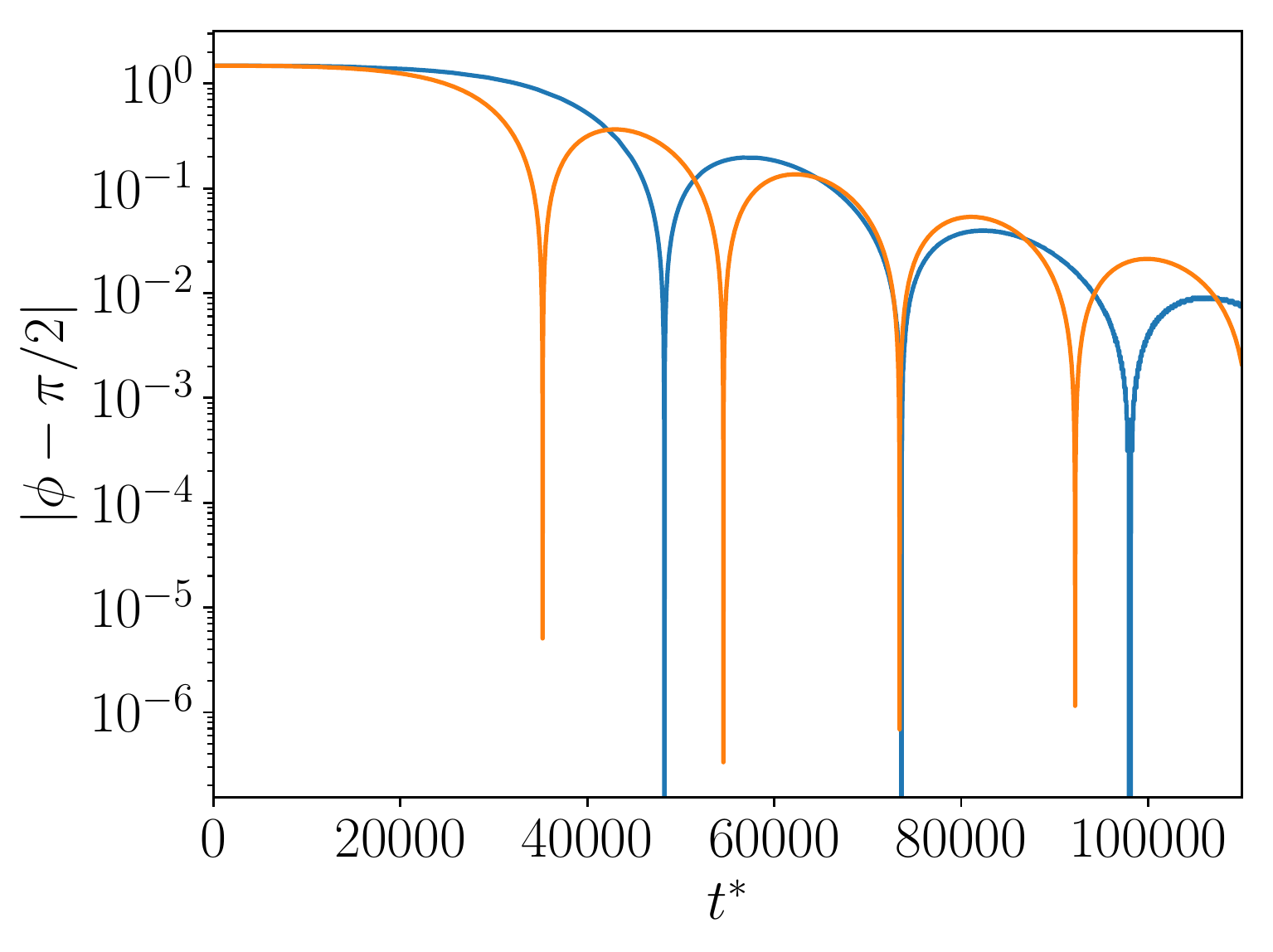}
\caption{Time evolution of the inclination angle ($Ar=96$, $\chi=4, \bar{\rho}=10$). $-$ : direct numerical simulation results, $--$ : prediction from equations \ref{eq:UpA**} - \ref{eq:OmegarA**}.}
\label{fig:underdamped_Ar96}
\end{figure}

We now consider much larger inertia effects ($Ar=96$). Within this regime, the Reynolds number is found to vary between $5.3$ and $6.2$ for $\chi=2$ and between $5.8$ and $8.6$ for $\chi=4$. Surprisingly, the quasi-steady model exhibits good agreement with numerical results for $\chi=2$ (as depicted in Figure \ref{fig:DNS_Ar96} (a)-(c)), despite the significant role played by inertia effects. However, for $\chi=4$, the model overestimates both the drift velocity and angular velocity (as shown in Figure \ref{fig:DNS_Ar96} (d)-(f)). We observe that all configurations result in underdamped regimes. This regime is much more pronounced for $\chi=4$ and $\bar{\rho}=10$ than for $\chi=2$ and $\bar{\rho}=1.5$ for which the amplitude of oscillation is small and nearly indistinguishable. Finally, we compare the numerical and model-based results (derived from Equations \ref{eq:UpA**} - \ref{eq:OmegarA**}) for $Ar=96$, $\chi=4$, and $\bar{\rho}=10$, by examining the oscillation period and the decay of amplitude. The results are displayed on Figure \ref{fig:underdamped_Ar96}. The model slightly underpredicts the amplitude decay as well as the oscillation period.

\section{Discussion and conclusion}
\label{sec:discussion}

Based on the results presented above, two main conclusions can be drawn. Firstly, it can be inferred that the quasi-steady assumption holds true across a broad range of dimensionless parameters, as it is supported by both the experiments conducted by \citet{cabrera2022} and \citet{roy2019} and by the direct numerical simulations. Although these experiments and simulations do not strictly adhere to the condition $Ar \ll 1/\chi$, the quasi-steady assumption is still valid for a significantly larger range of values than initially predicted. This can be attributed to the magnitude of the angular velocity $\Omega$, which has the correct scaling but is at least three order of magnitude smaller than the anticipated value. As a result, $\Omega L / U \ll 1$ for all configurations except the highest inertial simulations (see Table \ref{tab:omega}). Aditionnally, these findings support the notion that the quasi-steady assumption remains valid as long as $\Omega L / U \ll 1$, thereby strengthening our analysis.

\begin{table}[h]
\centering
   \begin{tabular}{| l |c c| c c | c c c r |}
     \hline
      & \multicolumn{2}{l|}{\citet{cabrera2022}}  & \multicolumn{2}{c|}{\citet{roy2019}} & \multicolumn{4}{c|}{Direct numerical simulations}\\ 
      & \multicolumn{2}{c|}{$Ar \approx 0.147$}  & \multicolumn{2}{c|}{$Ar \approx 0.76$}  & \multicolumn{2}{c}{$Ar =24$} & \multicolumn{2}{c|}{$Ar = 96$}\\
      & $\chi=8$ & $\chi=16$ & $\chi=20$ & $\chi=100$ & $\chi=2$ & $\chi=4$ & $\chi=2$ & $\chi=4$\\\hline
     max$(\Omega _r L / |\mathbf{U}|)$ & 0.024 & 0.035 & 0.12 & 0.11 & 0.13 & 0.25 & 0.2 & 0.33\\ \hline
   \end{tabular}
\caption{Values of max$(\Omega _r L / |\mathbf{U}|)$ in the experiments of \citep{cabrera2022}, \citep{roy2019} and in the simulation results.}
\label{tab:omega}
 \end{table}

\begin{figure}[h]
\centering
\includegraphics[width=5cm]{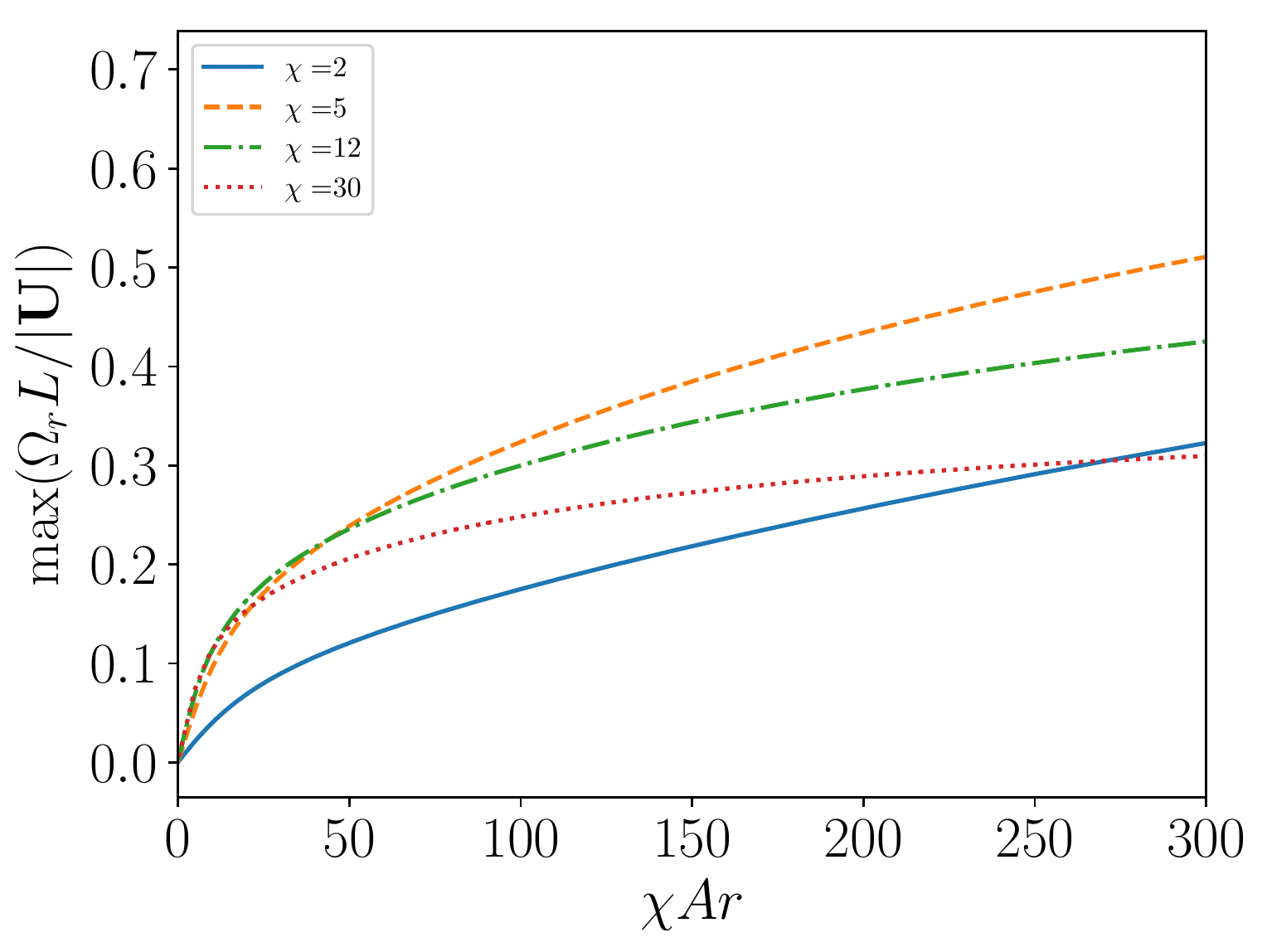} \includegraphics[width=5cm]{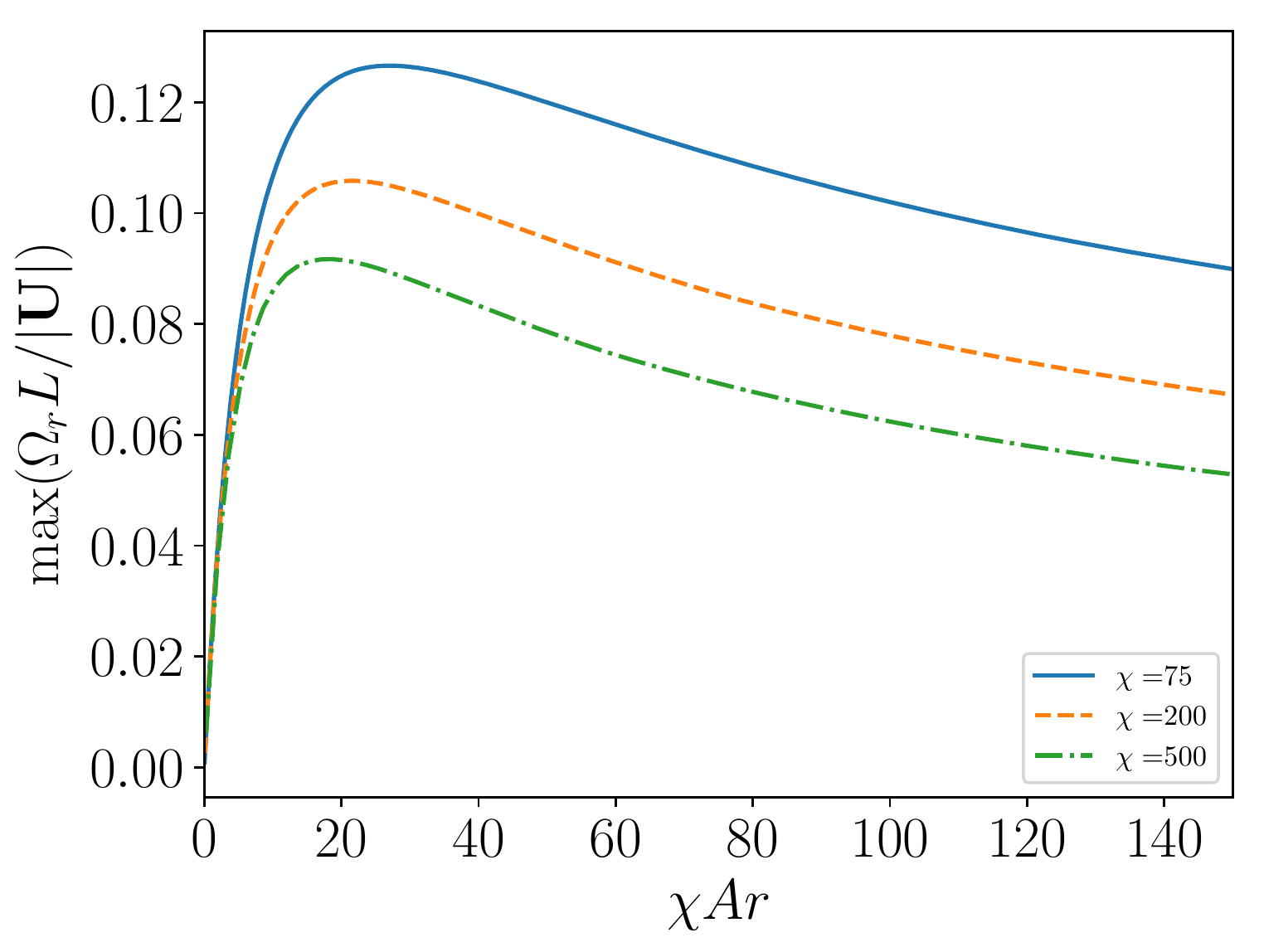}\\
\hspace{0.5cm}$(a)$ \hspace{5.5cm} $(b)$
\caption{Evolution of max($\Omega _r L/|\mathbf{U}|)$ with $\chi Ar$. Equations \ref{eq:UpA**qs} - \ref{eq:OmegarA**qs} are solved using a root-finding algorithm for 20 values of $\phi$ ranging between $0$ and $\pi/2$.}
\label{fig:OmegaLU}
\end{figure}

Secondly, it can be concluded based on the results of the previous section that particle inertia plays no significant role on the magnitude of the sedimenting velocities in the system under investigation. Hence, a correct estimate of $\Omega L / U$ can be obtained by disregarging the particle inertia in the equations of motion and considering the system made of equations \ref{eq:UpA**qs} - \ref{eq:OmegarA**qs}. We have solved this system for $\chi \in [2;500]$ and $Ar \in [0.001;150]$ by using the semi-empirical expression for the loads proposed by \citet{fintzi2023} for $\chi\leq 30$ and \citet{khayat1989} theory for $\chi > 30$. In each of the run we have computed the maximum value of $\Omega _r L / |\mathbf{U}|$ (Figure \ref{fig:OmegaLU}).




For $Ar \chi \leq 10$ except for the smallest aspect ratio $\Omega L/ U $ scales as $Ar\chi$ in agreement with our scaling analysis. For larger $\chi Ar$ the rate of increase of $\Omega L/U$ decreases strongly, especially for $\chi \geq 12$. Indeeed it can be noted that qualitatively different behavior may be observed for moderate inertial effects ($Re_L\sim 1$) and very large aspect ratios ($\chi \gg 1$). \citet{fintzi2023} observed that for $\chi = 30$, the inertial torque scales as $T_i /(\mu U L^2) \sim Re_L^{1/3}$, yielding $T_i \sim \rho^{1/3} \mu^{2/3}U^{4/3}L^{7/3}$. By balancing this inertial torque with the resistive torque, we obtain $\Omega \sim \rho^{1/3}\mu^{-1/3}U^{4/3}L^{-2/3}$, and $\Omega L/U \sim Re_L^{1/3}$. Using the same scaling as in section \ref{sec:scaling} for the velocity, we obtain $\Omega L/U \sim Ar^{1/3}\chi^{1/3}$. Therefore, the rate of increase of $\Omega L/U$ is slower as the Reynolds number increases for elongated particle. This trend is even more pronounced for very long fibers (Figure \ref{fig:OmegaLU} (b)). In this regime, \citet{khayat1989} demonstrated that the inertial torque decreases with the Reynolds number for $Re_L \geq 4$. However, the validity of their theory for such high Reynolds numbers may be questionable, as their asymptotic solution requires $Re_L \ll \ln(\chi)$. Nonetheless, simulations performed by \citet{khair2018} and \citet{shin2006} indicate that the solutions proposed by \citet{khayat1989} remain valid for $Re_L \approx 10$ and $Re_L \approx 5$, respectively, for the longitudinal force on a long spheroid aligned with the flow direction and the torque on a long fiber, with $\chi = 100$. Moreover, \citet{khayat1989} models are in very good agreement with \citet{roy2019} experimental results for $\chi=100$ and $Re_L \approx 7.6$. Consequently, the solution provided by \citet{khayat1989} is considered to provide quantitatively accurate results up to $Re_L \approx 10$ as long as the fiber is adequately elongated ($\chi \geq 100$). If we consider that the quasi-steady assumption fails for values of $\Omega L/U$ larger than $0.2$, then we can expect the quasi-steady models to be accurate for $\chi Ar \approx 200$ if $\chi=2$, and for $\chi Ar \approx 40$ if $2 < \chi \leq 30$. However, for more elongated fibers, the quasi-steady assumption should remain valid as long as the underlying assumptions made in the derivation of \citet{khayat1989} models are fullfilled, particularly if $Re\ll 1$.


This paper also focuses on the underdamped regime and its characteristics, such as oscillation period and amplitude decrease rate, in comparison to simplified models. It was found that the most simple model \textit{i. e.} the underdamped oscillator qualitatively reproduces numerical results but not quantitatively.  This may have significant implications for atmospheric flows where particle orientation is driven by this solution\citep{gustavsson2019,gustavsson2021}. To improve the model, particle inertia could be included in the parallel velocity equation, resulting in a third-order linear ordinary differential equation that can be easily solved. A natural perspective to this model might be to study rods wake instability, like fluttering motion \citep{toupoint2019}. However, in such high inertial flow, there is no reason for the quasi-steady assumption to remain valid \citet{fabre2011}. Moreover, even under the quasi-steady limit, the loads on such bodies at high Reynolds numbers are only known for very few configurations \citep{pierson2019,kharrouba2021}.

\section{Acknowledgements}
ANR MUSCATs financial support is greatly appreciated. We thank Bernhard Mehlig for stimulating discussions and for pointing out the former studies on the underdamped oscillator. The author is indebted to Greg Voth for providing the experimental data of \citet{roy2019}. We also thank Jacques Magnaudet for fruitful discussions on the added mass torque.

\appendix

\section{Added mass coefficients}
\label{app:added}
\begin{figure}[h!]
    \centering
        \includegraphics[height=0.22\textwidth]{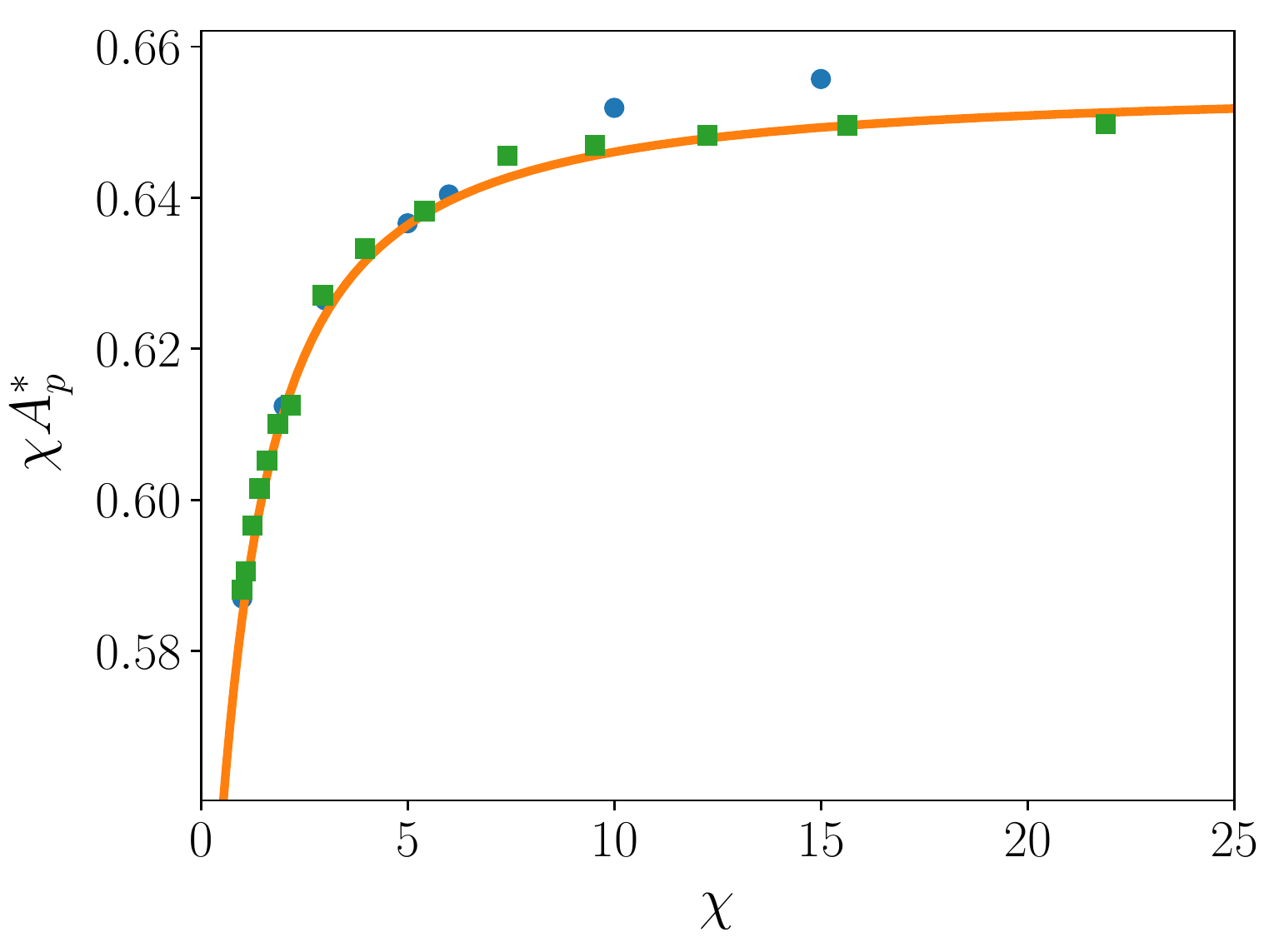}
        \includegraphics[height=0.22\textwidth]{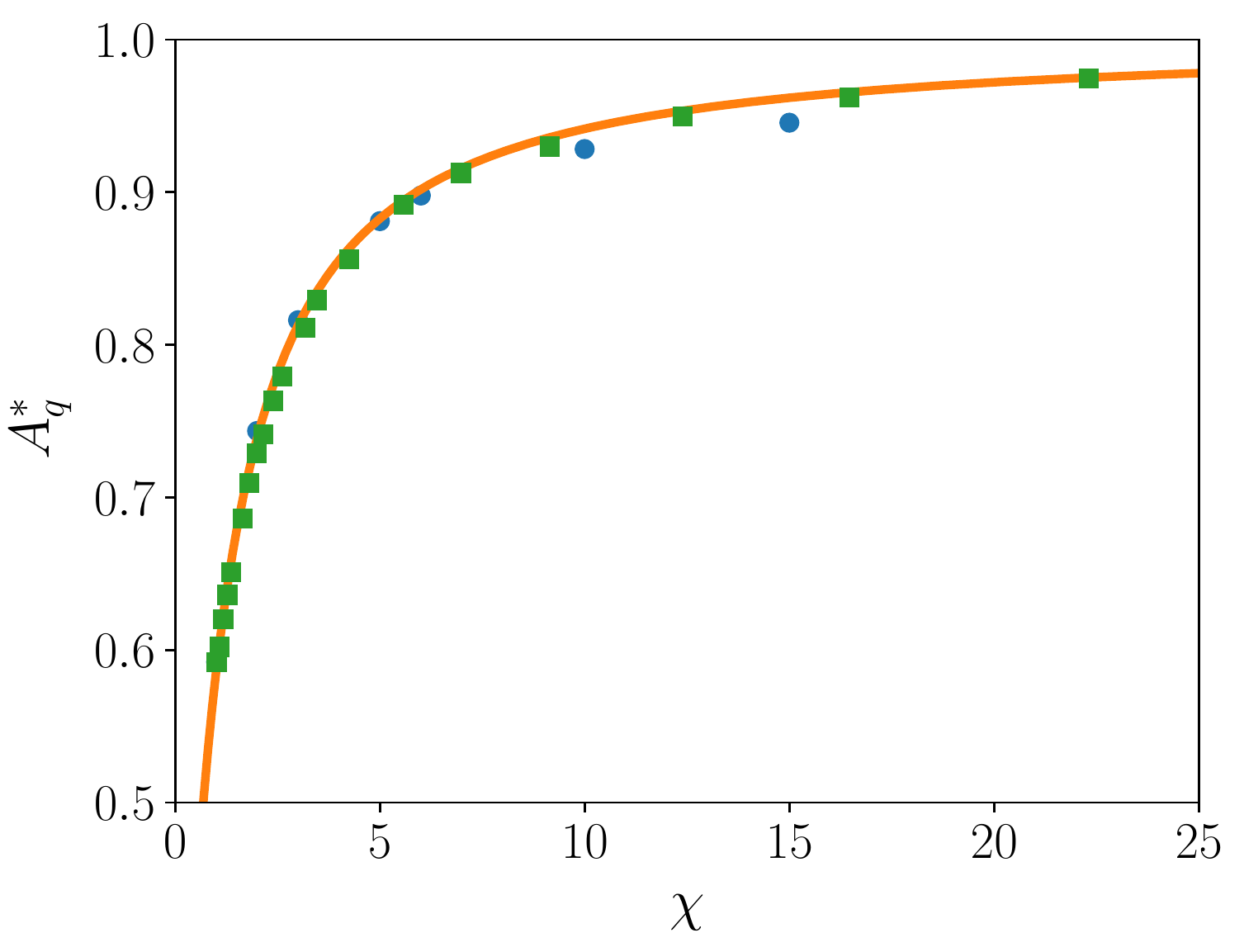}
        \includegraphics[height=0.22\textwidth]{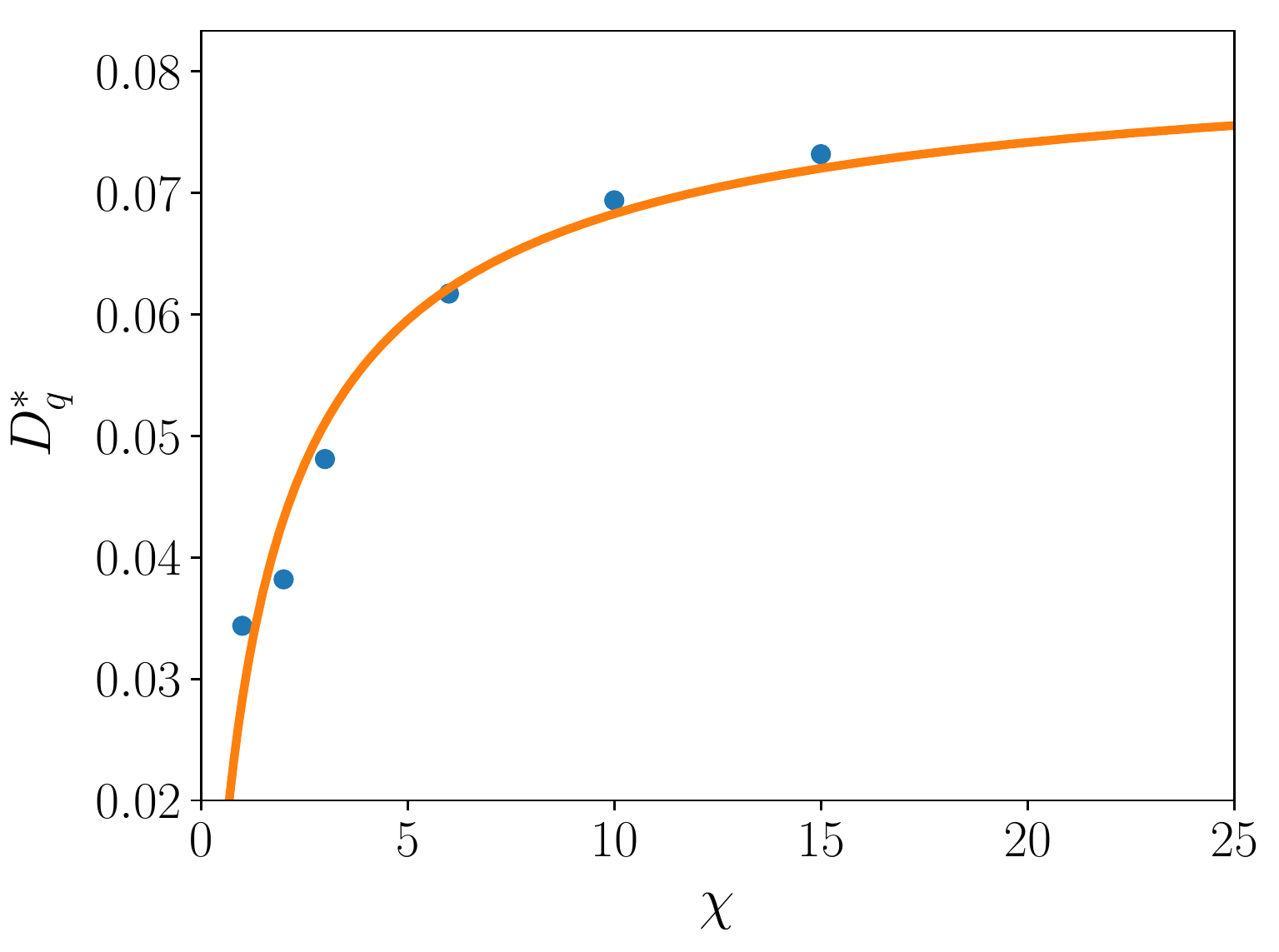}
    \caption{Dimensionless added mass coefficients as a function of $\chi$. ($\blacksquare$) : \citet{loewenberg1993} results, $\bullet$ : present results based on JADIM code \citep{kharrouba2020}.}
    \label{fig:added}
\end{figure}
In this appendix, we compute the coefficient $A_p$, $A_q$ and $D_q$ by using the JADIM code for $ 1 \leq \chi \leq 15$. The numerical details concerning the code as well as the mesh properties can be found elsewhere \citep{kharrouba2020,kharrouba2021,pierson2021}. At the time $t=0$ we impose a constant linear or angular acceleration to a cylinder initially at rest. Since in the very short time limit viscous and rotational contributions are negligible in comparison to potential flow contribution \citep{mougin2002} one can easily recover the added mass coefficients by computing the loads on the body. Figures \ref{fig:added} display the dimensionless added mass coefficient $A_p^* = A_p / (\rho V), A_q^*= A_q / (\rho V)$ and $D_q^* = D_q/ (\rho VL^2)$ as a function of the aspect ratio. A good agreement is observed between the present results and \citet{loewenberg1993} results obtained using potential flow calculations. In the limit $\chi \gg 1$, $\chi A_p^*$ and $A_q^*$ tend toward a constant value. For the particular case $A_q^*$ this constant is simply $1$, $\textit{i.e.}$ the added mass coefficient on an infinitely long cylinder perpendicular to the flow direction. As a result, we propose the following correlation


\begin{equation}
A_p^* = \frac{1}{\chi}\left(0.655-\frac{0.141}{1+\chi^{1.17}}\right),
\end{equation}
\begin{equation}
A_q^* = 1-\frac{0.828}{1+\chi^{1.12}}.
\end{equation}

To the best of our knowledge, the functional dependency of $D_q^*$ with respect to $\chi$ has not been published yet in the literature at least for $\chi \geq 1$. Figure \ref{fig:added} shows that in the limit $\chi \gg 1$ $D_q^*$ tends toward a constant value which is \textit{a priori} unknown. However, in the case of a Rankine ovoid, for $\chi \gg 1$, $D_q^*\approx 1/12$ \citet{howe2006}. Since in the limit of large aspect ratio one may assume that the rounded ends of the Rankine ovoid have little effect on the added mass coefficient, we propose the following empirical correlation  





\begin{equation}
D_q^* = \frac{1}{12}-\frac{0.11}{(1+\chi^{0.8})},
\end{equation}
which matches closely the numerical results.

\section{Short-time asymptotic expansion}
\label{app:short}
We calculate the leading-order terms of the short-time asymptotic expansion. Since $\cos(\phi^{(0)}+\epsilon \phi^{(1)}) \sim \cos\phi^{(0)}-\epsilon \phi^{(1)}\sin\phi^{(0)}$ and $\sin(\phi^{(0)}+\epsilon \phi^{(1)}) \sim \sin\phi^{(0)}+\epsilon \phi^{(1)}\cos\phi^{(0)}$ at zero-th order equations \ref{eq:Upvd}, \ref{eq:Uqvd}, \ref{eq:Omegarvd} and \ref{eq:Phivd} simplifies to
\begin{align}
 \frac{d U_p ^{*(0)}}{dt^*}
    &= \mathcal{A}\cos \phi^{(0)},\\
\frac{d U_q ^{*(0)}}{dt^*}
    &= -\mathcal{B}\sin \phi^{(0)},\\
    \frac{d\Omega _r^{*(0)}}{dt^*}
    &= -\mathcal{C}U_p^{*(0)}U_q^{*(0)} ,\\
\frac{d \phi^{(0)}}{dt^*} &=0.
\end{align}
 The zero-th order solution is easily obtained and reads 
\begin{align}
U_p ^{*(0)} &=   \mathcal{A}t^*\cos \phi^{(0)}, \\
U_q ^{*(0)} &=   -\mathcal{B}t^*\sin \phi^{(0)},\\
\Omega _r^{*(0)}
    &= \frac{\mathcal{A}\mathcal{B}\mathcal{C}}{3}t^{*3}\cos \phi^{(0)} \sin \phi^{(0)},\\
\phi^{(0)}&=\phi(t^*=0),
\end{align}
where we have assumed that the cylinder starts from rest. At first order equations \ref{eq:Upvd}, \ref{eq:Uqvd}, \ref{eq:Omegarvd} and \ref{eq:Phivd} give

\begin{align}
\frac{d U_p ^{*(1)}}{dt^*} 
    &= \frac{\mathcal{A}}{\mathcal{B}}\Omega _r ^{*(0)} U_q ^{*(0)}-\mathcal{A}\phi^{(1)}\sin\phi^{(0)}, \\
\frac{d U_q ^{*(1)}}{dt^*}    
    &=-\frac{\mathcal{B}}{\mathcal{A}}\Omega _r ^{*(0)} U_p ^{*(0)} - \mathcal{B}\phi^{(1)}\cos\phi^{(0)}, \\
    \frac{d\Omega _r^{*(1)}}{dt^*}
    &= 0 ,\\
\frac{d \phi^{(1)}}{dt^*} &=\Omega _r^{*(0)}, 
\end{align}
which leads to

\begin{align}
U_p ^{*(1)}
    &= -\frac{1}{12}\mathcal{A}^2\mathcal{B}\mathcal{C}t^{*5}\cos \phi^{(0)} \sin ^2\phi^{(0)}, \\
U_q ^{*(1)}    
    &=- \frac{1}{12}\mathcal{A}\mathcal{B}^2\mathcal{C}t^{*5}\cos ^2\phi^{(0)} \sin \phi^{(0)}, \\
\Omega _r^{*(1)}
    &= 0,\\
\phi^{(1)}&=\frac{\mathcal{A}\mathcal{B}\mathcal{C}}{12}t^{*4}\cos \phi^{(0)} \sin \phi^{(0)}.
\end{align}


\section{Quasi-steady loads}
\label{app:qsloads}


The expression of the quasi-steady loads used in this paper can be found below.

\subsection{Moderately long rods : $\chi \leq 30$}
\paragraph{Expression of $F_p$ : }
The expression of $F_p$ reads \citep{fintzi2023}
\begin{align}
F_p(Re_L^*,\chi,\theta) = -2\pi\mu |\mathbf{U}| L \cos \theta &\left(\frac{A_{Re=0}^{(1)}+A^{(1)}(Re_L^*)}{\ln(2\chi)}+\frac{A_{Re=0}^{(2)}+A^{(2)}(Re_L^*)}{\ln^2(2\chi)}+\frac{A_{Re=0}^{(3)}+A^{(3)}(Re_L^*)}{\ln^3(2\chi)} \right.\nonumber \\
    &\left. +\frac{A_{Re=0}^{(4)}+A^{(4)}(Re_L^*)}{\ln^4(2\chi)}+\frac{2.34}{\chi^{2/3}(\chi-\frac{1}{2})^{1.75}}\right),
    \label{eq:F_0all}
\end{align}
where $A_{Re=0}^{(1)} = 1$, $A_{Re=0}^{(2)} \approx 0.807$, $A_{Re=0}^{(3)} \approx 0.829$, $A_{Re=0}^{(4)} \approx 1.45$ \citep{kharrouba2021}. The first order inertial correction is null, $A^{(1)}(Re_L^*) = 0$ , while the second, third and fourth order inertial functions read
\begin{align}
    A^{(2)}(Re_L^*) &= \frac{1}{2}\left(\frac{E_1(2Re_L^*)+\ln(2Re_L^*)-e^{-2Re_L^*}+\gamma+1}{2Re_L^*}+E_1(2Re_L^*)+\ln(2Re_L^*)+\gamma-2 \right),\\
    A^{(3)}(Re_L^*) &= A_{A}^{(3)}(Re_L^*)+A_{B}^{(3)}(Re_L^*) +2A^{(2)}(Re_L^*)\ln{(2)},\\
    A^{(4)}(Re_L^*) &= 3\ln{(2)}\left(A_A^{(3)}(Re_L^*)+A_B^{(3)}(Re_L^*)\right) +3A^{(2)}(Re_L^*)\ln{(2)}^2   -0.636 Re_L^{*0.762},
\end{align}
where $\gamma$ is the Euler constant, $E_1(x) = \int_x^\infty \frac{e^{-t}}{t}dt$ the exponential integral function and here $Re_L^* = \rho |\mathbf{U}| L /(2\mu)$. One may observe that $|\mathbf{U}| \cos \theta = U_p$.

\paragraph{Expression of $F_q$ : }

The expression of $F_q$ can be found in \citep{fintzi2023} and reads
\begin{align}
    F_q(Re_L^*,\chi,\theta) = 4\pi\mu |\mathbf{U}| L \sin \theta &\left(\frac{B_{Re=0}^{(1)}+B^{(1)}(Re_L^*)}{\ln(2\chi)}+\frac{B_{Re=0}^{(2)}+B^{(2)}(Re_L^*)}{\ln^2(2\chi)}+ \frac{ B_{Re=0}^{(3)}+B^{(3)}(Re_L^*)}{\ln^3(2\chi)} \right.\\
&+\left.\frac{B_{Re=0}^{(4)}+B^{(4)}(Re_L^*)}{\ln^4(2\chi)} -\frac{0.568}{\chi^{2/3}(\chi-\frac{1}{2})^{1.75}}\right).
    \label{eq:FAllchiT90ReFit}
\end{align}
where $B_{Re=0}^{(1)} = 1$, $B_{Re=0}^{(2)} \approx -0.193$, $B_{Re=0}^{(3)} \approx 0.214$, $B_{Re=0}^{(4)} \approx 0.387$ \citet{kharrouba2021}.
$B^{(1)}(Re_L^*) = 0$ and 
\begin{align}
    B^{(2)}(Re_L^*) &= E_1\left(Re_L^*\right) +\ln{\left(Re_L^*\right)} - \frac{e^{-Re_L^*}-1}{Re_L^*} +\gamma-1\\,
    B^{(3)}(Re_L^*) &=  2\ln{(2)}B^{(2)}(Re_L^*) + B_e^{(3)}(Re_L^*),\\
    B^{(4)}(Re_L^*) &=  3\ln{(2)}^2B^{(2)}(Re_L^*) + 3 \ln{(2)} B_e^{(3)}(Re_L^*) + B_e^{(4)}(Re_L^*). 
\end{align} 
We have $|\mathbf{U}| \sin \theta = -U_q$.

\paragraph{Expression of $T_r^i$ : }
The expression of $T_r^i$ can be found in \citep{fintzi2023}
\begin{align}
    T_r^i(Re_L^*,\chi,\theta) = 
    \rho |\mathbf{U}|^2L^3\sin{(2\theta)}\frac{5\pi}{48(1+Re_L^{*1.991})^{0.331}}
    &\left(
        \frac{1}{\ln^2(3\chi)}
        +\frac{2.244 -1.813Re_L^{*0.543}}{\ln^3(3\chi)}
    \right.\\
    &\left.
        -\frac{3.603 + 8.854Re_L^{*0.538}}{\ln^4(3\chi)}
        -\frac{14.301(Re_L^*/\chi)^{0.448}}{\ln^5(3\chi)}
    \right) 
    \label{eq:torque_final}.
\end{align}
One may note that $|\mathbf{U}|^2\sin{(2\theta)} = -2U_pU_q$.
\paragraph{Expression of $T_r^\Omega$ : }

An expression for the resisting torque due to the particle rotation can be found in \citet{pierson2021} and may be expressed as
\begin{align}
T_r^\Omega=&-\frac{-\pi\mu\Omega _r L^3}{3}\left[\frac{1}{\ln(2\chi)}+\frac{1}{\ln^2(2\chi)}\left(\frac{11}{6}-\ln 2 + f(\chi,,Re_\Omega^*)\right)+\frac{1}{\ln^3(2\chi)}\left(\frac{161}{36}- \frac{\pi ^2}{12}-\frac{11}{3}\ln 2+(\ln 2)^2\right)\right. \nonumber\\
                             &\left. +\frac{1}{\ln^4(2\chi)}\left(1-\frac{1}{(2\chi)^{1.2}}\right)^5 \left(- \frac{5}{4}\zeta (3)+\frac{1033}{72}-\ln ^3 (2)+\frac{11}{2}\ln ^2(2) - \frac{161}{12}\ln 2-\pi ^2 \left(\frac{11}{24}-\frac{1}{4} \ln 2\right)\right)\right],
\label{eq:sl4_bism_re}
\end{align}
\noindent with $f(\chi,Re_\Omega^*) = 0.018\chi^{2.3}Re_\Omega^{*0.9}$ and $Re_\Omega^{*} = \rho |\boldsymbol{\Omega}| D^2 /\mu$.

\subsection{Long rods : $\chi > 30$}
For sufficiently long rods \citet{khayat1989} expressions for the loads are accurate. In the following we present the \citet{khayat1989} expression still making use of the linearization proposed by \citep{lopez2017} for the forces. Also we make use of the $1/\ln\chi$ original expansion proposed by \citet{khayat1989} rather than the $1/\ln(2\chi)$ expansion.  

\paragraph{Expression of $F_p$ : }
\begin{equation}
F_p(Re_L^*,\chi,\theta) = \frac{-2\pi\mu |\mathbf{U}| L \cos \theta}{\ln\chi} \left(1-\frac{A^{(2)}(Re_L^*)-4\ln 2+3}{\ln\chi}\right)^{-1}.
\label{eq:F_pKC}
\end{equation}

\paragraph{Expression of $F_q$ : }
\begin{equation}
F_q(Re_L^*,\chi,\theta) = \frac{4\pi\mu |\mathbf{U}| L \sin \theta}{\ln\chi} \left(1-\frac{B^{(2)}(Re_L^*)+1/2-\ln 4}{\ln\chi}\right)^{-1}.
\label{eq:F_qKC}
\end{equation}

\paragraph{Expression of $T_r^i$ : }
\begin{align}
    T_r^i(Re_L^*,\chi,\theta) = - \mu U L^2 \frac{\pi}{2}  \left(\frac{1}{\ln \chi}\right)^2 \left[\cos\theta \left(P(X) - Q(X) + P(Y) -Q(Y)\right) \right.\left.  + P(Y)- P(X)\right]\sin\theta
    \label{eq:torque_final},
\end{align}

with $Q(x) = \frac{E_1(x)+\ln(x)+\gamma}{x}$, $P(x) = \frac{2}{x}\left(1+\frac{e^{-x}-1}{x}\right)$, $X  = Re_L^* \left(1-\cos\theta\right)$ and $Y  = Re_L^* \left(1+\cos\theta\right)$.
\paragraph{Expression of $T_r^\Omega$ : }
Since to the best of our knowledge there is no expression for this torque for finite inertia effect and very long fibers we make use of expression \ref{eq:sl4_bism_re} neglecting the inertial correction $f$.

\bibliography{biblio}
\end{document}